\documentclass[12pt]{oxarticle}
\pdfoutput=1
\usepackage{amsmath, amssymb}
\usepackage{graphicx, array, float}
\usepackage{xspace}
\usepackage{ifthen}
\newboolean{usehyperref}
\setboolean{usehyperref}{true}
\ifthenelse{\boolean{usehyperref}}{
  \usepackage{hypbmsec}
  \usepackage[sort&compress]{natbib}
  \bibpunct{[}{]}{,}{a}{}{;} 
  \usepackage[plainpages=false]{hyperref}
  \newcommand{\fref}[1]{\autoref{#1}}
  \newcommand{\tref}[1]{\autoref{#1}}
  \newcommand{\sref}[1]{\autoref{#1}}
  \usepackage{twoopt}
  \newcommandtwoopt{\sectionhyp}[3][][]{\section[#1][#2]{#3}}
  \newcommandtwoopt{\subsectionhyp}[3][][]{\subsection[#1][#2]{#3}}
}{
  \newcommand{\fref}[1]{Figure~\ref{#1}}
  \newcommand{\tref}[1]{Table~\ref{#1}}
  \newcommand{\sref}[1]{Section~\ref{#1}}
  \newcommand{\sectionhyp}[1][]{\sectionhyprelay}
  \newcommand{\sectionhyprelay}[1][]{\sectionhyprelayrelay}
  \newcommand{\sectionhyprelayrelay}[1]{\section{#1}}
  \newcommand{\subsectionhyp}[1][]{\subsectionhyprelay}
  \newcommand{\subsectionhyprelay}[1][]{\subsectionhyprelayrelay}
  \newcommand{\subsectionhyprelayrelay}[1]{\subsection{#1}}
}
\let\SS=\S 
\renewcommand{\a}{\alpha}
\renewcommand{\b}{\beta}
\newcommand{\g}{\gamma}\newcommand{\G}{\Gamma}
\renewcommand{\d}{\delta}\newcommand{\D}{\Delta}
\newcommand{\e}{\epsilon}
\newcommand{\z}{\zeta}
\renewcommand{\th}{\theta}
\newcommand{\io}{\iota}

\renewcommand{\l}{\lambda}
\newcommand{\m}{\mu}
\newcommand{\n}{\nu}

\newcommand{\p}{\pi}
\renewcommand{\r}{\rho}
\newcommand{\s}{\sigma}\renewcommand{\S}{\Sigma}
\renewcommand{\t}{\tau}

\newcommand{\Ph}{\Phi}
\newcommand{\ch}{\chi}

\renewcommand{\o}{\omega}

\newcommand{\cD}{\mathcal{D}}

\newcommand{\cF}{\mathcal{F}}

\newcommand{\cO}{\mathcal{O}}

\newcommand{\cS}{\mathcal{S}}
\newcommand{\cT}{\mathcal{T}}

\newcommand{\cW}{\mathcal{W}}

\newcommand{\IC}{\mathbb{C}}

\newcommand{\IP}{\mathbb{P}}

\newcommand{\IZ}{\mathbb{Z}}
\font\twentyfourrm=cmr12 at 24pt
\font\eightrm=cmr8 at 8pt
\font\tenrm=cmr10 at 10pt
\font\csc=cmcsc10
\newcommand{\beq}{\begin{equation}}
\newcommand{\eeq}{\end{equation}}
\newcommand{\bea}{\begin{eqnarray}}
\newcommand{\eea}{\end{eqnarray}}
\newcommand{\bean}{\begin{eqnarray*}}
\newcommand{\eean}{\end{eqnarray*}}
\newcommand{\eref}[1]{(\ref{#1})}
\newcommand{\comment}[1]{}
\newcommand{\pd}[2]{\frac{\partial #1}{\partial #2}}

\newcommand{\+}{\phantom{-}}
\newcommand{\id}{{\bf 1}}
\font\frak eufm10 at 12pt
\font\fourteenfrak eufm10 at 14pt
\font\ninefrak eufm10 at 9pt
\newcommand{\euf}[1]{\text{\frak #1}}
\newcommand{\goth}{\euf{G}}
\newcommand{\biggoth}{\text{\fourteenfrak G}}
\newcommand{\smallgoth}{\text{\ninefrak G}}
\newcommand{\YG}{Y_{\hskip-1pt\smallgoth}}
\newcommand{\hatYG}{\widehat{Y}_{\hskip-1pt\smallgoth}}

\newcommand{\cy}{Calabi--Yau\xspace}
\newcommand{\cym}{Calabi--Yau manifold\xspace}
\newcommand{\cys}{Calabi--Yau manifolds\xspace}
\newcommand{\hodgenos}{(h^{11},\,h^{21})}
\newcommand{\quotient}[1]{_{\hskip-2pt\lower1pt\hbox{$/$}\lower2pt\hbox{\hskip-1pt$#1$}}}
\newcommand{\symm}[1]{_{\hskip-3pt\lower3pt\hbox{$\left\{#1\right\}$}}}

\newcommand{\cicystop}{~\lower8pt\hbox{.}}
\newcommand{\Bigcheck}{\lower3.8pt\hbox{\smash{\hbox{{\twentyfourrm \v{}}}}}}
\newcommand{\Xcheck}{\kern0pt\hbox{\Bigcheck\kern-12.5pt{$X$}}}
\newcommand{\cicy}[2]{\begin{matrix} #1\end{matrix}\!\left[\begin{matrix}#2 \end{matrix}\right]}
\newcommand{\ii}{\text{i}} 
\newcommand{\one}{{\hskip-0.75pt\bf 1}} 
\newcommand{\sfG}{\text{\sffamily G}}
\newcommand{\sfone}{\text{\sffamily 1}}
\newcommand{\spin}{\text{Spin}} 
\newcommand{\dP}{\text{dP}} 
\newcommand{\pt}{\text{pt}} 
\newcommand{\HG}{\euf{H}{\ltimes}\goth}                                       
\def\place#1#2#3{\vbox to0pt{\kern-\parskip\kern-7pt
                             \kern-#2truein\hbox{\kern#1truein #3}
                             \vss}\nointerlineskip}
\renewcommand{\baselinestretch}{1.1}
\numberwithin{equation}{section}
\proofmodefalse
\begin{document}
\begin{titlepage}
  \vspace*{-2cm}
  \hfill
  \vspace*{\stretch1}
  \begin{center}
     \Huge 
     A Three\,-\,Generation Calabi-Yau Manifold\\[5pt]
     with Small Hodge Numbers
  \end{center}
  \vspace*{5mm}
  \begin{center}
    \begin{minipage}{\textwidth}
      \begin{center}
        \csc 
        Volker Braun$^1$, 
        Philip Candelas$^2$,\\
        and\\
        Rhys Davies$^3$\\[5ex]
        \it
        ${}^1$Dublin Institute for Advanced Studies\hphantom{${}^1$}\\
        10 Burlington Road\\
        Dublin 4, Ireland\\[3ex]
      %
        ${}^2$Mathematical Institute\hphantom{${}^2$}\\
        University of Oxford\\
        24-29 St.\ Giles', Oxford OX1 3LB, UK\\[3.5ex]
      %
        ${}^3$Rudolf Peierls Centre for Theoretical Physics\hphantom{${}^3$}\\
        University of Oxford\\
        1 Keble Road, Oxford OX1 4NP, UK\\
      \end{center}
    \end{minipage}
  \end{center}
  \vfill
\begin{abstract}\vskip-5pt\noindent
We present a complete intersection Calabi-Yau manifold $Y$ that has Euler number $-72$ and which admits free actions by two groups of automorphisms of order $12$. These are the cyclic group $\IZ_{12}$ and the non-Abelian dicyclic group $\text{Dic}_3$. The quotient manifolds have $\chi=-6$ and Hodge numbers
$\hodgenos=(1,4)$. With the standard embedding of the spin connection in the gauge group, $Y$ gives rise to an $E_6$ gauge theory with 3~chiral generations of particles. The gauge group may be broken further by means of the Hosotani mechanism combined with continuous deformation of the background gauge field. For the non-Abelian quotient we obtain a model with 3 generations with the gauge group broken to that of the standard model. Moreover there is a limit in which the quotients develop 3 conifold points. These singularities may be resolved simultaneously to give another manifold with 
$\hodgenos=(2,2)$ that lies right at the tip of the distribution of \cys. This strongly suggests that there is a heterotic vacuum for this manifold that derives from the 3 generation model on the quotient of $Y$. The manifold $Y$ may also be realised as a hypersurface in the toric variety. The symmetry group does not act torically, nevertheless we are able to identify the mirror of the quotient manifold by adapting the construction~of~Batyrev.
\end{abstract}
\end{titlepage}
\pagestyle{empty}
{ 
  \setlength{\parskip}{2pt}
  \tableofcontents
}
\newpage
\setcounter{page}{1}
\pagestyle{plain}
\section{Introduction}
Here we report the existence of a \cym $Y$ with Euler number $-72$ that
admits freely acting symmetry groups of order $12$. Thus the quotients
are smooth and have Euler number $-6$. The two symmetry
groups\footnote{In the following, we will denote the cyclic group of order
$n$ as $\IZ_n$.} are $\IZ_{12}$ and a non-Abelian group $\goth$ which
is isomorphic to the dicyclic group\footnote{There are three non-Abelian groups of order 12. These are 
$\text{Dih}_6$, the group of symmetries of a regular hexagon, the alternating group $A_4$, which is also the symmetry group of a regular tetrahedron, and the  group $\text{Dic}_3$ which is neither of the foregoing. We will describe this group in detail in \sref{sec:Ymanifold}.}  $\text{Dic}_3$ and for both these groups
the quotient manifold has Hodge numbers $\hodgenos=(1,4)$. With the
standard embedding of the spin connection in the gauge group, the
quotient manifolds correspond to string theory vacua with gauge
symmetry $E_6$ and 3 net chiral generations of particles. 

Two avenues are open to break the gauge symmetry further. The first is the Hosotani 
mechanism~\cite{Hosotani1}, whereby one assigns vacuum expectation values to Wilson line
operators corresponding to homotopically non-trivial paths on the quotient manifold. The resulting gauge group is then the subgroup of $E_6$ that commutes with these Wilson line operators. The other avenue is to choose a heterotic vacuum by deforming the structure group of the background gauge field away from the $SU(3)$ that is provided by the standard embedding of the spin connection in the gauge group. Neither mechanism, on its own, is able to break the symmetry down to the group
$SU(3){\times}SU(2){\times}U(1)$ of the standard model \cite{WittenSBP,McInnes:1989rg}. However by combining the two mechanisms, for the quotient with non-Abelian fundamental group $\goth$, the gauge group may be reduced to that of the standard~model. The way this comes about is that we can start with the standard embedding, with $3$
net chiral families of $E_6$. The Hosotani mechanism can break the gauge group in basically two ways to either 
$SU(3){\times}SU(2){\times}U(1){\times}U(1)'$ or to $SU(4){\times}SU(2){\times}U(1)$, the reduction in the rank of the gauge group being due to the fact that the fundamental group of the quotient is non-Abelian.  To break the remaining unwanted gauge symmetry it is necessary to give a VEV to some standard model
singlet field(s), a process which we understand to being equivalent to deforming the background gauge field. This process is tightly constrained precisely because we are dealing with a manifold with few parameters. For the deformation to be possible requires the existence of a vector-like pair of such fields.  It transpires that these exist for the case of 
$SU(4){\times}SU(2){\times}U(1)$, where we can give a VEV to a field transforming in 
the~$({\bf 4},{\bf 1})_\frac{1}{2}$ representation of the gauge group.  This leads to a unique choice of Wilson lines, and hence massless spectrum.  If, however, we first break to 
$SU(3){\times}SU(2){\times}U(1){\times}U(1)'$ we find, despite the fact that this group appears more promising at the outset, that the deformations required to give the appropriate VEV's to the exotic standard model singlets do not, in fact,~exist. 

Heterotic vacua are of great interest in and of themselves. However we find it interesting also that, having started with a vacuum corresponding to the standard embedding of the spin connection in the gauge group, we have been led to deform to a heterotic vacuum. This point takes perhaps additional significance from the fact that the manifold $\YG$ lies almost at the tip of the distribution of \cys and is closely related to a manifold with $\hodgenos=(2,2)$ which occupies a remarkable position at the very tip of the distribution, in a way that we pause to~explain.

A \cym is partially characterised by the Hodge numbers $\hodgenos$. These are topological numbers which also count the number of parameters that deform the K\"{a}hler class and the complex structure of the manifold. For the purposes of the present discussion, let us call the sum $h^{11}+h^{21}$ the {\sl height} of the manifold.
The height of $\YG$ is~5 and, together with its mirror, these are the only manifolds known with this height. The only manifolds known with smaller height all have height 4; with this height manifolds are known with 
$\hodgenos=(1,3),\,(2,2)$ and~$(3,1)$. A notable feature of the model presented here is therefore that three generations are achieved in an economical manner, since the height is the least possible for a `$3$-generation' manifold with the standard embedding. The $3$-generation Yau manifold~\cite{Yau1,Yau2}, by contrast, has 
$\hodgenos=(6,9)$
(for a discussion of the phenomenology of this model see~\cite{Greene:1986bm,Greene:1986jb}), and a manifold presented in~\cite{SHN} has $\hodgenos=(5,8)$ so is little better, in this sense. There is no shortage, as such, of \cys with $|\chi|=6$ since there is also a manifold with Hodge numbers $(10,13)$ and many manifolds with Hodge numbers $(h,h+3)$ and $(h+3,h)$ for $h\geq 15$ that are provided by constructions of manifolds as hypersurfaces in weighted projective spaces and in toric 
varieties~\cite{Candelas:1989hd,Klemm:1992bx,KreuzerSkarkeReflexive}. \cys admitting freely-acting discrete symmetries seem to be rare, and those cases where the group is large seem to be very rare indeed. The remarkable features of the manifold presented here are the related facts that $|\chi|=6$ is achieved by Hodge numbers that are so small and that the fundamental group is large, by the standards of these groups.

The manifold $\YG$ has $4\,(=h^{21})$ parameters that
correspond to its complex structure. The covering manifold $Y$ has symmetries beyond those of $\goth$, these are not freely acting but are nevertheless symmetries, and some of these descend to the quotient $\YG$. For generic values of the complex structure parameters the symmetry is $\IZ_2$ but there are loci in the parameter space where the symmetry is enhanced. There is a two parameter subvariety, 
$\G$, of the parameter space for which the symmetry group is maximally enhanced to the dihedral group 
$\text{Dih}_6$. On $\G$ the generic variety $\YG$ is not smooth but has three conifold points.  
It is a striking fact that these singularities may be simultaneously resolved to give a new manifold $\hatYG$ with 
$\hodgenos=(2,2)$ which is, as we have noted,  currently at the very tip of the distribution. This transition suggests that there is a heterotic model associated with $\hatYG$, with three generations, that derives from the standard embedding of the gauge group in the tangent bundle via transgression \cite{Triadophilia}. The idea being that $\YG$ makes a conifold transition to $\hatYG$ while the bundle makes a `smooth' transition to a bundle on $\hatYG$ that derives from the tangent bundle on $\YG$. Thus we are led from a 3-generation model that derives from the standard embedding on $\YG$ to a 3-generation heterotic model on the 
manifold~$\hatYG$.
\begin{figure}[H]
\begin{center}
\includegraphics[width=6.5in]{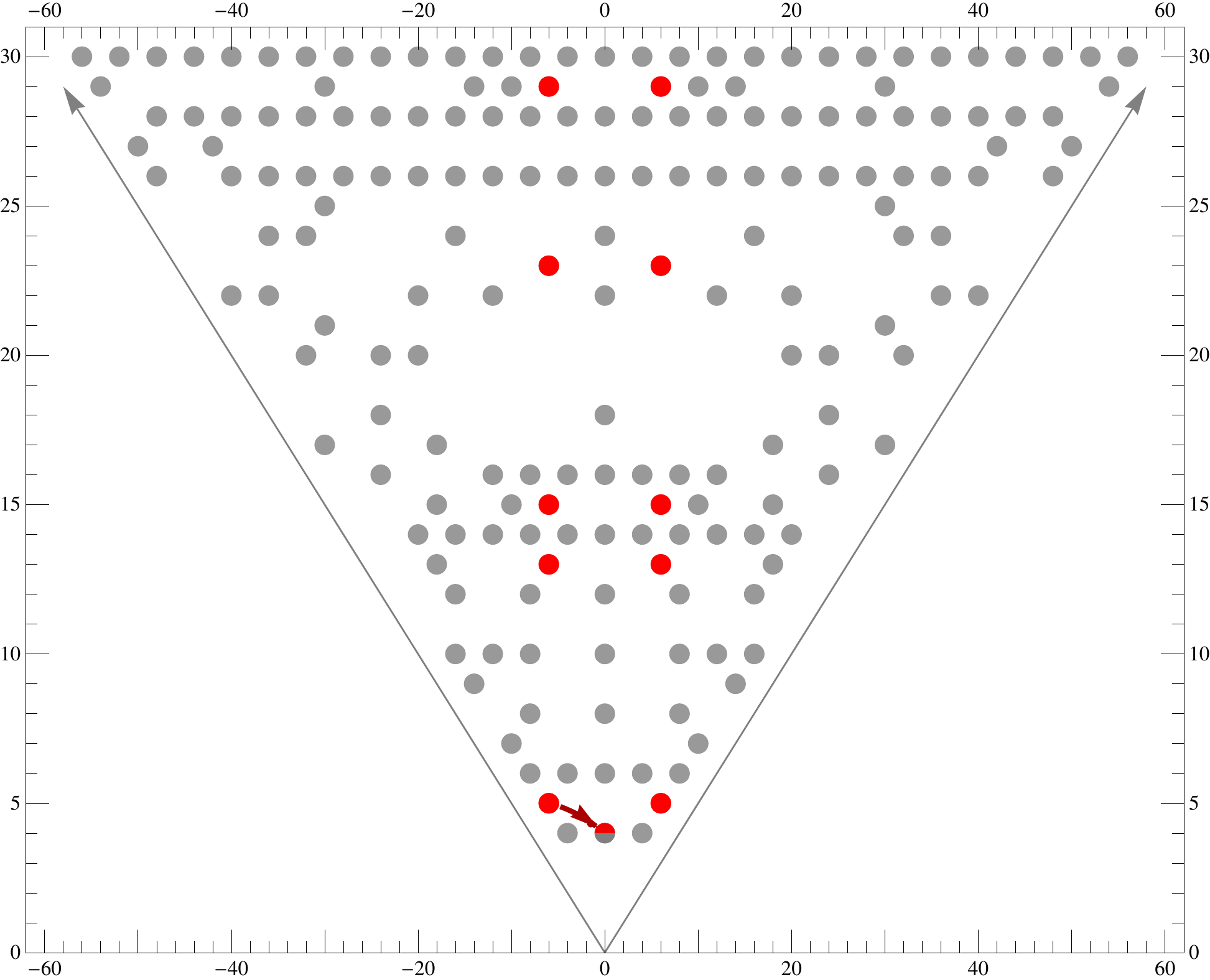}
\parbox{5.5in}{\caption{\label{fig:TransgressionBig}\small\it
The very tip of the distribution of \cys showing the manifolds that have $h^{11}{+}h^{21}\leq 30$. The Euler number \hbox{$\ch=2(h^{11}{-}h^{21})$} is plotted horizontally, the height $h^{11}{+}h^{21}$ is plotted vertically and the oblique axes bound the region $h^{11}\geq 0,\, h^{21}\geq 0$. The quotients $Y/\goth$ with $|\goth|=12$, that we discuss here, have $\hodgenos=(1,4)$ these and their corresponding mirrors are shown.
The conifold transition between the manifolds with $\hodgenos=(1,4)$ and $\hodgenos=(2,2)$ is indicated by the red arrow. Manifolds with $|\ch|=6$ are distinguished by red dots and there are many of these for 
$h^{11}+h^{21}>30$. The manifolds indicated in this plot with $h^{11}{+}h^{21}\leq 24$ are identified in \cite[Table 9]{SHN}.}}
\end{center}
\end{figure}
The  transgression that we propose is shown in \fref{fig:TransgressionBig} with an arrow. The figure illustrates the fact that the manifolds that we are discussing lie at the very tip of the distribution. 

There are many aspects of the phenomenology of these models that we do not discuss here. Notably absent is a discussion of the Yukawa couplings; however it is scarcely possible to discuss these without first achieving some understanding of the parameter space of the model and of the nature of the manifold that is the mirror to $\YG$ and this is the aim of the present work. A discussion of the Yukawa couplings and the quantum corrections to the picture presented here is a subject to which we hope to return.

The manifold $Y$ can be represented as a CICY (complete intersection \cym) in several ways, and for
two of these the group actions can be realised as linear transformations
of the embedding space.  The corresponding configuration matrices are
\begin{equation}
  \cicy{\IP^2\\ \IP^2\\ \IP^2\\ \IP^2}{  
    \one & \one  & \one & 0      & 0 \\
    0 & 0      & \one & \one & \one \\
    \one & \one  & \one & 0     & 0 \\
    0 & 0       & \one & \one & \one}~~,
\qquad\qquad
  \cicy{\IP^1\\ \IP^1\\ \IP^1\\ \IP^1\\ \IP^1\\ \IP^1\\}
  {\one&0&\one \\
    \one&0&\one \\
    \one&0&\one \\
    0&\one&\one \\
    0&\one&\one \\
    0&\one&\one}
\label{twoconfigs}
\end{equation}
The reader unfamiliar with the notation is referred to \sref{sec:nonabelianquotient} for an explanation. It~suffices here to recall that the statement that $Y$ corresponds to the first configuration is the statement that 
$Y$ can be realised in $(\IP^2)^4$ as the complete intersection of 5 polynomials whose multidegrees, in the homogeneous coordinates of the four $\IP^2$'s, are given by the columns of the matrix. The same manifold can also be realised in $(\IP^1)^6$ as the complete intersection of three polynomials with the stated multidegrees.
These configurations give us a very explicit description of the manifold $Y$ in terms of polynomials. 

Compactifications on \cys typically lead to a large numbers of massless neutral scalars in the low energy theory that are associated with the deformations of the gauge bundle on the manifold.  In this case we begin
with the gauge bundle equal to the tangent bundle, and have a heuristic argument based on the fact that the configuration matrices that describe $Y$ are all multilinear that suggests that the tangent bundle deforms with few parameters.  We can at least show that there are no deformations of the tangent bundle that arise as deformations of the differentials of the defining polynomials.

If one thinks carefully about the configurations \eqref{twoconfigs}, another picture of $Y$ emerges.  The manifold can be realised as a hypersurface, that is by a single equation, in a space $\cS{\times}\cS$ where $\cS$ is the twofold defined by two bilinear equations in 
$\IP^2{\times}\IP^2$ or a single trilinear equation in $\IP^1{\times}\IP^1{\times}\IP^1$. 
The surface specified in this way is the del~Pezzo\footnote{Note that we will be following the mathematical
  tradition and label the del Pezzo surfaces $\dP_d$ by their degree
  $d$. The blow-up of $\IP^2$ at $n$ generic points is a
  del Pezzo surface of degree $d=9-n$.} surface $\dP_6$.
The manifold $Y$ is then an anti-canonical hypersurface inside the fourfold $\cS{\times}\cS$. The surface 
$\cS=\dP_6$ is toric so we
find ourselves within the general framework of toric hypersurfaces and reflexive 
polyhedra \cite{Batyrev:1994hm}.
It turns out that the symmetries of $Y$ are also best understood as descending from symmetries of $\dP_6$.

For a review of constructions of \cys see~\cite{Triadophilia,hubsch},
which includes a description of the construction of CICY
manifolds. The construction of this class of manifolds was inspired by
its best known member, which is Yau's three-generation
manifold. A list of almost $8,000$ manifolds was
constructed in this way~\cite{Candelas:1987kf}.  Although a rather
special class of manifolds (the Euler numbers, for example, all fall
in the range $-200\leq \chi\leq 0$) these provide many
examples of \cys that admit freely acting symmetries. The list was
searched~\cite{Candelas:1987du,Aspinwall:1987cn} for manifolds that
admit a freely acting symmetry such that the quotient has $\chi=-6$
and in these searches, to the chagrin of one of the present authors,
the manifold presented here, which occurs $3$ times in the list
of~\cite{Candelas:1987kf} (two of these presentations are the configurations displayed above, the third is a hybrid of the two), was wrongly
rejected. The manifold has subsequently lain, largely unremarked, in
the list for more than 20 years.  In~\cite{Candelas:1987du} one of the
presentations of this manifold was recognised as admitting a symmetry group
$\IZ_3$. More recently~\cite{SHN} it was recognised that the symmetry
group could be promoted to $\IZ_6$. The groups of order $12$ finally
came to light in the course of a recent project to classify all the
freely acting symmetries for the manifolds of the CICY list. It
transpires that, apart from the covering manifold of Yau's
three-generation manifold, the manifold presented here is the only one
to admit a smooth quotient with~$\chi=-6$.

The layout of the paper is as follows. In \sref{sec:nonabelianquotient} we describe the covering manifold, the
free group action by the group $\text{Dic}_3$, and the relation to the del Pezzo surface $\dP_6$.  
In \sref{sec:phenom} we use this information to examine possible gauge symmetry breaking patterns and the resulting 4D theories.  We turn in \sref{sec:extrasymmetries} to explore the existence of additional discrete symmetries of the covering manifold which are not freely acting but nevertheless play a role in the low energy theory. Our main interest here is to identify the symmetries that survive as symmetries of the quotient manifold. One conclusion is that the quotient manifolds have a $\IZ_2$-symmetry for all values of the complex structure parameters. Additional symmetries arise at various loci in the parameter space. The symmetry group of the quotient is a subgroup of $\text{Dih}_6$ (the group corresponding to the symmetries of a regular hexagon) and is another non-Abelian group of order 12. There is a 2-dimensional subspace of the parameter space where the symmetry group of the quotient is the full group $\text{Dih}_6$. The corresponding quotient varieties are, however, all singular and, generically, have three conifold points. These three-nodal varieties are especially interesting since they offer the possibility of transgression to heterotic models with 3 generations based on manifolds with $\hodgenos=(2,2)$ as discussed~above.

We take up in detail the issue of the 3-nodal varieties with symmetry $\text{Dih}_6$ in 
\sref{sec:transgression}. The nodes of the quotient derive from a very beautiful configuration of 36 nodes on the covering space. The singularities of a nodal variety may be resolved locally to give a smooth complex manifold but it is not generally the case that the nodes may be resolved such that the resulting manifold is K\"ahler. In the present case however there is a configuration of non-Cartier divisors that contain the nodes and by blowing up along these divisors we are able to show that there is a \cy resolution in this case.
In \sref{sec:toric} we use the formalism of toric geometry to describe $\YG$ and its mirror, and demonstrate that, rather unexpectedly, the mirror also admits an action by the group $\goth$. We are fortunate to be able to do this since, although the covering space $Y$ is a hypersurface in a toric variety the symmetries of $\goth$ do not act torically. Nevertheless, the structure of the reflexive polyhedra associated with $Y$ is such that we are able to identify the mirror manifold of $\YG$.

While our main interest is with the case of the quotient by the non-Abelian group we present in
\sref{sec:abelianquotient} a brief account of the analysis for the $\IZ_{12}$ quotient. 
Two appendices deal with an alternative presentation of the covering manifold and a demonstration that the tangent bundle of $Y$ does not have any deformations that correspond to deformations of the differentials of the defining polynomials. 

An account of the computer search for free
group actions on CICY's and their classification will be presented elsewhere.
\newpage
\sectionhyp
[{\boldmath A Manifold with a $\chi=-6$ Quotient}]
[A Manifold with a chi=-6 Quotient]
{\boldmath A Manifold with a $\chi=-6$ Quotient}
\label{sec:nonabelianquotient}
\subsectionhyp
[The manifold $Y^{8,44}$]
[The manifold Y(8,44)]
{\boldmath The manifold $Y^{8,44}$}
\label{sec:Ymanifold}
The manifold that we shall denote by $Y^{8,44}$ is specified by the configuration
\begin{equation*}
  Y^{8,44}~=~~
  \cicy{\IP^2\\ \IP^2\\ \IP^2\\ \IP^2}{  
    \one & \one  & \one & 0      & 0 \\
    0 & 0      & \one & \one & \one \\
    \one & \one  & \one & 0     & 0 \\
    0 & 0       & \one & \one & \one}^{8,44}_{-72}
  \hskip0.6in
  \vcenter{
    \hbox{\parbox{2.2in}{
        \vskip0.16in
        \includegraphics[width=2.2in]{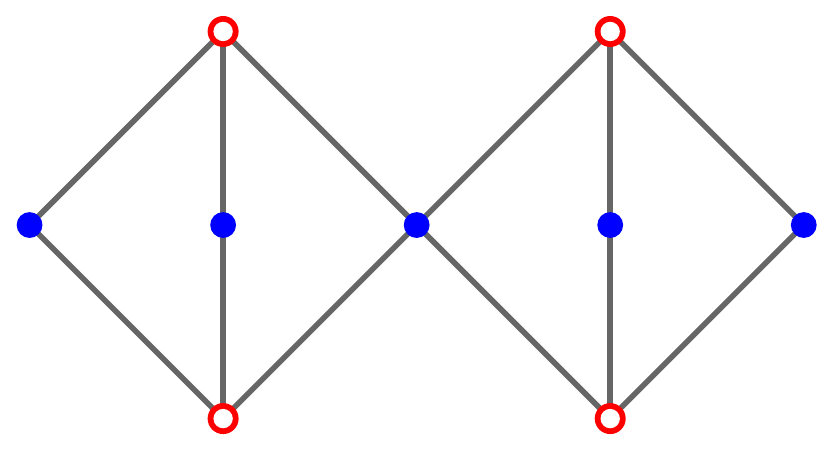}
        \vskip-0.16in
        \place{0.55}{1.25}{$x_{1j}$}
        \place{0.55}{-0.05}{$x_{3j}$}
        \place{1.55}{1.25}{$x_{2j}$}
        \place{1.55}{-0.05}{$x_{4j}$}
        \place{-0.15}{0.67}{$p^1$}
        \place{0.4}{0.67}{$q^1$}
        \place{1.08}{0.77}{$r$}
        \place{1.7}{0.67}{$q^2$}
        \place{2.23}{0.67}{$p^2$}
        \vskip0.1in
      }}
  }
\end{equation*}
The manifold is defined by 5 polynomials, that we denote by
$p^1,\,p^2,\, q^1,\,q^2$, and $r$ that act in $(\IP^2)^4$. We take
coordinates $x_{\a j}$ for the four $\IP^2$, where the indices $\a$
and $j$ take values in $\IZ_4$ and $\IZ_3$, respectively. The columns
of the matrix correspond to the degrees of the polynomials in the
coordinates of each space, in the order $\{p^1,\,q^1,\, r,\,
q^2,\,p^2\}$. The diagram on the right encodes the same information
and shows how the polynomials, represented by the blue dots, depend on
the variables of the four $\IP^2$'s, which correspond to the open red
dots. The fact that the dots are all connected by single lines in the
diagram corresponds to the fact that the polynomials are all
multilinear.
We begin by seeking equations that are covariant under the cyclic
permutation of the four $\IP^2$'s,
\begin{equation*}
g_4:~x_{\a j}~\to~x_{\a+1,j}~,~~p^1\leftrightarrow p^2~,~~q^1\leftrightarrow q^2~,~~r~\to~r~.
\end{equation*}
For $w_j$ and $z_k$ homogeneous coordinates on $\IP^2{\times}\IP^2$ define bilinear polynomials $p$ and $q$ by
\begin{equation*}
p(w,z)~=~\sum_{jk} A_{jk}\, w_j z_k~,\qquad q(w,z)~=~\sum_{jk} B_{jk}\, w_j z_k~,
\end{equation*}
where the coefficient matrices $A_{jk}$ and $B_{jk}$ are symmetric. Define also $g_4$-invariant polynomials
\begin{equation*}
m_{ijk\ell}~=~\frac{1}{4}\,\sum_\a x_{\a,i}\, x_{\a+1,j}\, x_{\a+2,k}\,  x_{\a+3,\ell}~.
\end{equation*}
In terms of these quantities we may take defining polynomials, for $Y$, of the form
\begin{equation*}
  \begin{split}
    p^1~&=~p(x_1,x_3)~,\qquad p^2~=~p(x_2,x_4)\\[5pt]
    q^1~&=~q(x_1,x_3)~,\qquad q^2~=~q(x_2,x_4)\\[5pt]
    &\hskip30pt r~=~\sum_{ijk\ell}C_{ijk\ell}\, m_{ijk\ell}~.
  \end{split}
\end{equation*}
Note that the quantities $m_{ijk\ell}$ are cyclically symmetric in their indices so, in the definition of $r$, we take the sum over indices $i,j,k,\ell$ to run over combinations that are identified up to cyclic permutation and the coefficients $C_{ijk\ell}$ to be cyclically symmetric.
Consider now a second symmetry
\begin{equation*}
  g_3:~
  x_{\a j}~\to~\z^{(-1)^\a j}\, x_{\a j}
  ,\quad
  p^1~\to~p^1
  ,\quad
  p^2~\to~p^2
  ,\quad
  q^1~\to~\z\,q^1
  ,\quad
  q^2~\to~\z^2\,q^2
  ,
\end{equation*}
where $\z$ is a nontrivial cube root of unity. Covariance under $g_3$ restricts the coefficients that can appear in the defining polynomials. We see that
\begin{equation*}
  \begin{split}
    A_{jk}~&=~0~~\text{unless}~~j+k\equiv 0 \mod3\\[3pt]
    B_{jk}~&=~0~~\text{unless}~~j+k\equiv 2 \mod3\\[3pt]
    C_{ijk\ell}~&=~0~~\text{unless}~~~i+k\equiv j+\ell \mod3~.
  \end{split}
\end{equation*}
Thus, removing overall scales, $p$ and $q$ are of the form
\begin{equation*}
  p(w,z)~=~w_0 z_0 + a\,(w_1 z_2 + w_2 z_1)~,~~~q(w,z)~=~w_1 z_1 +
  b\,(w_0 z_2 + w_2 z_0)
  ,
\end{equation*}
while $r$ is a linear combination of $9$ of the $m_{ijk\ell}$
\begin{equation}
  \label{eq:rlong}
  \begin{split}
    r~=~
    c_0\,m_{0000}& \,+\, c_1\,m_{1111} \,+\, c_2\,m_{2222} \,+\,
    c_3\,m_{0011} \,+\, c_4\,m_{0212} \\  
    &\,+\, c_5\,m_{0022} \,+\, c_6\,m_{1122} \,+\, c_7\,m_{0102} \,+\,
    c_8\,m_{0121} 
    .
  \end{split}
\end{equation}
The freedom to redefine the coordinates $x_{\a j}$ is restricted by the action of $g_3$ and $g_4$. The remaining freedom allows only a two-parameter redefinition
\begin{equation}
(x_{\a 0},\,x_{\a 1},\,x_{\a 2})~\to~(x_{\a 0},\,\l\, x_{\a 1},\,\m\, x_{\a 2})
\label{lammutransf}
\end{equation} 
which we may use to set the constants $a$ and $b$ that appear in $p$ and $q$ to~unity.
We may also redefine $r$ by multiples of the other polynomials. Let $\tilde p$ and $\tilde q$ be generic polynomials with the same degrees and covariance properties as $p$ and $q$, that is
\begin{equation*}
\tilde{p}(w,z)~=~ a_1 w_0 z_0 + a_2 (w_1 z_2 + w_2 z_1)~,~~~
\tilde{q}(w,z)~=~ b_1 w_1 z_1 + b_2 (w_0 z_2 + w_2 z_0)~.
\end{equation*}
These provide a four-parameter freedom to redefine $r$:
\begin{equation*}
r~\to~r + \Big(\tilde{p}(x_1,x_3)\, p(x_2,x_4) + p(x_1,x_3)\, \tilde{p}(x_2,x_4)\Big)
+ \Big(\tilde{q}(x_1,x_3)\, q(x_2,x_4) + q(x_1,x_3)\, \tilde{q}(x_2,x_4)\Big)~.
\end{equation*}
The coefficients $c_2,\,c_3$ and $c_4$ are unaffected by this process but we may use this freedom to eliminate the terms in eq.~\eqref{eq:rlong} with coefficients  $c_5,\,c_6,\, c_7$ and $c_8$, say. An overall scale is of no consequence so the resulting equation has four parameters. We summarise the present form of the polynomials as
\begin{equation} 
  \label{eq:finalpolys}
  \begin{split}
    p^1~&=~x_{10}\,x_{30} + x_{11}\,x_{32} + x_{12}\,x_{31}~,\qquad
    q^1~=~x_{11}\,x_{31} + x_{10}\,x_{32} + x_{12}\,x_{30}~,\\[8pt]
    p^2~&=~x_{20}\,x_{40} + x_{21}\,x_{42} + x_{22}\,x_{41}~,\qquad
    q^2~=~x_{21}\,x_{41} + x_{20}\,x_{42} + x_{22}\,x_{40}~,\\[8pt]
    &\hskip20pt r~=~ c_0\,m_{0000} + c_1\,m_{1111} + c_2\,m_{2222} + c_3\,m_{0011} + c_4\,m_{0212}~.
\end{split}
\end{equation}
Consider again the coordinate transformations \eref{lammutransf}. We
see that the equations $p^1,\,p^2,\,q^1$ and $q^2$ are invariant
provided $\l\m=1$ and $\m^3=1$. We take, therefore, $\l=\z$ and
\hbox{$\m=\z^2$}. The polynomial $r$, however is not invariant since,
under the transformation,
$m_{ijk\ell}\to\z^{i+j+k+\ell}\,m_{ijk\ell}$. The effect is equivalent
to changing the coefficients. In this way we see that there is a
$\IZ_3$-action on the coefficients, and that we should identify
\begin{equation}
(c_0,\, c_1,\,c_2,\,c_3,\, c_4)~\simeq~(c_0,\, \z\,c_1,\,\z^2 c_2,\,\z^2 c_3,\, \z^2 c_4)~.
\label{eq:Z3symmetry}\end{equation}
Returning to the symmetries $g_4$ and $g_3$: these generate a group that we shall denote by $\goth$. Note that
\begin{equation*}
g_4\, g_3 = g_3^2\, g_4~,
\end{equation*}
so the group is non-Abelian. This relation, however, permits the enumeration of the elements of the group as 
$g_3^m\,g_4^n$, $0\leq m\leq 2$, $0\leq n\leq 3$. Thus $\goth$ has order 12 and is isomorphic to 
the dicyclic group $\text{Dic}_3$. A fact that will be useful shortly is that the element $g_6=g_3^2 g_4^2$ generates a $\IZ_6$ subgroup of $\goth$ and that the elements of $\goth$ may also be enumerated as 
$g_6^m\, g_4^n$ with $0\leq m\leq 5$ and $n=0$ or 1.
The commutation relation above may be expressed as $g_4 g_3 g_4^{-1} = g_3^2$, so the group contains $\IZ_3$ as a normal subgroup, and can be thought of as a semi-direct product 
$\IZ_3 \rtimes \IZ_4$.
\vskip15pt
\begin{table}[H]
  \setlength{\doublerulesep}{1pt}
  \def\str{\vrule height14pt depth8pt width0pt}
  \begin{center}
    \newcolumntype{C}{>{$}c<{$}}
    \begin{tabular}{| C || C | C | C | C | C | C | C | C | C | C | C | C |}\hline
     \str g_3^m\, g_4^n & 1 & g_4 & g_4^2 & g_4^3 & g_3 & g_3\, g_4 & g_3\, g_4^2 & g_3\, g_4^3 & g_3^2 
      & g_3^2\, g_4 & g_3^2\, g_4^2 & g_3^2\, g_4^3 \\ \hline
      \str g_6^m\, g_4^n &1 & g_4 & g_6^3 & g_6^3\, g_4 & g_6^2 & g_6^2\, g_4 & g_6^5 & g_6^5\, g_4 
      & g_6^4  & g_6^4\, g_4 & g_6 & g_6\, g_4 \\ \hline
      \str \text{Order} & ~~1~~ & ~~4~~ & ~~2~~ & ~~4~~ & ~~3~~ & ~~4~~ & ~~6~~ & ~~4~~ 
      & ~~3~~ & ~~4~~ & ~~6~~ & ~~4~~\\ \hline
    \end{tabular}
  \parbox{5.7in}{\caption{\label{DicThree}\small\it
    The elements of the group $\text{Dic}_3$ presented in the form $g_3^m\, g_4^n$, for 
    $0\leq m\leq 2$, $0\leq n\leq 3$,
    and $g_6^m\, g_4^n$, for $0\leq m\leq 5$, $n=0,1$, together with the order of
    each element.}}
\end{center}
\end{table}
\vskip-10pt
In order to check for fixed points of $\goth$ note that if an element $h$ has a fixed point then so has $h^m$ for each $m\geq 1$. The order of an element of $\goth$, that is not the identity, must divide the order of the group so can be $2$, $3$, $4$, or $6$. If an element $h$ has order~$4$ then $h^2$ has order~$2$ and if it has order~$6$ then $h^3$ has order~$2$. Hence it is enough to check the elements of order 2 and 3 for fixed points. The only elements of order 3 are $g_3$ and $g_3^2$ and if $g_3^2$ has a fixed point then so has $g_3^4=g_3$. Thus it suffices to check $g_3$ and $g_4^2$, the latter being the unique element of order~$2$. 

A fixed point of $g_3$ is such that $x_{\a j}$ takes one of the values 
$\{(1,0,0),\,(0,1,0),\,(0,0,1)\}$, for each~$\a$. Thus there are $3^4$ fixed points in the embedding space 
$(\IP^2)^4$. It is an easy check to see that these points do not
coincide with simultaneous zeros of the defining polynomials provided
none of the coefficients $c_2$, $c_3$, and $c_4$ vanish.

A~fixed point of $g_4^2$ is such that $x_{1j}=x_{3j}=w_j$ and $x_{2j}=x_{4j}=z_j$ for some $w_j$ and 
$z_j$ that satisfy the equations
\begin{equation*}
  \begin{split}
    p(w,w)~&=~0~,\qquad p(z,z)~=~0\\[3pt]
    q(w,w)~&=~0~,\qquad q(z,z)~=~0\\[3pt]
    &\hskip-5pt r(w,z,w,z)~=~0
  \end{split}
\end{equation*}
and it is easily checked that these five equations do not have a solution in $\IP^2{\times}\IP^2$ for generic values of the parameters. We pause to do this explicitly since the values of the parameters for which there 
{\em are}\/ fixed points will be of interest later. The equations $p(w,w)=q(w,w)=0$ have four solutions for $w$. These are $w=(0,0,1)$ and $w=(1,\o,-\o^2/2)$, for $\o^3=1$, and the solutions for $z$ are the same, giving rise to 16 points in $\IP^2{\times}\IP^2$. The polynomial $r(w,z,w,z)$ does not vanish on any of these points unless at least one of the quantities
\vspace{-4pt}
\beq
c_2~,~~c_2 + 2c_4~,~~~c_0^3  +  c_1^3  + d_2^3 - 3\, c_0 c_1 d_2~,
\vspace{-4pt}
\label{d2eqn}\eeq
where $16 d_2=c_2  + 16 c_3 + 4 c_4$, vanishes.

\begin{figure}[H]
  \parbox{6.5in}{\begin{center}
      \framebox[2.1in][c]{\vrule width0pt height 1.75in depth 0.25in\includegraphics[height=1.5in]{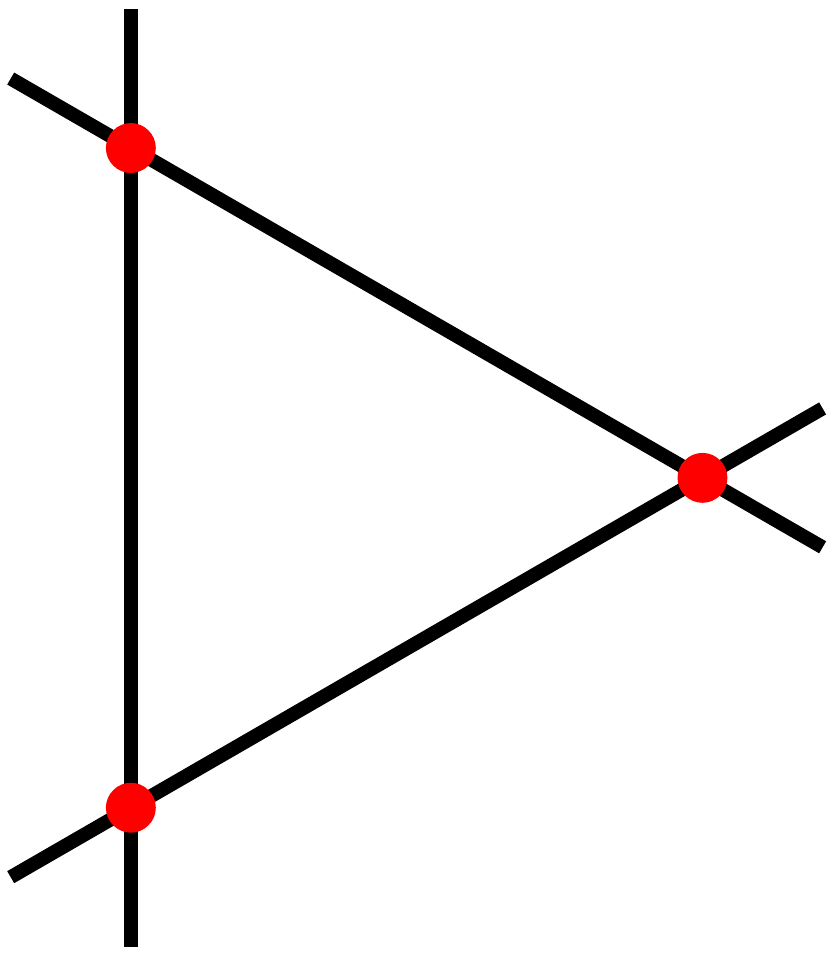}}\hfill
      \framebox[2.1in][c]{\vrule width0pt height 1.75in depth 0.25in\includegraphics[height=1.5in]{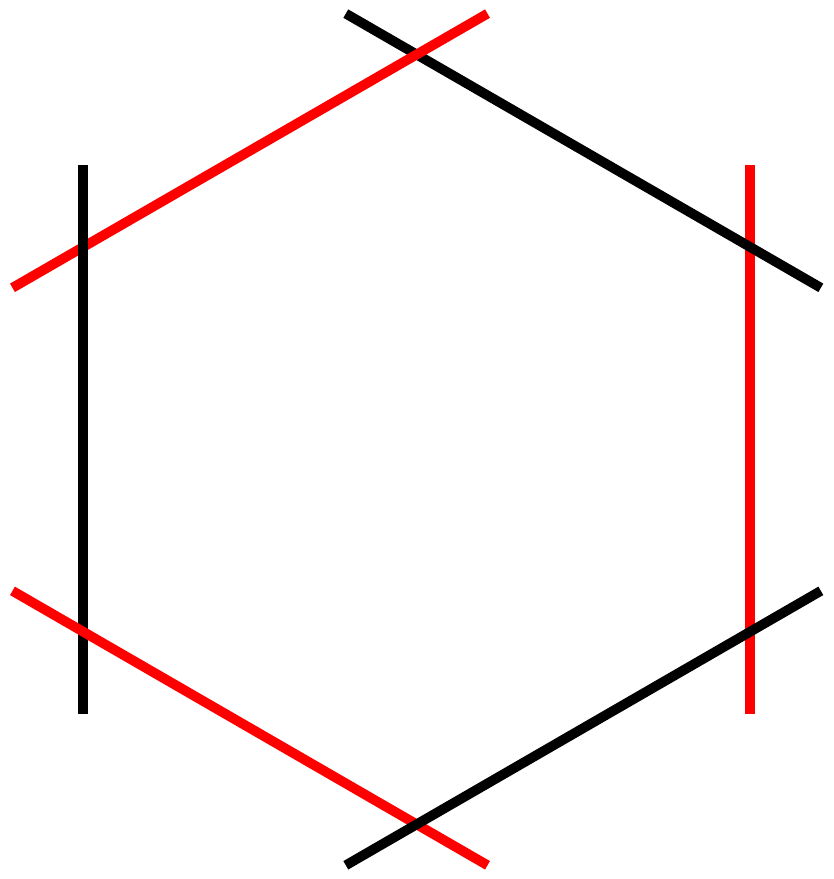}}\hfill
      \framebox[2.1in][c]{\vrule width0pt height 1.75in depth 0.25in\includegraphics[height=1.5in]{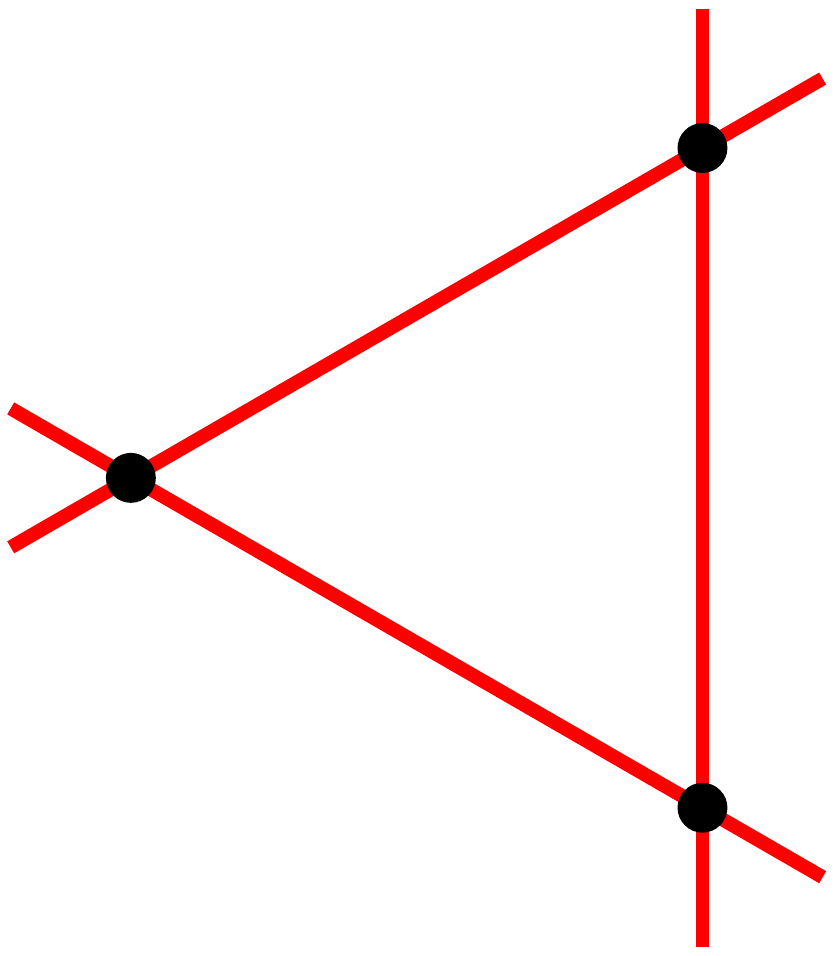}}
    \end{center}}
  \vskip0pt
  \place{0.3}{1.22}{\small $L_{12}$}
  \place{1.1}{0.82}{\small $L_{23}$}
  \place{1.1}{1.6}{\small $L_{31}$}
  \place{2.82}{1.85}{\small $E_1$}
  \place{2.39}{1.22}{\small $L_{12}$}
  \place{2.82}{0.6}{\small $E_2$}
  \place{3.55}{0.6}{\small $L_{23}$}
  \place{3.9}{1.22}{\small $E_3$}
  \place{3.55}{1.85}{\small $L_{31}$}
  \place{5.3}{1.6}{\small $E_1$}
  \place{5.3}{0.82}{\small $E_2$}
  \place{5.98}{1.22}{\small $E_3$}
  \vspace{-20pt}
\begin{center}
\parbox{5.6in}{\caption{\label{fig:dP6}\small\it 
    A sketch of $\dP_6$, in the centre, showing the hexagon
    formed by the six (-1)-lines. This surface may be realised in
    $\IP^2{\times}\IP^2$ as the locus $p(w,z)=q(w,z)=0$ defined by two
    bilinear polynomials in the coordinates $w_j$ and $z_j$ of the two
    $\IP^2$'s. If we project to the first $\IP^2$ then the image is as
    in the sketch on the left, in which the three lines $E_i$ project
    to points. If, instead, we project to the second $\IP^2$ then the
    image is as in the sketch on the right, in which the three lines
    $L_{ij}$ have been projected to points.}}
\end{center}
\end{figure}

We have checked that the polynomials \eref{eq:finalpolys} are transverse following the methods of~\cite{SHN}.  We conclude that there exists a smooth quotient Calabi-Yau manifold, which we will denote by
\vspace{-4pt}
\begin{equation*}
\YG := Y/\goth .
\vspace{-4pt}
\end{equation*}
Now we may regard the manifold $Y^{8,44}$ as a hypersurface in $\cS{\times}\cS$, where $\cS$ is the surface
\vspace{-2pt}
\begin{equation}
\cS~=~\cicy{\IP^2\\ \IP^2}{1&1\\ 1&1}
\vspace{-4pt}
\label{Srepone}\end{equation}
which has Euler number $6$ and is the del Pezzo surface $\dP_6$ obtained by blowing up three points of $\IP^2$ that are in general position. It is instructive to verify this explicitly and to locate the three blown up points by considering the defining equations~\eqref{eq:finalpolys}.  The polynomials $p^1$ and $q^1$, that define the first copy of $\cS$, are $p^1=x_1^T A\, x_3=0$ and $q^1 = x_1^T B\, x_3 = 0$ where
\begin{equation*}
A ~=~ \left( \begin{array}{cccc}
 1 & 0 & 0 \\
 0 & 0 & 1 \\
 0 & 1 & 0 \end{array} \right)
\qquad\text{and}\qquad
B ~=~ \left( \begin{array}{cccc}
 0 & 0 & 1 \\
 0 & 1 & 0 \\
 1 & 0 & 0 \end{array} \right) 
\end{equation*}
Given $x_1 \in \IP^2$, consider the corresponding values of $x_3$
which solve these equations.  For generic values, $x_3$ is determined
up to scale, hence uniquely as a point of $\IP^2$, as the vector
orthogonal to $x_1^T A$ and $x_1^T B$, but at the three points coming
from left eigenvectors of $AB^{-1}$ we have $x_1^T A\propto x_1^T B$,
and there is a whole $\IP^1$ of solutions for $x_3$.  These three
points are $x_1 =(1,1,1),\, (1,\z,\z^2)$ and $(1,\z^2,\z)$, with $\z$
again a non-trivial cube root of unity. The corresponding $\IP^1$'s
are the exceptional curves $E_1,\, E_2,\, E_3$. Also important to us
are the three lines $L_{ij}$ in $\cS$ that correspond to the lines in
the $\IP^2$ that join the points that are blown up to $E_i$ and
$E_j$. The three $L_{ij}$ together with the three $E_i$ form a hexagon
in $\cS$, as sketched in \fref{fig:dP6}. We will see presently that the
order $6$ symmetry $g_6$ acts on this hexagon by~rotation.
\subsectionhyp
[The group representations of $\text{Dic}_3$]
[The group representations of Dic3]
{\boldmath The group representations of $\text{Dic}_3$}
\label{sec:representations}
Before proceeding we pause to describe the representations of the
group $\goth \cong \text{Dic}_3$.  There are four one-dimensional
representations of $\text{Dic}_3$, in which $g_3$ acts trivially and
$g_4$ is multiplication by one of the fourth roots of unity.  We will
denote these, in an obvious notation, by $R_1, R_\ii, R_{-1}$ and
$R_{-\ii}$.  These are the homomorphisms of $\text{Dic}_3$ to its
Abelianisation $\IZ_4$. There are also two distinct two-dimensional
representations, distinguished in a coordinate-invariant way by
$\text{Tr}(g_4^2) = \pm 2$.  These we will denote by $R^{(2)}_\pm$.
For completeness we display the corresponding matrices in a basis for
which $g_3$ is diagonal:
\begin{equation*}
\begin{split}
&R^{(2)}_{+}: \quad g_3 \to \left( \begin{array}{cc}
\z & 0 \\
0 & \z^2 \end{array} \right)\, ,
\quad g_4 \to \left( \begin{array}{cr}
0 & \!\!\+1 \\
1 & 0 \end{array} \right)\, ,\\[5pt]
&R^{(2)}_{-}: \quad g_3 \to \left( \begin{array}{cc}
\z & 0 \\
0 & \z^2 \end{array} \right)\, ,
\quad g_4 \to \left( \begin{array}{cr}
0 & \!\! -1 \\
1 & 0 \end{array} \right)\, .
\end{split}
\end{equation*}
The non-obvious tensor products of representations are as follows
\begin{align*}
R_1 \otimes R^{(2)}_\pm \hskip11pt&=~ R_{-1} \otimes R^{(2)}_\pm ~=~ R^{(2)}_\pm\, ,  &\quad
R_\ii \otimes R^{(2)}_\pm \hskip11pt&=~ R_{-\ii} \otimes R^{(2)}_\pm ~=~ R^{(2)}_\mp \\[8pt]
R^{(2)}_{+} \otimes R^{(2)}_{+} ~&=~ R_1 \oplus R_{-1} \oplus R^{(2)}_{+}\, , &\quad
R^{(2)}_{-} \otimes R^{(2)}_{-} ~&=~ R_1 \oplus R_{-1} \oplus R^{(2)}_{+} \\[10pt]
&\hskip60pt R^{(2)}_{+} \otimes R^{(2)}_{-} ~=~ R_{\ii} \oplus R_{-\ii} \oplus R^{(2)}_{-}\hskip-120pt
\end{align*}
\subsection{Group action on homology} \label{sec:homology}
The cohomology group $H^2(Y^{8,44})$ descends from that of $H^2(\cS{\times}\cS)=H^2(\cS){+}H^2(\cS)$. For each $\cS$ we have $h^{11}(\cS)=4$ and the cohomology group is spanned by (the duals of) the hyperplane class, $H$, and the three  $E_i$.
The intersection numbers of the four classes $\{H, E_1, E_2,E_3\}$ are given by a matrix $\eta$
\vskip-25pt
\begin{equation*}
H\cdot H~=~1~,~~~H\cdot E_i ~=~ 0~,~~~E_i\cdot E_j~=~-\d_{ij}~~~\text{so}~~~
\eta~=~\left(\begin{array}{rrrr}1&0&0&0\\
                                                 0&-1&0&0\\
                                                 0&0&-1&0\\
                                                 0&0&0&-1\end{array}\right)\cicystop
\end{equation*}
By considering the intersection numbers of $L_{ij}$ with $H$ and the $E_k$ one sees that
\begin{equation*}
L_{ij}~=~H - E_i - E_j
\end{equation*}
and that the $L_{ij}$ are also $(-1)$-lines, that is $L_{ij}\cdot L_{ij}=-1$.  Of course, everything said above also applies on the second copy of $\cS$.  To distinguish the cohomology classes coming from this copy, we denote them by $\widetilde{H}, \widetilde{E}_1, \widetilde{E}_2, \widetilde{E}_3$.
\vskip5pt
\begin{table}[H]
\def\str{\vrule height16pt width0pt depth10pt \hskip0.5cm}
\newcolumntype{C}{@{\hskip10pt}>{$}c<{$}@{\hskip10pt}}
\newcolumntype{L}{@{\hskip10pt}>{$}l<{$}@{\hskip10pt}}
\begin{center}
\begin{tabular}{| L | C | C |}
\hline
\vrule height15pt width0pt depth8pt (-1)\text{-line} & w & z \\ 
\hline\hline 
\str E_1      & (1,1,1)                                  & z_0 + z_1 + z_2 = 0             \\ \hline 
\str L_{12} & w_0 + \z w_1 + \z^2 w_2 = 0 & (1, \z^2, \z)                          \\ \hline
\str E_2      & (1, \z, \z^2)                          & z_0 + \z^2 z_1 + \z z_2 = 0  \\ \hline  
\str L_{23} & w_0 + w_1 + w_2 = 0             & (1, 1, 1)                                \\ \hline
\str E_3      & (1, \z^2, \z)                          & z_0 + \z z_1 + \z^2 z_2 = 0  \\ \hline 
\str L_{31} & w_0 + \z^2 w_1 + \z w_2 = 0 & (1, \z, \z^2)                           \\ \hline    
\end{tabular}
\vspace{5pt}
\parbox{3.5in}{\caption{\label{lines}\small\it The equations that define the six $(-1)$-lines in the coordinates $(w_i,\,z_j)$ of $\IP^2{\times}\IP^2$.}}
\end{center}
\end{table}
\vskip-20pt
We know the action of the group generators on the spaces, and this allows us to calculate the induced action on $H^2(\cS{\times}\cS)$.  Choosing the ordered basis 
$\{ H, E_1, E_2, E_3, \widetilde H, \widetilde E_1, \widetilde E_2, \widetilde E_3 \}$, we can write 
$8 {\times} 8$ matrices $U(g)$ representing the action of $g\in\goth$. It is clear that $g_3$ preserves the hyperplane classes and rotates the exceptional curves into each other, so that
\begin{equation*}
U(g_3)~=~ 
\left ( \begin{array}{cc}
\sfG_3 & 0 \\
0 & \sfG_3^2
\end{array} \right )
\quad \text{with} \quad
\sfG_3 =
\left ( \begin{array}{cccc}
1 & 0 & 0 & 0 \\
0 & 0 & 0 & 1 \\
0 & 1 & 0 & 0 \\
0 & 0 & 1 & 0
\end{array} \right )
\end{equation*}
The action of $g_4$ is slightly more complicated to read off.  The coordinates for the two copies of  $\cS$ are 
$(x_{1j},\, x_{3k})$ and $(x_{2j},\, x_{4k})$ and $g_4$ acts by mapping
\begin{equation*}
(x_1,\, x_3)~\to~(x_2,\, x_4)~, \quad \text{while}\quad (x_2,\, x_4)~\to~(x_3,\, x_1)~. 
\end{equation*}
We can think of this action as being an exchange of the two copies of $\cS$ followed by the involution 
$x_1 \leftrightarrow x_3$ on the first copy.  We need to calculate the action of this involution on $H^2(\cS)$.
To this end, choose one of the exceptional curves, say $E_1$, which lies above $(1,1,1)$ in $\IP^2_{x_1}$.  Then it is described in $\IP^2_{x_3}$ by the line $x_{30} + x_{31} + x_{32} = 0$.  On the coordinates, the involution acts as $x_1 \leftrightarrow x_3$, so it maps $E_1$ to the curve described by $x_{10} + x_{11} + x_{12} = 0$ and $(x_{30}, x_{31}, x_{32}) = (1,1,1)$.  Since this line passes through both $(1,\z,\z^2)$ and $(1,\z^2,\z)$, which are the other points that are blown up, we see that it is actually the line referred to earlier as $L_{23}$. Thus the action of the involution is 
$E_1 \leftrightarrow L_{23}$, or more generally $E_i \leftrightarrow L_{i+1,i+2}$.  This is enough information to work out the action of $g_4$ on $H^2(\cS \times \cS)$, with respect to the basis given above.  The result is
\begin{equation*}
U(g_4)~=~ 
\left ( \begin{array}{cc}
0        & \sfG_2 \\
\sfone & 0
\end{array} \right )
\quad \text{with} \quad
\sfG_2 ~=~
\left ( \begin{array}{rrrr}
2 & 1 & 1 & 1 \\
-1 & 0 & -1 & -1 \\
-1 & -1 & 0 & -1 \\
-1 & -1 & -1 & 0
\end{array} \right )
\end{equation*}
It is a quick check that $\sfG_2^2 = \sfG_3^3 = \sfone$ and that $\sfG_2$ and $\sfG_3$ commute and preserve the intersection matrix $\eta$.
It is useful also to express the transformations in terms of the $(-1)$-lines of the two copies of $\cS$, eliminating  the explicit reference to $H$. Denote by $D_a$ and $\widetilde{D}_b$, $a,b\in\IZ_6$, the six $(-1)$-lines on the two copies of $\cS$
\beq
D_a~=~(E_1,\, L_{12},\, E_2,\, L_{23},\, E_3,\, L_{31})~~~\text{and}~~~
\widetilde{D}_b~=~
(\widetilde{E}_1,\, \widetilde{L}_{12},\, \widetilde{E}_2,\, \widetilde{L}_{23},\, \widetilde{E}_3,\, 
\widetilde{L}_{31})~.
\label{DefDs}\eeq
In terms of these the action of the generators $g_6$ and $g_4$ is
\beq
g_6:~D_a{\times}\widetilde{D}_b~\to~D_{a+1}{\times}\widetilde{D}_{b-1}~~~\text{and}~~~
g_4:~D_a{\times}\widetilde{D}_b~\to~D_{b+3}{\times}\widetilde{D}_a~.
\label{GensOnDs}\eeq
If we change our basis for $H^2(\cS \times \cS)$ such that $U(g_3)$ becomes diagonal, we can compare with \sref{sec:representations} and see that the eight-dimensional representation decomposes into the sum
\begin{equation*}
R_1 \oplus R_{-1} \oplus R_{\ii} \oplus R_{-\ii} \oplus R^{(2)}_{+} \oplus R^{(2)}_{+}
\end{equation*}
In particular, there is only a single invariant, implying $b_2 = 1$ for the quotient.  This invariant corresponds to the canonical class, as it must, which we can see explicitly as follows.
The group element $g_6=g_3^2\,g_4^2$ generates a $\IZ_6$ subgroup, as noted previously. We have
\begin{equation*}
U(g_6)~=~\left(\begin{array}{cc} \sfG_6 & 0 \\ 0 & \sfG_6^{-1} \end{array}\right)
\qquad \text{where}\qquad 
\sfG_6~=~\sfG_2\,\sfG_3^2~=~
\left(\begin{array}{rrrr}
  2 &  1 &  1 & 1 \\
 -1 & -1 &  0 & -1 \\
 -1 & -1 & -1 & 0 \\
 -1 &  0 & -1 & -1 
\end{array}\right)
\end{equation*}
which acts separately on the two homology bases. The two canonical classes
\begin{equation*}
K_\cS~=~3H - E_1 - E_2 - E_3\qquad\text{and}\qquad
K_{\widetilde{\cS}} ~=~3\widetilde{H} - \widetilde{E}_1 - \widetilde{E}_2 - \widetilde{E}_3
\end{equation*}
are invariant under $g_6$ and since the eigenvalues of $\sfG_6$ are
$\{1,\,-1,\,\z,\,\z^2\}$ we see that these are the only two invariant homology classes. A class that is invariant under $g_6$ is invariant under $\goth$ if it is also invariant under $g_4$ and we immediately see that the only such invariant combination is $K_\cS + \widetilde{K}_\cS$.
The fact that $b_2 = 1$ for the quotient variety implies $h^{11} = 1$
for the hypersurface $\YG$.  Since the Euler number divides by the order of the group,
$\chi=-72/12=-6$ and it follows that $h^{21}=4$, in agreement with our
count of parameters in the defining polynomials.
\newpage
\section{Rudiments of the Phenomenology} \label{sec:phenom}
Compactification of $E_8{\times}E_8$ heterotic string theory on the manifold $\YG$, with the standard embedding, leads to a 4D effective theory with unbroken gauge group $E_6{\times}E_8$ with four chiral multiplets in the $\bf 27$ of $E_6$ and one in the $\bf\overline{27}$.  We believe there are few $E_6$ singlets, since these would correspond to rank-three deformations of the gauge bundle, and we argue in Appendix 
\ref{sec:deformations} that there are few of these.  We wish to break the $E_6$ gauge symmetry further to achieve, if possible, the gauge group $G_{\text{SM}} = SU(3){\times}SU(2){\times}U(1)$ of the standard model. Given that the unbroken gauge group is the commutant in $E_8$ of the holonomy group of the gauge connection, there are two related mechanisms at our disposal for gauge symmetry breaking.  The first is to continuously deform the internal gauge field, which corresponds to the Higgs mechanism in the 4D theory.  We may take the vector bundle corresponding to the background gauge field to be a deformation of $\cT\!\oplus\cO$ or of $\cT\!\oplus\cO\oplus\cO$, where $\cT$ is the tangent bundle of $\YG$ and $\cO$ is a trivial line bundle. In this way the structure group of the bundle becomes, in many cases, $SU(4)$, for deformations of $\cT\!\oplus\cO$, or $SU(5)$, for the case of $\cT\!\oplus\cO\oplus\cO$. The commutants of these groups in $E_8$ are $\spin(10)$ and $SU(5)$, respectively, which are attractive groups for phenomenology.  In a specific model, however, it requires checking that suitably deformations of the bundles exist in order to break the symmetry as desired.  We will see in \sref{sec:gaugedeform} that on $\YG$ it is possible to obtain $\spin(10)$ but not $SU(5)$ in this way.

The second possibility is to invoke the Hosotani mechanism, whereby we avail ourselves of the fact that $\YG$ is multiply connected to give non-zero values to Wilson lines around homotopically nontrivial paths in $\YG$. This amounts to taking the gauge bundle to be $\cT \oplus \cW$, where $\cW$ is a non-trivial flat bundle.  $\cW$ is specified by choosing a group homomorphism $\Ph: \pi_1\left( \YG\right) \cong \goth \to E_6$ and letting the holonomy of the connection on $\cW$ around a path $\g$ be given by $\Ph([\g])$.  The holonomy group therefore acquires an extra discrete factor $\Ph(\goth)$.  The smallest resulting unbroken gauge group containing $G_{\text{SM}}$ is $SU(3){\times}SU(2){\times}U(1){\times}U(1)$.  In~principle the extraneous $U(1)$ can be broken by a vacuum expectation value for a standard model singlet, corresponding to a continuous deformation of the gauge bundle away from $\cT\oplus\cW$.  In fact in this model all such fields have a $D$-term potential forcing their VEV to be zero.  We therefore consider a less minimal choice of Wilson lines, for which the larger group $SU(4){\times}SU(2){\times}U(1)$ is unbroken.  In this case there are flat directions in the low energy theory along which the group is Higgsed to exactly $G_{\text{SM}}$.

In the following we make extensive use of standard group theory, for which the comprehensive reference is the review by Slansky~\cite{Slansky}.
\subsection{Deforming the gauge bundle} \label{sec:gaugedeform}
We obtain unbroken $E_6$ gauge symmetry by choosing the non-trivial
part of the gauge bundle to be equal to the tangent bundle $\cT$ of the
manifold, with structure group $SU(3)$.  We will now consider, as
discussed above, taking instead a non-trivial deformation of
$\cT \oplus \cO$ or $\cT \oplus \cO \oplus \cO$, with structure group
$SU(4)$ or $SU(5)$, respectively.  Since deformation is a continuous
process, it must correspond to the Higgs mechanism in the low energy
theory, whereby the vector multiplets lying outside the unbroken
sub-algebra gain mass by eating chiral multiplets with the same charges.
The gauge bosons transform in the adjoint representation of $E_6$,
while the families and antifamilies transform in the $\bf 27$ and
$\bf\overline{27}$, respectively.  These representations decompose
under the $\spin(10)$ subgroup as:
\begin{equation*}
{\bf 78}~=~ {\bf 45} \oplus {\bf 16} \oplus \overline{\bf 16} \oplus {\bf 1}~,\qquad
{\bf 27}~=~ {\bf 16} \oplus {\bf 10} \oplus {\bf 1}~,\qquad
\overline{\bf 27}~=~ \overline{\bf 16} \oplus {\bf 10} \oplus {\bf 1}~.
\end{equation*}
So if $E_6$ is Higgsed to $\spin(10)$, the vector multiplets in ${\bf
  16} \oplus \overline{\bf 16} \oplus {\bf 1}$ will eat corresponding
chiral multiplets, leaving the following representation
\begin{equation*}
(3\times {\bf 16}) \oplus (5\times {\bf 10}) \oplus (4\times {\bf 1})
\end{equation*}
Notice that all anti-generations are gone.  Although ${\bf 10}$'s
and singlets can obtain mass through the Yukawa couplings
when $E_6$ is broken, one would hope that at least one of the
${\bf 10}$'s remains massless, since the standard model Higgs lives
in this multiplet.
Now suppose we want to go further, and deform to an $SU(5)$ bundle, in
order to break $\spin(10)$ to $SU(5)$.  Once again, this would be a
continuous process, and so correspond to the Higgs mechanism in the
low energy theory, so we can repeat the analysis above.  The relevant
representations decompose as follows:
\begin{equation*}
{\bf 45} ~=~ {\bf 24} \oplus {\bf 10} \oplus \overline{\bf 10} \oplus {\bf 1}~,\qquad
{\bf 16} ~=~ {\bf 10} \oplus \overline{\bf 5} \oplus {\bf 1}~,\qquad
{\bf 10} ~=~ {\bf 5} \oplus \overline{\bf 5}~.
\end{equation*}
So if such a Higgsing were possible, there would have to be chiral
multiplets transforming as ${\bf 10} \oplus \overline{\bf 10} \oplus
{\bf 1}$ to be eaten by the corresponding vector fields.  But there
are no chiral multiplets transforming as $\overline{\bf 10}$, so we
conclude that this cannot happen. Notice that this is a consequence
of having only one anti-generation of $E_6$ to begin with.
The simple analysis above relies on general features of supersymmetric
physics in 4D, but the complete mathematical answer to whether one
can deform $\cT \oplus \cO^{\oplus r}$ to a slope-stable rank $3+r$
bundle on any Calabi-Yau threefold $X$ was given in~\cite{Li:2004hx}.
Roughly, this involves three ingredients:
\newpage
\begin{enumerate}
\item A choice of $r$ linearly independent classes $\a_i \in
  H^1\big(X,\cT^\vee\big)$ corresponding to an extension
  \begin{equation*}
    0 \longrightarrow 
    \cT \longrightarrow 
    \cF_1 \longrightarrow
    \cO_X^{\oplus r} \longrightarrow
    0.
  \end{equation*}
  This amounts to the choice of $r$ $\overline{\mathbf{27}}$ that we want
  to use in the Higgsing of $E_6$.
\item A choice of $r$ linearly independent classes $\b_i \in
  H^1\big(X,\cT\big)$ corresponding to the opposite extension
  \begin{equation*}
    0 \longrightarrow 
    \cO_X^{\oplus r} \longrightarrow
    \cF_2 \longrightarrow
    \cT \longrightarrow 
    0.
  \end{equation*}
  This amounts to the choice of the corresponding $\mathbf{27}$s.
\item A family of vector bundles interpolating between $\cF_1$ and
  $\cF_2$. This ensures that, to all orders in perturbation theory,
  there is no superpotential term that forbids the VEVs.
\end{enumerate}
We again notice that there is no deformation at $r=2$ because one
would need linearly independent $\alpha_1$, $\alpha_2 \in
H^1(\YG,\cT)$. But $\dim H^1(\YG,\cT) = h^{11}(\YG) = 1$,
so there are no two linearly independent cohomology classes.  This is a
simple example of the beautiful interplay between mathematical features
of the compactification and low energy supersymmetric field theory; for a
recent analysis of vector bundle stability in this spirit see
\cite{Anderson:2009sw,Anderson:2009nt}. 

In summary, deforming the tangent bundle can at most break the gauge
group to $\spin(10)$, but nothing smaller. We might then consider further
breaking via the Hosotani mechanism, but as explained in
\cite{McInnes:1989rg}, it is not possible to break
$\spin(10)$ directly to $G_{\text{SM}}$ this way.  It can however be broken to
$G_{\text{SM}}{\times}U(1)$, as in \cite{Braun:2004xv}, with the extra $U(1)$
corresponding to baryon number minus lepton number.  This is an option
for the Abelian quotient of $Y^{8,44}$, with fundamental group $\IZ_{12}$,
but not for $\YG$.
\subsectionhyp
[Breaking $E_6$ by the Hosotani mechanism]
[Breaking E6 by the Hosotani mechanism]
{\boldmath Breaking $E_6$ by the Hosotani mechanism}
\label{sec:Hosotani}
In the previous section we separated Wilson lines and gauge bundle
deformations into commuting subgroups of $E_6$, and found we could not
get all the way to the standard model gauge group $G_{\text{SM}}$.  We now
consider relaxing this restriction.  First we will choose values for the Wilson
lines, to partially break $E_6$, and then determine whether we can deform the
resulting bundle to obtain just $G_{\text{SM}}$.  The maximal symmetry breaking
achievable with $\goth$-valued Wilson lines is $E_6 \to G_{\text{SM}}\times U(1)'$,
which we describe below.  Unfortunately, in this case it is impossible to
remove the extraneous $U(1)'$ by a deformation.  In the following subsection
we therefore describe a model in which the Wilson lines leave
$SU(4){\times}SU(2){\times}U(1)$ unbroken, and a bundle deformation
can break this to precisely $G_{\text{SM}}$.
\newpage
\subsubsection{Model 1}\vskip-10pt
First we follow the original approach of~\cite{WittenSBP} and
take the image of $\Ph:\pi_1(\YG)\to E_6$ to lie in the maximal subgroup
$SU(3)_C {\times} SU(3)_L {\times} SU(3)_R$. There is in fact
a unique choice of such an embedding that achieves the maximal breaking
of $E_6$ to $SU(3)_C{\times}SU(2)_L{\times}U(1)_Y{\times}U(1)'$,
namely
\begin{equation}
  \label{eq:SU3cubed} 
  \begin{split}
    \Ph(g_3) \;=& \left ( \begin{array}{rrr}
        1 &\+ 0 &\+ 0 \\
        0 & 1 & 0 \\
        0 & 0 & 1
      \end{array} \right )
    \times
    \left ( \begin{array}{rrr}
        1 & \+0 & \+0 \\
        0 & 1 & 0 \\
        0 & 0 & 1
      \end{array} \right )
    \times
    \left ( \begin{array}{rrr}
        1 & \+0 & \+0 \\
        0 & \z & 0 \\
        0 & 0 & \z^2
      \end{array} \right )
    ,
    \\[2ex]
    \Ph(g_4) \;=& \left ( \begin{array}{rrr}
        1 &\+ 0 &\+ 0 \\
        0 & 1 & 0 \\
        0 & 0 & 1
      \end{array} \right )
    \times
    \left ( \begin{array}{rrr}
        i & \+0 & 0 \\
        0 & i & 0 \\
        0 & 0 & -1
      \end{array} \right )
    \times
    \left ( \begin{array}{rrr}
        1 & \+0 & 0 \\
        0 & 0 & -1 \\
        0 & 1 & 0
      \end{array} \right ) 
    .
  \end{split}
\end{equation}
In particular, the rank of the unbroken gauge group is reduced from
$6$ down to $5$, which would not be possible with Abelian Wilson
lines. 
The next step is to compute the massless particle spectrum. Note that
the $\bf{27}$ of $E_6$ decomposes as
\begin{equation*}
  {\bf 27} ~=~ 
  ({\bf 3}, {\bf 3}, {\bf 1}) \oplus 
  (\overline{\bf 3}, {\bf 1}, \overline{\bf 3}) \oplus 
  ({\bf 1}, \overline{\bf 3}, {\bf 3})
\end{equation*}
and, therefore, under the unbroken gauge group as
\begin{multline*}
    {\bf 27} =
    \underbrace{ ({\bf 3}, {\bf 2})_{\frac{1}{3}, 2}  \oplus
      (\overline{{\bf 3}}, {\bf 1})_{-\frac{4}{3}, 2}  \oplus
      ({\bf 1}, {\bf 1})_{2, 2} }_{\bf 10} \oplus
    \underbrace{ (\overline{{\bf 3}}, {\bf 1})_{\frac{2}{3}, -1} \oplus 
      ({\bf 1}, {\bf 2})_{-1, -1}  }_{\bf\overline 5}
    \\[1.5ex] \oplus
    \underbrace{ ({\bf 3}, {\bf 1})_{-\frac{2}{3}, -4} \oplus 
      ({\bf 1}, {\bf 2})_{1, -4} }_{\bf 5}\, \oplus
    \underbrace{ (\overline{{\bf 3}}, {\bf 1})_{\frac{2}{3}, -1} \oplus 
      ({\bf 1}, {\bf 2})_{-1, -1}  }_{\bf\overline 5}\oplus 
    \underbrace{ ({\bf 1}, {\bf 1})_{0, 5\vphantom{\frac13}} }_{\bf 1} 
    \oplus
    \underbrace{ ({\bf 1}, {\bf 1})_{0, 5\vphantom{\frac13}} }_{\bf 1}
    .
\end{multline*}
Here, the subscripts represent the charges under $U(1)_Y$ and $U(1)'$,
respectively and the underbraces gather together the $SU(5)$ representations.
The first line is the field content of a single
standard model generation, and the rest consists of exotics.  There
are two potentially important things to notice.  First, two of the
exotics are standard model singlets, but charged under $U(1)'$, so if
some combination of these develops a VEV it will break the symmetry to
exactly $G_{\text{SM}}$.  Second, if $U(1)'$ does get
broken, there is nothing forbidding mass terms for all of the exotics,
since they can pair up to give standard model singlets.
Unfortunately, the unique choice eq.~\eqref{eq:SU3cubed} of
$\goth$ Wilson lines also projects out the one vector-like pair of
standard model singlets (the calculation proceeds as described below for
the second choice of embedding). Therefore, it is impossible to Higgs the
remaining~$U(1)'$.
\newpage
\subsubsection{Model 2}\vskip-10pt
We must therefore turn to an embedding of $\goth$ in $E_6$ which leaves
a larger unbroken gauge symmetry.  This time we will choose a different
maximal subgroup in which to embed $\goth$, which will turn out to be
equivalent to using $SU(3)^3$, but is clearer in this case.
Therefore consider the maximal subgroup $SU(6)\times SU(2)$ of $E_6$.  We
can define the map $\Ph$ by giving the images of $g_3, g_4$ in this
maximal subgroup:
\begin{equation*}
  \begin{split}
    \Ph(g_3) ~&=~ \id_6 \times 
    \left( \begin{array}{cc}
        \z & 0 \\
        0 & \z^2 
      \end{array} \right) \\
    \Ph(g_4) ~&=~
    \text{diag}(~\ii,~ \ii,~ \ii,~ \ii, -1, -1) \times 
    \left( \begin{array}{cr}
        0 & -1 \\
        1 & 0 
      \end{array} \right) \\
  \end{split}
\end{equation*}
The generators of $E_6$ which do not lie in $SU(6) {\times} SU(2)$
transform as $({\bf 20}, {\bf 2})$, so none of these are invariant
under $\Ph(\goth)$, and the unbroken gauge group is the commutant of
$\Ph(\goth)$ in $SU(6) {\times} SU(2)$.  It is easy to check that this
is $SU(4){\times} SU(2) {\times} U(1) \subset SU(6)$.
We now work out the representations of the unbroken group into which
the matter fields fall. The $\bf{27}$ of $E_6$ decomposes under
$SU(6) {\times} SU(2)$ as
\begin{equation*}
{\bf 27}~=~ ({\bf 15}, {\bf 1}) \oplus (\overline{\bf 6}, {\bf 2})
\end{equation*}
and therefore under the unbroken gauge group as follows:
\begin{equation*}
{\bf 27}~=~ ({\bf 4}, {\bf 2})_{\frac 12} \oplus ({\bf 6}, {\bf 1})_{-1} \oplus ({\bf 1}, {\bf 1})_2
\oplus \big[ 2\times (\overline{\bf 4}, {\bf 1})_{\frac 12} \big] \oplus \big[ 2\times ({\bf 1}, {\bf 2})_{-1} \big]
\end{equation*}
where the subscripts denote the $U(1)$ charges with a convenient normalisation.  We pause momentarily to describe how the structure of the standard model appears in this setup.  $SU(4)$ has $SU(3) {\times} U(1)$ as a maximal subgroup, and the $SU(3)$ is that of colour, while a combination of this $U(1)$ with the explicit one gives the hypercharge group $U(1)_Y$.  The explicit $SU(2)$ is weak $SU(2)$.
We can also describe the representations in terms of the quantum numbers of left-handed standard model fields.  The $({\bf 1}, {\bf 2})_{-1}$ and $({\bf 1}, {\bf 1})_2$ transform like the lepton doublet $l$ (or down-type Higgs $H_d$) and the anti-electron $e^c$ respectively.  The $({\bf 4}, {\bf 2})_{\frac 12}$ contains the quark doublet $Q$ and an up-type Higgs $H_u$.  The $(\overline{\bf 4}, {\bf 1})_{\frac 12}$ contains the anti-down quark $d^c$ and a standard model singlet (or `right-handed neutrino') $N$.  Finally, the $({\bf 6}, {\bf 1})_{-1}$ contains the anti-up quark $u^c$ and an exotic colour triplet.
It is clear from the above that the extended symmetry can be broken to that of the standard model by giving a VEV to a field in the $(\overline{\bf 4}, {\bf 1})_{\frac 12}$ representation, as this is the only one containing a standard model singlet.  To decide whether or not this is possible, we must first determine which fields are massless.
In the presence of discrete holonomy (VEVs for Wilson lines), it is no longer necessarily true that the light charged multiplets will be the content of $h^{11}$ $\overline{\bf 27}$'s and $h^{21}$ ${\bf 27}$'s.  This is so because the fields no longer transform simply as cohomology classes of the manifold; this representation of 
$\goth$ is tensored with the representation of $\Ph(\goth)\subset E_6$.  Calculating the light spectrum in this case proceeds as follows:
In the absence of flux lines the structure group of the gauge bundle is SU(3), and an index theorem tells us that the difference between the number of zero modes for a ${\bf 3}$ and a $\overline{\bf 3}$ is half the Euler number of $\YG$.  This corresponds to the net number of generations, since ${\bf 3}$ and $\overline{\bf 3}$ come paired with ${\bf 27}$ and $\overline{\bf 27}$ of $E_6$.  Discrete holonomy valued in $E_6$ does not change the curvature of the connection, so because the index is calculated via a curvature integral, it does not change the net number of generations, and we still have 3 chiral ${\bf 27}$s.  However, the vector-like pairs no longer need to form complete $E_6$ multiplets, since $E_6$ is broken.  If we can explicitly calculate the zero modes coming from the $\overline{\bf 27}$, we know that these must be paired with their conjugates from a ${\bf 27}$ so as not to violate the index theorem.
We have already calculated the representation of $\goth$ on $(1,1)$-forms in \sref{sec:homology}, so we just need to work out the action of $\Ph(\goth)$ on the $\overline{\bf 27}$, tensor these together and pick out the invariants.  It will be simpler to calculate how $\Ph(\goth)$ acts on the ${\bf 27}$ and then conjugate this, since we have written everything so far in terms of the ${\bf 27}$.  We know the transformation properties of the ${\bf 27}$ under $SU(6) {\times }SU(2)$, and thus under $\Ph(\goth)$.  The results are shown 
in~\tref{holonomyreps}.
\begin{table}[bt]
  \def\str{\vrule height15pt width0pt depth8pt \hskip1.4cm}
  \begin{center}
    \begin{tabular}{|l| l | c | c |}
      \hline
      \parbox{3.8cm}{\begin{center}\vskip-6pt $SU(4){\times}SU(2){\times}U(1)$\\[-3pt] representation
        \end{center}\vskip-6pt }
      & $\Ph(\goth)$ rep. 
      & \parbox{2.1cm}{\begin{center}\vskip-6pt No.\ of light\\[-3pt] copies \end{center}\vskip-6pt } 
      & \parbox{2.1cm}{\begin{center}\vskip-6pt No.\ of light\\[-3pt] conjugates \end{center}\vskip-6pt } \\ 
      \hline\hline
      \str $({\bf 4}, {\bf 2})_{\frac 12}$
      & \hskip16pt $R_{-\ii}$
      & 4 & 1 \\ \hline
      \str $({\bf 6}, {\bf 1})_{-1}$
      & \hskip16pt$ R_{-1}$
      & 4 & 1 \\ \hline
      \str $({\bf 1}, {\bf 1})_2$
      & \hskip16pt $R_1$
      & 4 & 1 \\ \hline
      \str $(\overline{\bf 4}, {\bf 1})_{\frac 12}$
      & \hskip16pt $R^{(2)}_+$
      & 8 & 2 \\ \hline
      \str $({\bf 1}, {\bf 2})_{-1}$
      & \hskip16pt $R^{(2)}_-$
      & 6 & 0 \\ \hline
    \end{tabular}
  \vspace{5pt}
  \parbox{5.4in}{\caption{\label{holonomyreps}\small\it The transformation properties of the ${\bf 27}$ fields
    under the unbroken gauge group and $\Ph(\goth)$, the discrete part
    of the holonomy group.  The last two columns give the number of
    zero modes transforming in each representation and its
    conjugate.}}
 \end{center}
\end{table}
The most important outcome is the appearance of two vector-like pairs of the $(\overline{\bf 4}, {\bf 1})_{\frac 12}$.  On general grounds, these fields correspond to classical D-flat directions breaking the gauge symmetry to that of the standard model.  Furthermore, the (renormalisable) superpotential contains only terms linear in the singlet components, meaning that at this level the F-flat conditions $W = dW = 0$ are also still satisfied by giving this component an expectation value.

Suppose then that we give a supersymmetric VEV to some linear combination of the $(\overline{\bf 4}, {\bf 1})_{\frac 12}$ fields (and those transforming in the conjugate representation).  The discussion is then familiar:  by a gauge transformation, we can take the VEV to be $\langle \phi \rangle~=~v\, (0, 0, 0, 1)$, which is clearly invariant under $SU(3) \subset SU(4)$.  It is also invariant under a particular $U(1)$ subgroup of $SU(4){\times}U(1)$, which commutes with $SU(3)$ and is given by
\begin{equation*}
  e^{\ii \th} ~\to~ \left( \begin{array}{c@{}c@{}c@{}c@{~}c@{~}c@{}}
      e^{-\ii \th} &&&& \vrule height 12pt depth 5pt &\\
      & e^{-\ii \th} &&& \vrule height 12pt depth 5pt &\\
      && e^{-\ii \th} && \vrule height 12pt depth 5pt &\\
      &&& e^{3\ii \th} &\vrule height 12pt depth 7pt &\\ \hline
      &&&                   &\vrule height 14pt depth 5pt & e^{3\ii \th}
    \end{array}\right)
\end{equation*}
The unbroken gauge group will therefore be $SU(3){\times}SU(2){\times}U(1)$, and with an appropriate normalisation of the $U(1)$ generator the representations break down as follows:
\begin{equation*}
\begin{split}
&\hskip 45pt ({\bf 4}, {\bf 2})_{\frac 12} = ({\bf 3}, {\bf 2})_{\frac 13} \oplus ({\bf 1}, {\bf 2})_1 ~,~~
(\overline{\bf 4}, {\bf 1})_{\frac 12} = (\overline{\bf 3}, {\bf 1})_{\frac 23} \oplus ({\bf 1}, {\bf 1})_0 \\[5pt]
&({\bf 6}, {\bf 1})_{-1} = ({\bf 3}, {\bf 1})_{-\frac 23} \oplus (\overline{\bf 3}, {\bf 1})_{-\frac 43}~,~~ 
({\bf 1}, {\bf 1})_2 = ({\bf 1}, {\bf 1})_2 ~, ~~ ({\bf 1}\, , {\bf 2})_{-1} = ({\bf 1}, {\bf 2})_{-1}
\end{split}
\end{equation*}
We therefore see explicitly that the unbroken $U(1)$ is hypercharge.  We can now ask what the light matter content of the theory will be after this Higgsing of the extra symmetry.  Certainly chiral multiplets in the representation $({\bf 3}, {\bf 1})_{-\frac 23} \oplus (\overline{\bf 3}, {\bf 1})_{\frac 23} \oplus ({\bf 1}, {\bf 1})_0$ will be absorbed into massive vector multiplets, but deciding which other fields gain masses will require detailed knowledge of the superpotential.
Above we have embedded the Wilson lines in the $SU(6){\times} SU(2)$ maximal subgroup of $E_6$.  In fact they also lie in the $SU(3){\times}SU(3){\times}SU(3)$ maximal subgroup, and our choice corresponds to one of the cases described in~\cite{WittenSBP}.  In this case the preceding discussion is more transparent than that if one proceeds via $SU(3)^3$, but it is easy to check that the results are the same.
\newpage
\sectionhyp
[{\boldmath Symmetries of $Y^{8,44}$ and $Y^{8,44}/\goth$}]
[Symmetries of Y(8,44) and Y(8,44)/G]
{\boldmath Symmetries of $Y^{8,44}$ and $Y^{8,44}/\biggoth$}
\label{sec:extrasymmetries}
\subsectionhyp
[Symmetries of $\dP_6{\times}\dP_6$]
[Symmetries of dP6 x dP6]
{\boldmath Symmetries of $\dP_6{\times}\dP_6$}
\label{sec:SymmetriesdPsix}
The study of possible symmetries of the quotient manifold is necessarily related to a study of the parameter space of the manifold since such symmetries will exist only for special values of the parameters. For the manifold $Y^{8,44}$  and its quotient the symmetries originate in the symmetries of $\cS\cong\dP_6$. We summarise these following~\cite[pp38-39]{dolgachev-2006}. 
The $\dP_6$ surface contains, as remarked previously, six $(-1)$-curves that intersect in a
hexagon. It turns out that the entire symmetry group of the hexagon, that is, the dihedral group
\begin{equation*}
  \text{Dih}_6 = \IZ_2' \ltimes \Big( \IZ_2 \times \IZ_3 \Big)
\end{equation*}
acts on the $\dP_6$. In addition, the $\dP_6$ surface is toric, corresponding to the fan over the polygon shown in \fref{fig:dP6toric}, and so is acted on by the torus $(\IC^*)^2$. 

As previously, we realise the del Pezzo surface as the complete intersection in $\IP^2{\times}\IP^2$ of the polynomials
\begin{equation*}
p~=~w_0 z_0 + w_1 z_2 + w_2 z_1~,~~~q~=~w_1 z_1 + w_0 z_2 + w_2 z_0~.
\end{equation*}
In these coordinates, the symmetry group acts as follows:
\begin{itemize}
\item $\IZ_2$ acts via the coordinate exchange
  \begin{equation}
   (w_0,w_1,w_2) \leftrightarrow  (z_0,z_1,z_2)~.
  \label{z2transf}
  \end{equation}
  
\item $\IZ_3$ acts as the phase rotation 
  \begin{equation}
    \Big( (w_0,w_1,w_2),\, (z_0,z_1,z_2) \Big) \mapsto 
    \Big( (w_0,\z w_1,\z^2 w_2),\, (z_0,\z z_1,\z^2 z_2) \Big)~.
  \label{z3transf}
  \end{equation}
 
\item $\IZ_2'$ acts via the coordinate exchange
  \begin{equation}
    \Big(  (w_0,w_1,w_2),\, (z_0,z_1,z_2) \Big) \mapsto \Big( (w_1,w_0,w_2) ,\, (z_1,z_0,z_2) \Big)~.
  \label{z2primetransf}
  \end{equation}

\item The toric action will be described in \sref{sec:toric}, where we discuss the toric point
         of view in detail.

\end{itemize}

\subsection{Symmetries of the quotient}
If we refer to the group above as $\text{Aut}(\cS)$ then the symmetries of $\cS{\times}\cS$ are 
\begin{equation*}
\text{Aut}(\cS{\times}\cS)~=~\IZ_2\ltimes \big(\text{Aut}(\cS)\times \text{Aut}(\cS)\big)~.
\end{equation*}
Not all of these symmetries descend to symmetries of the quotient $\YG$ since an element of 
$\text{Aut}(\cS{\times}\cS)$ can (i) fail to be a symmetry of the covering manifold $Y$ owing to the fact that it does not preserve the hypersurface $r=0$, or (ii) it may fail to commute appropriately with~$\goth$. We shall see presently that the symmetry group of the quotient $\YG$ is reduced to a subgroup of (a single copy of) the dihedral group $\text{Dih}_6$. This subgroup is $\IZ_2$ for generic values of the parameters, 
$\IZ_2{\times}\IZ_2$ on a certain $3$ parameter family of manifolds and the full group $\text{Dih}_6$ for an interesting $2$ parameter family of singular varieties.
The condition that, to be a symmetry of the quotient, the symmetry must commute appropriately with $\goth$ is very restrictive so we postpone testing for the preservation of the hypersurface $r=0$ and begin with a discussion parallel to that of~\cite{WittenSBP}. Let $\p$ denote the operation of taking the quotient by $\goth$ then the condition that a linear symmetry $h$ be a symmetry of the quotient is that $h\,\p(x)=\p(hx)$, which is the condition that for each $g\in\goth$ there is an element $g'\in\goth$ such that $gh=hg'$. Since our group 
$\goth$ is generated by $g_3$ and $g_4$ this is the condition that, for each $g$ and $h$
\begin{equation*}
h g h^{-1}~=~g_3^m\, g_4^n
\end{equation*}
for some integers $0\leq m\leq 2,~0\leq n\leq 3$. It suffices to apply this condition to the generators $g_3$ and $g_4$ themselves. The form of the right hand side of this relation is restricted by the fact that $h g_3 h^{-1}$ must be an element of $\goth$ that is of order $3$ and $h g_4 h^{-1}$ must be of order $4$. There are just two elements of order $3$, which are $g_3$ and $g_3^2$, and six of order $4$, which are the elements $g_3^m\,g_4$ and $g_3^m\,g_4^3$ for $0\leq m\leq 2$. Thus we may write
\begin{equation}
h g_3 h^{-1}~=~g_3^{1+k}~~,~~~h g_4 h^{-1}~=~g_3^m\,g_4^{1+2n}~~;
~~~k~=~0,\,1~,~~0\leq m\leq 2~~,~n=0,\,1~.
\label{symmconds}\end{equation}
Consider first the case that $k$, $m$ and $n$ all vanish; these are the symmetries $h$ that commute with 
$\goth$. A symmetry that commutes with both $g_3$ and $g_4$ must be of the form $h_3^\ell$ or 
$g_4^2\, h_3^\ell$, $0\leq \ell\leq 2$ where
\begin{equation*}
h_3:~x_{\a\, j}~\to~\z^j\, x_{\a\, j}~.
\end{equation*}
This is the $\IZ_3$ symmetry of \eref{z3transf}, understood as applying symmetrically to the two copies of 
$\cS$, and also a symmetry operation that we first saw in \sref{sec:Ymanifold}.
Consider now the symmetry of \eref{z2transf}, which we take to act on the first copy of $\cS$ only since acting equally on both copies is equivalent to the operation $g_4^2$,
\begin{equation*}
h_2:~x_{1,\, j}~\leftrightarrow~x_{3,\, j}~,~~x_{2,\, j}~\to~x_{2,\, j}~,
~~x_{4,\, j}~\to~x_{4,\, j}~.
\end{equation*}
It is an easy check that $h_2$ satisfies \eref{symmconds} with $(k,m,n)=(0,0,1)$.
Next we take a symmetry $h_2'$ corresponding to \eref{z2primetransf}, which we take to act symmetrically on the two copies of $\cS$
\begin{equation*}
h_2':~x_{\a,\, j}~\to~x_{\a,\, 1-j}
\end{equation*}
This satisfies \eref{symmconds} with $(k,m,n)=(1,0,0)$.
We note also that $h=g_3^2$ satisfies \eref{symmconds} with $(k,m,n)=(0,1,0)$. In terms of these symmetries we can give a transformation 
\begin{equation*}
h~=~(h_2')^n\, g_3^{2m}\, h_2^k
\end{equation*}
that satisfies \eref{symmconds} for general values of the integers $k$, $m$ and $n$.
Furthermore this solution to the conditions is uniquely determined modulo $h_3$ and $g_4^2$ since if $h$ and $\tilde{h}$ both satisfy \eref{symmconds}, for given $k$, $m$ and $n$, then $\tilde{h}h^{-1}$ commutes with $\goth$.
We have shown that the symmetry group of $\YG$ is a subgroup of 
$\euf{H}=\langle h_2,\, h_2',\, h_3\rangle$. One sees that
\begin{equation*}
h_2 h_3~=~h_3 h_2~,~~h_2 h_2'~=~h_2' h_2~~~\text{and}~~~h_2'\, h_3~=~h_3^2\, h_2'~.
\end{equation*}
We set $h_6 = h_2 h_3^2$ so that $h_2=h_6^3$ and $h_3=h_6^2$. Thus $\euf{H}$ is generated by $h_6$ and $h_2'$ and we note that
\begin{equation*}
h_2'\, h_6~=~h_6^5\, h_2'~.
\end{equation*}
We see that $\euf{H} \cong \text{Dih}_6$ and we recover, in this way, the dihedral group. 
\vskip15pt
\begin{table}[hb]
\def\str{\vrule height18pt width0pt depth10pt \hskip20pt}
\newcolumntype{C}{>{$}c<{$}}
\newcolumntype{L}{>{$}l<{$}}
\begin{center}
\begin{tabular}{| L | L | C | L | L |}
\hline
\vrule height18pt width0pt depth10pt \text{Symmetry} & \text{Relation} & x_{\a j} &~ D_a 
&~ \widetilde{D}_b \\ 
\hline\hline
\str g_3 &\hskip12pt g_6^2 & \z^{(-1)^\a j} x_{\a j} &~ D_{a+2}  &~  \widetilde{D}_{b-2}\\
\hline 
\str g_4 &            & x_{\a+1,\,j} &~ \widetilde{D}_a &~ D_{b+3} \\
\hline
\str g_6&\hskip12pt g_3^2 g_4^2 &\z^{(-1)^{\a+1} j}x_{\a+2,\, j} &~ D_{a+1} &~ \widetilde{D}_{b-1} \\
\hline
\str h_2&\hskip12pt h_6^3& x_{1j}\leftrightarrow x_{3j},~ x_{2j}\to x_{2j},~ x_{4j}\to x_{4j} &~ D_{a+3} 
&~ \widetilde{D}_b\\
\hline
\str h_2' &           & x_{\a,\, 1-j} &~ D_{6-a} &~ \widetilde{D}_{6-b} \\
\hline
\str h_3 &\hskip12pt h_6^2 & \z^j x_{\a j} &~ D_{a+2} &~ \widetilde{D}_{b+2} \\
\hline
\hskip20pth_6 &\hskip12pt h_2 h_3^2 
&\parbox{2in}{\begin{center}\vskip-5pt 
$ x_{1j}\to \z^2 x_{3j},~ x_{2j}\to \z^2 x_{2j},$\\[3pt]
$x_{3j}\to \z^2 x_{1j},~ x_{4j}\to x_{4j}$ \end{center}\vskip-5pt} 
&~ D_{a+1} &~ \widetilde{D}_{b-2}\\
\hline 
\end{tabular}
\vskip5pt
\parbox{4.6in}{
\caption{\label{tab:symmetryoperations}\small\it 
A table of the various symmetry operations that we have met with their actions on the coordinates and the 
$(-1)$-lines $D_a$ and $\widetilde{D}_b$.}}
\end{center}
\end{table}
We have met a number of symmetry operations in the course of this discussion and we gather these together in 
\tref{tab:symmetryoperations} for reference.
We now examine which of the symmetries of $\euf{H}$ preserve the
hypersurface $r=0$. The transformation $h_2$ affects only the
coordinates $x_{1j}$ and $x_{3k}$. The effect on the polynomials
$m_{ijkl}$ is $m_{ijkl}\to m_{kjil}$ and this clearly preserves the
polynomials $m_{iiii}$ and it also preserves $m_{0011}$ and $m_{0212}$
since these are transformed to polynomials $m_{ijkl}$ whose indices
are related to the original polynomials by cyclic permutation. Thus
$\YG$ is invariant under $h_2$ for all values of the parameters.
The transformation $h_2'$ transforms $m_{ijkl} \to m_{1-i,\,1-j,\,1-k,\,1-l}$. The effect is to interchange the polynomials $m_{0000}$ and $m_{1111}$, the other terms in $r$ being invariant since the indices transform by a cyclic permutation. Thus $r$ is invariant if $c_0 = c_1$.  For generic coefficients satisfying this condition, the quotient variety is smooth and is invariant under the $\IZ_2 {\times}\IZ_2$ action generated by $h_2'$ and $h_2$.  
The transformation $h_3$ has been discussed in \sref{sec:Ymanifold} and can be understood as inducing the transformation 
$(c_0,\, c_1,\,c_2,\,c_3,\,c_4) \to (c_0,\, \z c_1,\,\z^2 c_2,\,\z^2 c_3,\,\z^2 c_4) $
on the parameters.
If we make the choice $c_0 = c_1 = 0$ then $r$ also transforms homogeneously, $r \to \z^2 r$.
Thus there is a $\IP^2$ within the parameter space corresponding to parameters $(0,0,c_2,c_3,c_4)$ for which each quotient $\YG$ has a group $\text{Dih}_6$ of automorphisms. These varieties are, however, all singular and, for generic values of $c_2$, $c_3$ and $c_4$, have 3 nodes.  These nodal varieties will be studied in detail in \sref{sec:transgression}, where we demonstrate that the nodes can be resolved to obtain a new family of Calabi-Yau manifolds. There are also two isolated points 
$c_j = (1,0,0,0,0)$ and $c_j = (0,1,0,0,0)$ that correspond to $\text{Dih}_6$-invariant varieties that are very singular.

Although we have not yet described the torus action, we note here that for the special parameter choice $c_j=(0,0,1,4,4)$, the hypersurface $r=0$ is invariant under the complete $(\IC^*)^4$ action.  This is a point
corresponding to a very singular variety.

\newpage
\sectionhyp
[{\boldmath Conifold Transition to a Manifold with $\hodgenos=(2,2)$}]
[Conifold Transition to a Manifold with Hodge numbers (2,2)]
{\boldmath Conifold Transition to a Manifold with $\hodgenos=(2,2)$}
\label{sec:transgression}
We have noted that when $c_0=c_1=0$ there is a two-parameter family of $\text{Dih}_6$ invariant varieties 
$\YG$ and that these are all singular, the generic member having 3 nodes. These arise as 36 nodes on the covering manifold $Y^{8,44}$ which form three $\goth$-orbits, or a single $\IZ_3{\times}\goth$ orbit where the $\IZ_3$ is generated by $h_3$. The fact that there are 36 singularities for generic $(c_2,c_3,c_4)$ is best checked by a Gr\"{o}bner basis calculation. The location of the singularities will be given presently and once it is known that these are the only singularities then it is easy to check that these are nodes by expanding the equations in a neighborhood of a singular point. Owing to the $\HG$ action it is sufficient to examine any one of the singularities locally.
We describe the resolution of the three nodes on $\YG$ in two steps.  First we demonstrate that the nodal varieties $Y$ admit K\"ahler small resolutions, by identifying smooth divisors which intersect the nodes in an appropriate way, and blowing up along these divisors.  We then show that such a resolution is $\goth$-equivariant, and therefore yields a resolution of $\YG$.
\subsectionhyp
[The parameter space of 3-nodal quotients $\YG$]
[The parameter space of 3-nodal quotients Y/G]
{\boldmath The parameter space of 3-nodal quotients $\YG$}
\label{sec:3NodalParameters}
Before proceeding we pause to describe the parameter space $\G$ of 3-nodal, $\text{Dih}_6$-invariant quotients $\YG$. The generic quotient with $c_0=c_1=0$ has three nodes however on a certain locus within this space there are more severe degenerations. We find this locus to consist of the components listed in 
\tref{tab:DiscriminantComponents}.

\begin{figure}[!t]
\begin{center}
\framebox[6.5in][c]{\vrule width0pt height 2.75in depth 0.05in
\includegraphics[width=6.4in]{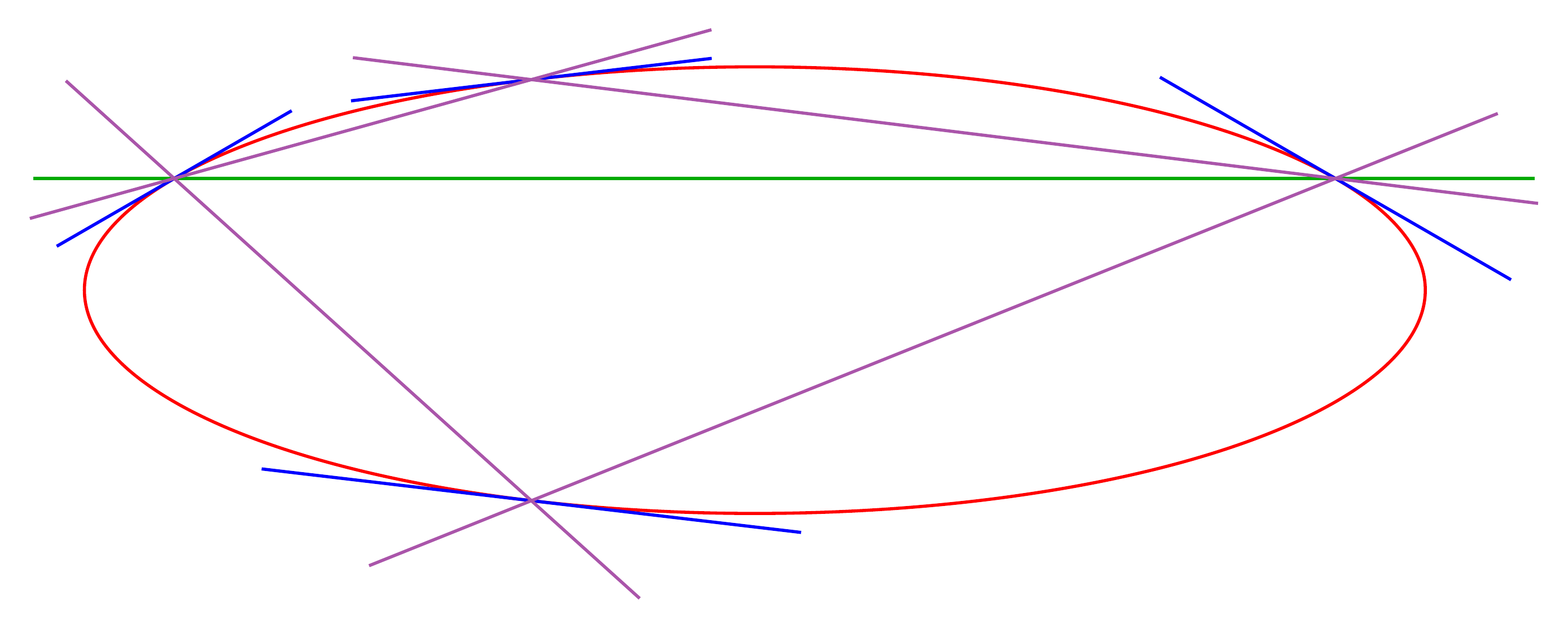}}
\vskip0pt
\place{0.6}{2.3}{\tenrm (0,1,0)}
\place{5.35}{2.3}{\tenrm (1,0,0)}
\place{1.9}{0.4}{\tenrm (16,1,-8)}
\place{2.0}{2.7}{\tenrm (1,4,4)}
\place{4.5}{0.65}{\small $(\text{o})$}
\place{3}{2.58}{\small $(\text{i})$}
\place{3.4}{0.5}{\small $(\text{ii})$}
\place{0.1}{1.65}{\small $(\text{iii})$}
\place{6.2}{1.5}{\small $(\text{iv})$}
\place{3.2}{1.9}{\small $(\text{v})$}
\place{3.9}{1.3}{\small $(\text{vi})$}
\place{1.6}{2.17}{\small $(\text{vii})$}
\place{2.7}{2.23}{\small $(\text{viii})$}
\place{1.1}{1.4}{\small $(\text{ix})$}
\framebox[6.5in][c]{\vrule width0pt height 2.6in depth 0in
\includegraphics[width=6.5in]{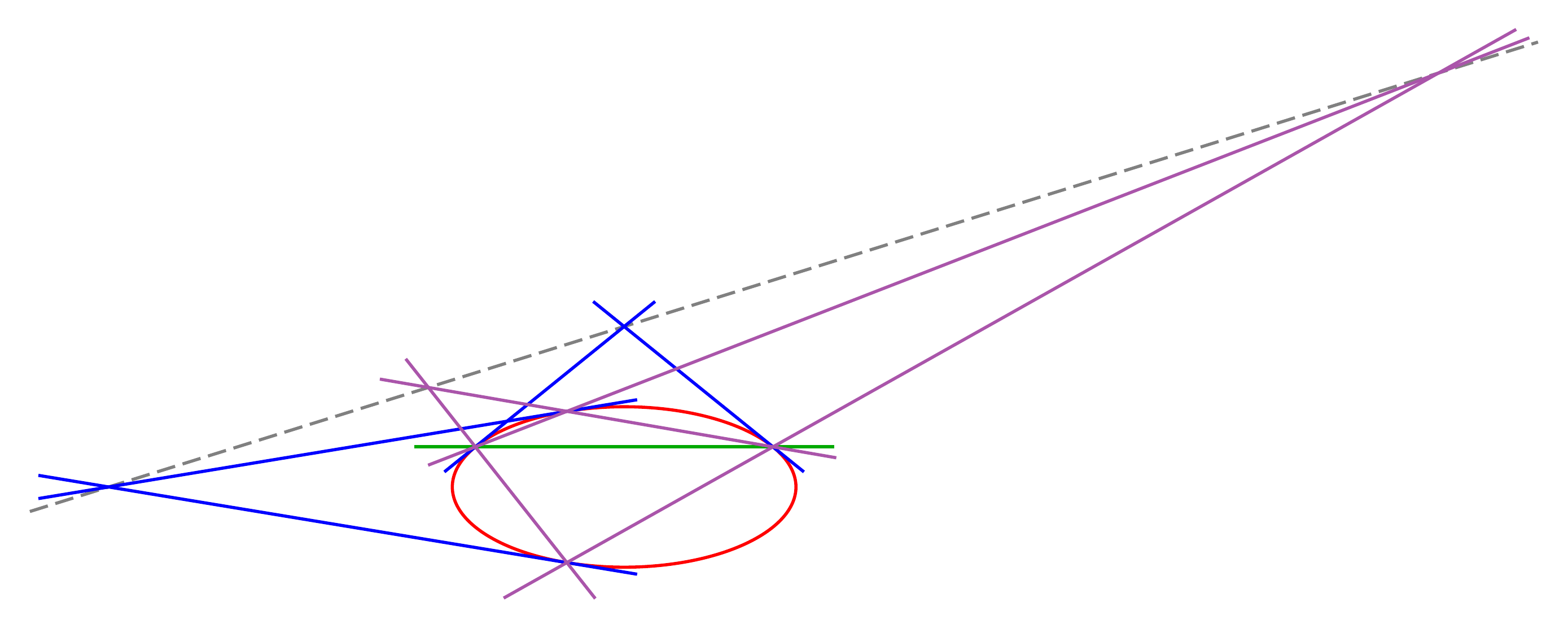}}
\parbox{6.25in}{\caption{\label{fig:Dih6Parameters}\small\it
Two sketches of the surface\/ $\G$, the locus of $\text{Dih}_6$-invariant varieties showing the discriminant of the space of 3-nodal varieties. The components of the discriminant locus are labeled according to 
\tref{tab:DiscriminantComponents}. For the resolved manifold with Hodge numbers $\hodgenos=(2,2)$ this is the space of complex structures. The second sketch zooms out to show how the components intersect. The four intersections of the pairs of blue and purple lines lie on the dashed line~$\G^{\rm (x)}$.}}
\end{center}
\end{figure}

These loci arise in the Groebner basis calculation that finds the nodes. We have seen most of these conditions at some point in this paper. For example several of the linear conditions are those for elements of $\goth$ to have fixed points on $Y^{8,44}$.  When this occurs, the covering space $Y^{8,44}$ necessarily has a node at the fixed point, so $\YG$ develops an isolated singularity which is a quotient of a node \cite{Davies}.  The remaining linear conditions lead to varieties with additional nodes and/or orbifolds of nodes.  We shall see the significance of the initial, quadratic, condition shortly.  A sketch indicating the intriguing manner in which these components intersect is given in 
Figures~\ref{fig:Dih6Parameters}.

The curves listed in \tref{tab:DiscriminantComponents} show an intriguing symmetry under the $\IZ_2$-automorphism
\begin{equation*}
c_2~\to~2\,c_3~,~~~c_3~\to~\frac{1}{2}c_2~,~~~c_4~\to~-c_4~.
\end{equation*}
This operation fixes the curves $\G^{\text{(o)}}$, $\G^{\text{(v)}}$ and $\G^{\text{(x)}}$ and interchanges the curves
\begin{equation*}
\G^{(\text{i})}~\leftrightarrow~\G^{(\text{ii})}~,~~~
\G^{(\text{iii})}~\leftrightarrow~\G^{(\text{iv})}~,~~~
\G^{(\text{vi})}~\leftrightarrow~\G^{(\text{vii})}~,~~~
\G^{(\text{viii})}~\leftrightarrow~\G^{(\text{ix})}~.
\end{equation*}
We are unclear whether this is a genuine symmetry of the geometry of the parameter space. It is not a symmetry at the same level as the $\IZ_3$-symmetry \eqref{eq:Z3symmetry} which arises from a coordinate transformation that preserves the form of the defining polynomials of $Y$. We can see this by noting that the transformation interchanges, for example, varieties on curve $\G^{\text{(i)}}$, which have 6 nodes, with varieties on curve $\G^{\text{(ii)}}$ which have singularities of a different type. The curve $\G^\text{(x)}$ is not a curve of varieties that have more severe singularities since the number of nodes is, generically, three. We include it in the discriminant as a curve that, being fixed under the automorphism, may require special consideration. 
\vskip15pt
\begin{table}[t]
\def\str{\vrule height18pt width0pt depth10pt \hskip25pt}
\newcolumntype{L}{>{$}l<{$}}
\newcolumntype{T}{>{\quad} l<{\quad}}
\newcolumntype{R}{>{\quad $}r<{$\quad}}
\centering
\begin{tabular}{| L | R | T |}
\hline
\vrule height18pt width0pt depth10pt ~\text{Component}~ & \multicolumn{1}{c|}{Equation} 
& \multicolumn{1}{c|}{Type of Singularity} \\ 
\hline\hline
\str \G^\text{(o)} & 4\,c_2 c_3 - c_4^2~=~0           & 3 higher order singularities \\
\hline
\str \G^\text{(i)} & 4\,c_2 + c_3 - 2\,c_4~=~0        & 6 nodes \\
\hline
\str \G^\text{(ii)} & c_2 + 16\, c_3 + 4\, c_4~=~0   & 3 nodes, 1 $g_4^2$-node, 1 $g_4$-node \\
\hline
\str \G^\text{(iii)} & c_2~=~0                                 & 3 nodes, 1 $\goth$-node \\
\hline
\str \G^\text{(iv)} & c_3~=~0                                 & 3 nodes, 1 $g_3$-node \\
\hline
\str \G^\text{(v)} & c_4~=~0                                  & 3 nodes, 1 $g_3$-node \\
\hline
\str \G^\text{(vi)} & 8\,c_3+c_4~=~0                      & 4 nodes \\
\hline
\str \G^\text{(vii)} & 4\,c_2 - c_4~=~0                    & 4 nodes \\
\hline
\str \G^\text{(viii)} & c_3 - c_4~=~0                       & 5 nodes \\
\hline
\str \G^\text{(ix)} & c_2 + 2\,c_4~=~0                   & 3 nodes, 1 $g_4^2$-node \\
\hline
\str \G^\text{(x)} & c_2 + 2\,c_3~=~0                    & 3 nodes \\
\hline
\end{tabular}
\vskip5pt
\parbox{5.0in}{\caption{\label{tab:DiscriminantComponents}\small\it
The generic member of\/ $\G$ is a variety with 3 nodes but along these curves the varieties develop more severe singularities. Some of these extra singularities are orbifolds of nodes as indicated. A $g_3$-node, for example, is a node fixed by the symmetry $g_3$.}}
\end{table}
\subsection{The nodes and their resolution}\label{sec:nodes}
To describe the location of the singularities we refer to the $(-1)$-lines on the two copies of $\cS$ labelled and ordered as in \eref{DefDs}. On each $D_a$ we specify a point, $\pt(D_a)$ by the condition 
$x_{12}\, x_{32}=0$. Thus $\pt(E_1)$, for example, is specified by the coordinates 
$x_{ij}=(1,1,1)$ and $x_{3j}=(1,-1,0)$ and $\pt(L_{12})$ has, as we see from \tref{lines}, 
$x_{1j}=(1,-\z^2,0)$ and $x_{3j}=(1,\z^2,\z)$. In an analogous way we define on each 
$\widetilde{D}_a$ a point, $\pt(\widetilde{D}_a)$, by the condition $x_{22}\, x_{42}=0$. We often abbreviate $\pt(D_a)=\pt_a$ and $\pt(\widetilde{D}_a)=\widetilde{\pt}_a$ respectively. The 36 nodes are the points
$$
\pt(D_a)\times\pt(\widetilde{D}_b)~~;~~~a,b\in\IZ_6~.
$$
The action of the symmetries $\euf{H}{\ltimes}\goth$ on the nodes follows from the action of the symmetries on the $(-1)$-lines. For $g_6$ and $g_4$ these are given by \eref{GensOnDs} and the action of $h_3$ is easily read off from \tref{lines}. We have
\begin{alignat}{2}
g_6:&~\pt_a{\times}\widetilde{\pt}_b~\to~\pt_{a+1}{\times}\widetilde{\pt}_{b-1}~,~~~
&&g_4:~\pt_a{\times}\pt_b~\to~\pt_{b+3}{\times}\widetilde{\pt}_a\notag\\[7pt]
h_6:&~\pt_a{\times}\widetilde{\pt}_b~\to~\pt_{a+1}{\times}\widetilde{\pt}_{b-2}~,~~~
&&h_2':~\pt_a{\times}\widetilde{\pt}_b~\to~\pt_{6-a}{\times}\widetilde{\pt}_{6-b~.}
\label{eq:ActionNodes}\end{alignat}
Notice that the generators of $\goth$ preserve the sum $a+b$ mod 3 so this sum distinguishes the $\goth$-orbits. The transformation $h_3$ permutes these orbits so the 36 nodes form a single orbit under 
$\HG$. 
The points $\pt_a\in\cS$ have the property that if the coordinates $x_{1j}$ and $x_{3j}$ are restricted to these values then the equation $r=0$ is identically satisfied for all $(x_{2\,j},x_{4\,k})\in\widetilde{\cS}$. Thus $\cD_a = \pt_a{\times}\widetilde{\cS}$ is a Weil divisor in $Y$, as is $\widetilde \cD_b = \cS{\times}\widetilde{\pt}_b$.  The six divisors $\cD_a$ are mutually disjoint, and each contains six nodes.  The same applies to the six divisors $\widetilde\cD_b$, and the two collections intersect precisely in the 36 nodes $\pt_a{\times}\widetilde{\pt}_b$. The configuration of the divisors and nodes is sketched in \fref{fig:NodesGOrbits}. 
As we will see below, the given divisors are non-Cartier in a neighbourhood of each node, and we can blow up $Y$ along such a divisor to obtain a small resolution of each node it contains.  We may therefore resolve all 36 nodes by blowing up each of the `horizontal' divisors $\cS{\times}\widetilde{\pt}_b$. In this way we obtain a K\"ahler manifold $\widehat{Y}$ that has vanishing first Chern class and $\chi=0$. Alternatively, we can blow up each of the `vertical' divisors $\pt_a{\times}\widetilde{\cS}$ but as we will see, this gives the same variety.
\vskip10pt
\begin{figure}[!b]
\begin{center}
\framebox[5.0in][c]{\vrule width0pt height 2.8in depth 0.1in
\includegraphics[width=2.7in]{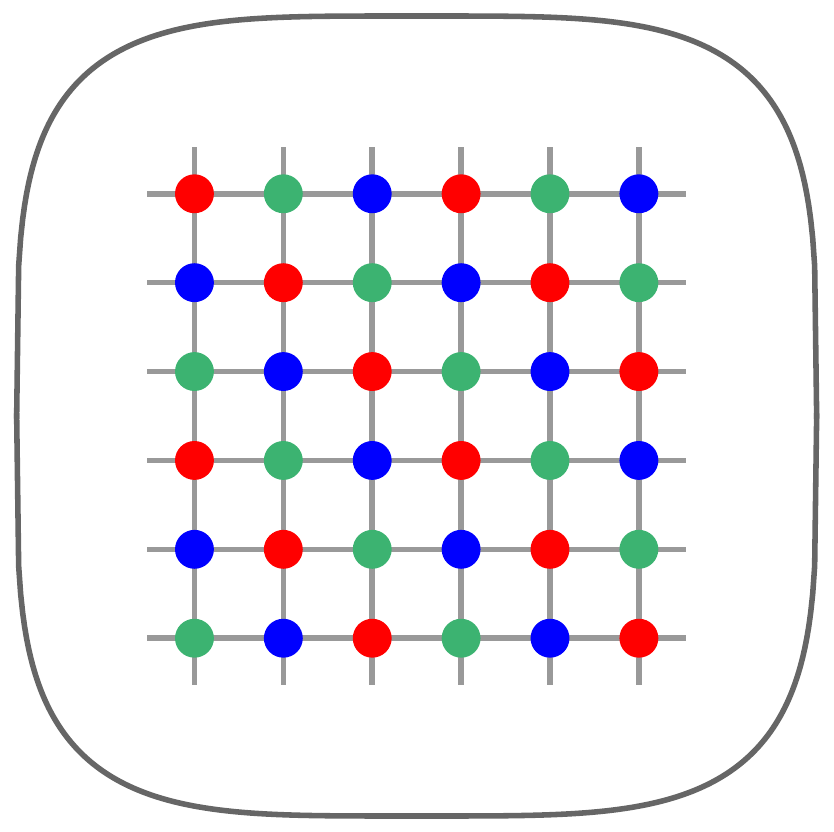}}
\parbox{4.4in}{\caption{\label{fig:NodesGOrbits}\small\it
The divisors $\pt_a{\times}\widetilde{\cS}$ and $\cS{\times}\widetilde{\pt}_b$. These intersect in the 36 nodes which form three $\goth$-orbits that are distinguished by colour.}}
\end{center}
\end{figure}
We may examine the singularities and their resolutions locally since, as remarked previously, all the singularities are related by the group $\HG$. We expand about a singularity by writing 
$x_{\a 0}=1$ and $x_{\a j} = x_{\a j}^\sharp + \e_{\a j}$ for $j=1,2$, 
where the $x_{\a j}^\sharp$ are the coordinates of a singularity. We have 8 coordinates $\e_{\a j}$ and we may solve the four equations $p^1=p^2=q^1=q^2=0$ for the $\e_{\a 2}$ as functions of the $\e_{\a 1}$. We are left with the constraint $r=0$ which, for the point 
$x^\sharp=(1,1,1){\times}(1,-1,0){\times}(1,1,1){\times}(1,-1,0)$, becomes 
\beq
\e_1( A\, \e_2 - B\, \e_4) - \e_3(B\, \e_2 - C\, \e_4)~=~0
\label{eq:conifold}\eeq
where
$$
A~=~2(c_2 + c_3 +c_4)~~,~~~B~=~4c_2 - 2c_3 + c_4~~,~~~C~=~2(4c_2 + c_3 - 2c_4)~.
$$
We see that the singularity is indeed a node provided that the determinant of the matrix associated to the quadratic form does not vanish. This determinant is proportional to $(AC{-}B^2)^2 = 3^4(c_4 - 4 c_2 c_3)^2$, so this requirement provides an understanding of the quadratic condition in
\tref{tab:DiscriminantComponents}. 
We now see that the `vertical' divisor $\cD_0=\pt_0{\times}\widetilde{\cS}$ is non-Cartier in a neighbourhood of the node, being given by the two local equations $\e_1 = \e_3 = 0$ (by definition, a Cartier divisor is given by a single polynomial in each affine patch). We may blow up $Y$ along this divisor by introducing a $\IP^1$ with coordinates $(s_1,\,s_2)$, and considering the following equation in 
$Y{\times }\IP^1$:
\beq
\e_1\, s_1 + \e_3\, s_2~=~0
\label{eq:vertical}\eeq  
If $\pi:Y{\times}\,\IP^1 \to Y$ is projection onto the first factor, then the blow up of $Y$ is given by 
$$
\widehat Y = \overline{\pi^{-1}(Y\setminus \cD_0)}~,
$$
that is, $\widehat{Y}$ is the closure of the preimage. 
One can check that $\widehat Y$ defined in this way is indeed just $Y$ with each node on $\cD_0$ replaced by a $\IP^1$. Alternatively we may blow up along the `horizontal' divisor 
$\widetilde{\cD}_0 = \cS{\times}\widetilde{\pt}_0$;  the discussion is the same, but instead of \eqref{eq:vertical} we take
\beq
(B\, \e_2 - C\, \e_4)\, s_1 + ( A\, \e_2 - B\, \e_4)\, s_2~=~0~.
\label{eq:horizontal}\eeq
The vanishing of the combinations $B\, \e_2 - C\, \e_4$ and $A\, \e_2 - B\, \e_4$ is equivalent to the vanishing of $\e_2$ and $\e_4$, provided $AC - B^2 \neq 0$.  We can see that the two resolutions are identical by observing that they are each given by the following matrix equation in $\IC^4{\times}\,\IP^1$, where $(\e_1,\e_2,\e_3,\e_4) \in \IC^4$:
$$
\begin{pmatrix} \e_1 & \e_3 \\  
    B\, \e_2 - C\, \e_4 & A\, \e_2 - B\, \e_4 \end{pmatrix}
\begin{pmatrix} s_1 \\ s_2 \end{pmatrix}~=~0~.
$$
We can now argue that our resolution of the 36 nodes on $Y$ is $\goth$-equivariant, and therefore gives a resolution of the 3 nodes on $\YG$.  Suppose we blow up the six divisors $\cD_a$.  These are just permuted by $g_6$, so the resolution is manifestly $\IZ_6$ invariant.  The element $g_4$ on the other hand interchanges the six $\cD_a$ for the six $\widetilde \cD_a$, but we have shown above that at each node $\pt_a{\times}\widetilde\pt_b$ we obtain the same resolution whether we blow up $\cD_a$ or $\widetilde\cD_b$.  The resolution is therefore also equivariant under the $g_4$ action, and thus under the action of the whole 
group~$\goth$.
Finally we can ask about the Hodge numbers of $\hatYG$.  The space of complex structures is two-dimensional, so $h^{21}(\hatYG)=2$.  To obtain $h^{11}$ note that we blow up a single divisor on $Y/\IZ_6$, and this resolution happens also to be $g_4$-covariant, so we simply have $h^{11}(\hatYG)= h^{11}(\YG)+1 = 2$.  This is consistent, because we obtain $\hatYG$ by resolving 3 nodes, which gives 
$\chi(\hatYG) = \chi(\YG)+6 = 0$.

The considerations above suggest, along the lines of~\cite{Triadophilia}, that there is a 3-generation heterotic model on $\hatYG$ that derives from the model that we have presented on $\YG$.  This is an intriguing possibility, not least because the automorphism group of $\YG$ is at least $\text{Dih}_6$ at all points in its moduli space, so any such theory will feature this quite large discrete symmetry group.
\newpage
\section{Toric Considerations and the Mirror Manifold}\label{sec:toric}
\subsection{Toric generalities}
Let us briefly review the construction of a toric variety in terms of homogeneous coordinates, as described by Cox \cite{Cox:1993fz} (there are several reviews of toric geometry to be found in the physics literature 
\cite{Aspinwall:1993nu,Bouchard:2007ik,Hori:2003ic}, see also the text by Fulton \cite{Fulton}).  Abstractly, an $n$-dimensional toric variety is an algebraic variety $X$ containing the algebraic torus $\mathbf{T}^n \cong (\IC^*)^n$ as a dense open subset, such that the group action of the torus on itself extends to an action on $X$.  Each such $X$ is given by a `fan', which consists of a collection of strictly convex, rational, polyhedral cones in an $n$-dimensional lattice $N \cong \IZ^n$.  Each face of a cone is also considered to be a cone and forms part of the fan.  Furthermore two cones can intersect only in a cone which is a face of each.  Let 
$\{v_\r\}_{\r=1\ldots d}$ be the set of lattice vectors generating the one-dimensional cones, and denote by $\nabla$ the polyhedron given by the convex hull of $\{v_\r\}$.  These vectors will generally satisfy $d-n$ relations
\begin{equation}\label{eq:toricrels}
\sum_{\r=1}^d\; Q_{r\r}\, v_\r = 0 \quad r=1,\ldots,d-n
\end{equation}
We associate a complex variable $x_\r$ with each one-dimensional cone; these will be our homogeneous coordinates.  Our first step is to delete a subset of the space $\IC^d$ spanned by these variables:  if some set $v_{\r_1},\ldots,v_{\r_k}$ does \emph{not} span a cone in the fan, we remove the set $\{x_{\r_1} = \ldots = x_{\r_k}=0 \}$.  The toric variety $X$ is then a quotient of this space\footnote{We ignore possible discrete factors in the quotient group, which will not be relevant here.} by $(\IC^*)^{d-n}$, where for each relation in \eqref{eq:toricrels} we impose
$$
(x_1,\ldots,x_d) \sim (\l^{Q_{r1}} x_1,\ldots, \l^{Q_{rd}} x_d) \quad \l \in \IC^*
$$
Batyrev described a general method for constructing mirror pairs of Calabi-Yau manifolds as hypersurfaces in toric varieties \cite{Batyrev:1994hm}.  Let $M$ be the dual lattice to $N$, and define the polyhedron $\D \subset M$ dual to $\nabla$ as follows
\beq
\D = \{ u\in M ~|~ u\cdot v \geq -1 ~~ \forall~ v \in \nabla \}
\eeq
Points in the dual lattice $M$ give rise rise to holomorphic functions on $\mathbf{T}^n \subset X$ via
\beq \label{eq:toricmonos}
  u \, \mapsto \prod_i \, t_i^{u_i}
\eeq
\vskip-15pt
where $\{u_i\}$ are the components of $u$ relative to the standard basis for $N$.  When the polyhedra are reflexive, the integral points of the dual polyhedron $\D$ correspond in this way to sections of the anticanonical bundle of $X$, so the closure of the vanishing locus of a linear combination of these monomials is a Calabi-Yau variety.  We can also dualise this construction:  taking cones over the faces of $\D$ gives the fan of another toric variety $X^*$, and points in $\nabla$ correspond to sections of its anti-canonical bundle.  In this way we obtain the mirror family.
\subsection{The Newton polyhedron and its dual}\label{sec:DeltaNabla}
\begin{figure}[!b]
  \begin{center}
    \framebox[4.5in][c]{\vrule width0pt height 2.25in depth 0.5in\includegraphics[height=1.75in]{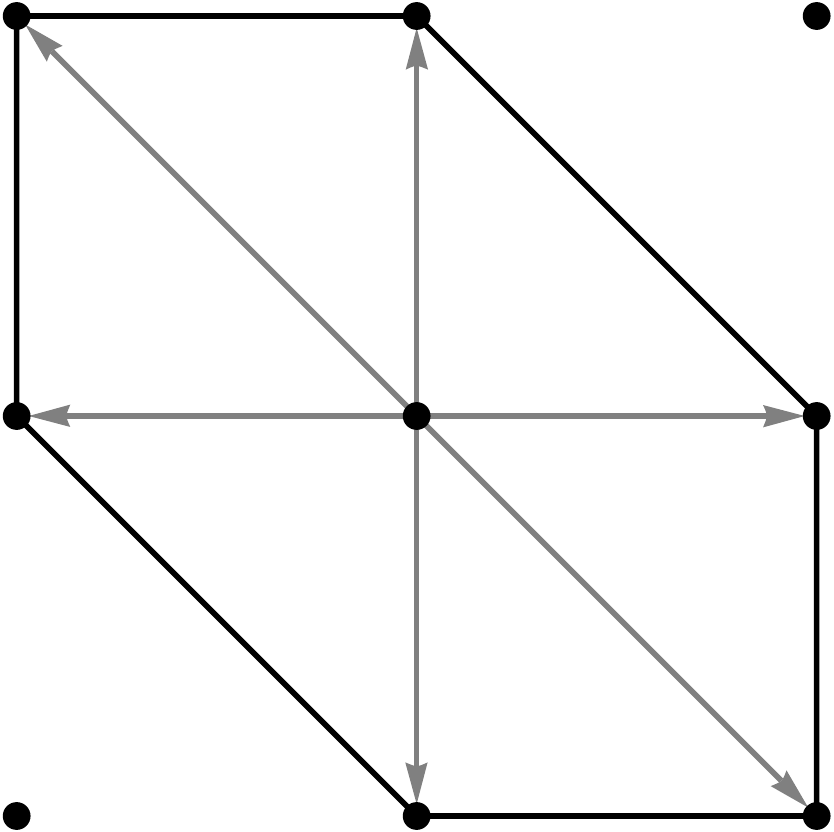}}
  \vskip0pt
\place{3.15}{2.42}{\small $E_1$} 
\place{2.1}{2.4}{\small $L_{12}$}
\place{2.15}{1.4}{\small $E_2$}
\place{3.15}{0.4}{\small $L_{23}$} 
\place{4.15}{0.42}{\small $E_3$}
\place{4.2}{1.42}{\small $L_{31}$}
 \vskip0pt
\parbox{4.0in}{\caption{\label{fig:dP6toric}\small\it 
The fan and polygon for $\dP_6$. The one-dimensional cones can be taken to correspond to the divisors as indicated.}}
  \end{center}
\end{figure}
It is a felicitous fact that the del-Pezzo surface $\dP_6$ is toric and has a fan with six one-dimensional cones $\{v_a\}$ as shown in 
\fref{fig:dP6toric}. These correspond to toric divisors and so are in correspondence with the~$D_a$. Let $\S$ denote this collection of one-dimensional cones
$$
\S~=~\{v_a\} ~=~ \big\{\,(0,1),\,(-1,1),\,(-1,0),\,(0,-1),\,(1,-1),\,(0,1)\,\big\}~.
$$
We can also think of $\S$ as the polygon over the fan, which in the present case is a hexagon.  The dual polygon is na\"ively $\S$ rotated by $90^\circ$, but by appropriate choice of coordinates in $M$ we can simply identify the two.
We have previously been specifying $\cS \cong \dP_6$ as a complete intersection in $\IP^2{\times}\IP^2$, as described in \sref{sec:nonabelianquotient}.  In order to describe the toric action it is convenient to first make a change of coordinates to
\begin{equation}\label{eq:coordchange}
  y_{\a j} = \sum_k M_{jk}\, x_{\a k} \quad \text{where} \quad
  M = \frac{1}{\sqrt{3}}\left( \begin{array}{lll}
  1 & 1 & 1 \\
  1 & \z & \z^2 \\
  1 & \z^2 & \z \end{array}\right)
\end{equation}
This brings the polynomials $p^1$ and $q^1$ to the form
\begin{equation*}
p^1 = y_{10}\,y_{30} + y_{11}\,y_{31} + y_{12}\,y_{32}~,~~
q^1 = y_{10}\,y_{30} + \z\, y_{11}\,y_{31} + \z^2\, y_{12}\,y_{32}
\end{equation*}
In these coordinates the blown-up points in the $y_1$ plane are $(1,0,0),\, (0,1,0)$ and $(0,0,1)$.  It is easy to see that for any $(\m,\l) \in (\IC^*)^2$, the polynomials are invariant under
$$
(y_{10},y_{11},y_{12})\times(y_{30},y_{31},y_{32}) \to
(y_{10},\l\,y_{11},\m\,y_{12})\times(y_{30},\l^{-1}\,y_{31},\m^{-1}\,y_{32})
$$
The torus $(\IC^*)^2$ can then be explicitly embedded in $\dP_6$ follows:
\begin{equation} \label{eq:dP6torus}
(t_1,t_2) ~\to~ (1, t_1\, \z^2, t_2\, \z)\times(1, t_1^{-1}\, \z^2, t_2^{-1}\, \z) \subset \IP^2{\times} \IP^2 .
\end{equation}
We can do the same for the second copy $\widetilde \cS$, and call the extra toric variables $t_3,t_4$.
The polyhedron, $\nabla$, over the fan for $\cS{\times}\widetilde{\cS}$ is obtained as the convex hull of the union of $\S$ with a second orthogonal copy $\widetilde{\S}$, corresponding to $\widetilde{\cS}$
\beq
\nabla~=~\text{Conv}(\S,\,\widetilde{\S})~=~
\big\{\,(v_a,\, 0) ~\big|~ v_a \in \S \big\}\cup
\big\{\,(0,\, v_b) ~\big|~ v_b \in \widetilde\S \big\}
\label{eq:nabla}\eeq 
The Newton polyhedron, $\D$, is dual to $\nabla$, in the sense of \sref{sec:DeltaNabla}, and is given by the Minkowski sum of $\S$ and $\widetilde{\S}$,
\beq
\D~=~\text{Mink}(\S,\,\widetilde{\S})~=~
\big\{\,(v_a,\, v_b)  ~\big|~ v_a \in \S \, ,\, v_b \in \widetilde\S\big\}~.
\label{eq:delta}\eeq
The polyhedron $\nabla$ has 12 vertices, which are the points given explicitly by \eref{eq:nabla}, and no other integral points apart from the origin. It has 36 three-faces that are tetrahedra.
The polyhedron $\D$ has 36 vertices and 12 three-faces that are all hexagonal prisms, these each have two hexagonal two-faces and six rectangular two-faces. The 36 points given explicitly by \eref{eq:delta} are the vertices. The polyhedron also contains the origin and 12 additional integral points, one interior to each of the hexagonal two-faces.  These 12 additional points are in fact the vertices of $\nabla$, so we have $\nabla \subset \D$.
\begin{figure}[!b]
  \begin{center}
    \framebox[6.5in][c]{\vrule height7.5cm depth0.5cm width0pt
\raise0.5cm\hbox{\includegraphics[height=2.5in]{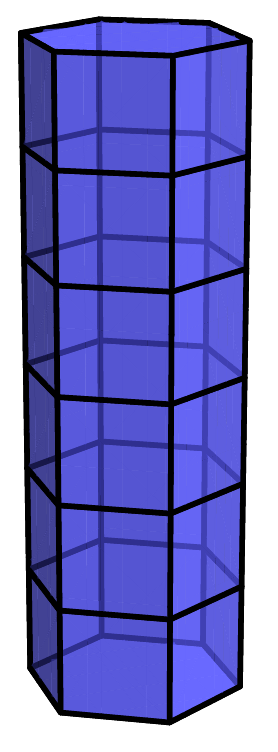}}\hskip-5pt
\raise0.5cm\hbox{\includegraphics[height=2.5in]{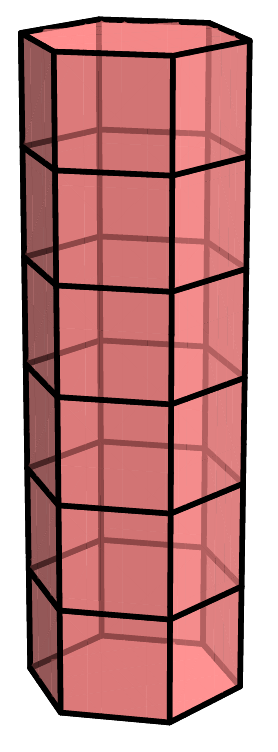}}
\hskip1cm\includegraphics[width=3.75in]{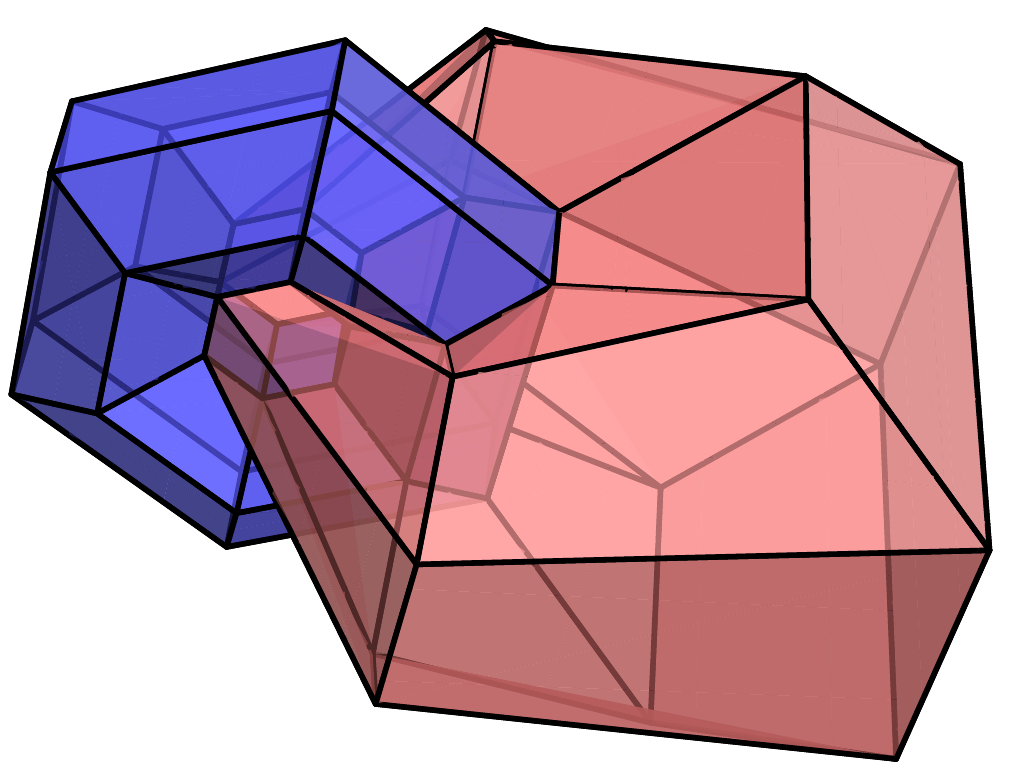}}
 \vskip0pt
\parbox{5.5in}{\caption{\label{fig:hexagonalprisms}\small\it 
The three-faces of the polyhedron $\D$. Six hexagonal three-faces are stacked to make the blue prism on the left and the remaining six are stacked to make the pink prism. The top and bottom faces of these prisms are identified to make the two intersecting rings shown on the right. The faces that are depicted as exposed are identified between the two rings with the result that there is in fact no boundary.}}
  \end{center}
\end{figure}
These facts are most quickly established by having recourse to a programme such as POLYHEDRON or PALP \cite{Kreuzer:2002uu} that analyse reflexive polyhedra. We can, with benefit of hindsight, get useful insight into this structure and understand $\D$ and $\nabla$ rather simply in terms of the divisors $D_a$ and $\widetilde{D}_b$. We will not rederive the structure of the polyhedra but will content ourselves with giving a description using only what we know about these divisors.
It is easy to see from the relation \eref{eq:delta} that there is a one-to-one correspondence between vertices of $\D$ and pairs of divisors
$$
\n_{a\,b}~\leftrightarrow~ (D_a,\widetilde{D}_b)~,~~~a,b\in\IZ_6~.
$$
The integral points $\tilde \io_b$ that are interior to the six blue hexagonal two-faces each correspond to a divisor 
$\widetilde{D}_b$ and the vertices of this hexagonal face are the $\n_{a\,b}$ as $a$ varies. The same vertices arise in both the blue and pink three-faces and this gives the correspondence between the blue and pink 
two-faces in \fref{fig:hexagonalprisms}. The rectangular two-faces contain the vertices 
$\{\n_{a,\,b},\, \n_{a+1,\, b},\,\n_{a+1,\, b+1},\,\n_{a,\, b+1}\}$.

Having associated the points of $\D$ with the divisors $D_a$ and
$\widetilde{D}_b$ we see that there is a natural $\HG$ action on the
points. Let $\r$ and $\s$ denote the matrices
$$
\r~=~\begin{pmatrix} 0 & -1\\ 1 &\+ 1\end{pmatrix}~,~~~
\s~=~\begin{pmatrix} -1~ & 0\\ \+1~ &1\end{pmatrix}
$$
and note\footnote{As a $P\hskip-1pt S\hskip-1pt L(2,\IZ)$ matrix $\r$
  is often understood to have order 3 however in the present context
  an overall sign is significant since $-v_a = v_{a+3}$.} that
$\r^6=\one$ and $\r^3=-\one$. It is an immediate check that $\r\, v_a
= v_{a+1}$ and that $\s\,v_a = v_{6-a}$.  In virtue of the polyhedron of \fref{fig:dP6toric} it comes as no surprise that $\r$ and $\s$ furnish a representation of $\text{Dih}_6$.
We know the action of $\HG$
on the $D_a$ and $\widetilde{D}_b$ from \eref{eq:ActionNodes} and in
this way we see that the action of the generators on the points of
$\D$ is given by a linear action on the lattice $M$, with matrices
\begin{alignat*}{2}
  g_6~&=~\begin{pmatrix} \r & \+0~~\\ 0 & \+\r^{-1}\!{}\end{pmatrix}~,\qquad
  &&g_4~=~\begin{pmatrix} 0 & -\one\\ \one &\+ 0\end{pmatrix} \\[10pt]
  h_6~&=~\begin{pmatrix} \r & \+0\\ 0 & -\r\end{pmatrix}~,\qquad
  &&h_2'~=~\begin{pmatrix} \s\hskip7pt{}  & 0\\ 0\hskip7pt{} & \s\end{pmatrix}
  \hskip-2pt\raisebox{-10pt}{.} 
\end{alignat*}
The method we have used relies on the fact that the polygon $\S$ is
self-dual; a more general approach would be to use \eqref{eq:dP6torus}
to find the action of $\HG$ on the toric coordinates, and translate
this to an action on $M$ by utilising the relationship in
\eqref{eq:toricmonos}.

So far we have described $\D$.  The polyhedron $\nabla$ is
simpler. The 12 vertices of $\nabla$ are the points $\io_a$ and
$\tilde\io_b$. Thus $\nabla$ is contained in $\D$ and the vertices of
$\nabla$ are the points interior to the two-faces of $\D$. The
three-faces are the 36 tetrahedra with vertices
$\{\io_a,\,\io_{a+1},\,\tilde\io_b,\,\tilde\io_{b+1}\}$. Owing to the
fact that $\nabla$ is contained in $\D$, the group acts on the points
of $\nabla$ in the same representation as the action on $\D$.
\begin{figure}[H]
\begin{center}
\framebox[6.5in]{\parbox{6.5in}{
\begin{center}\vskip-8pt
\includegraphics[width=3.0in]{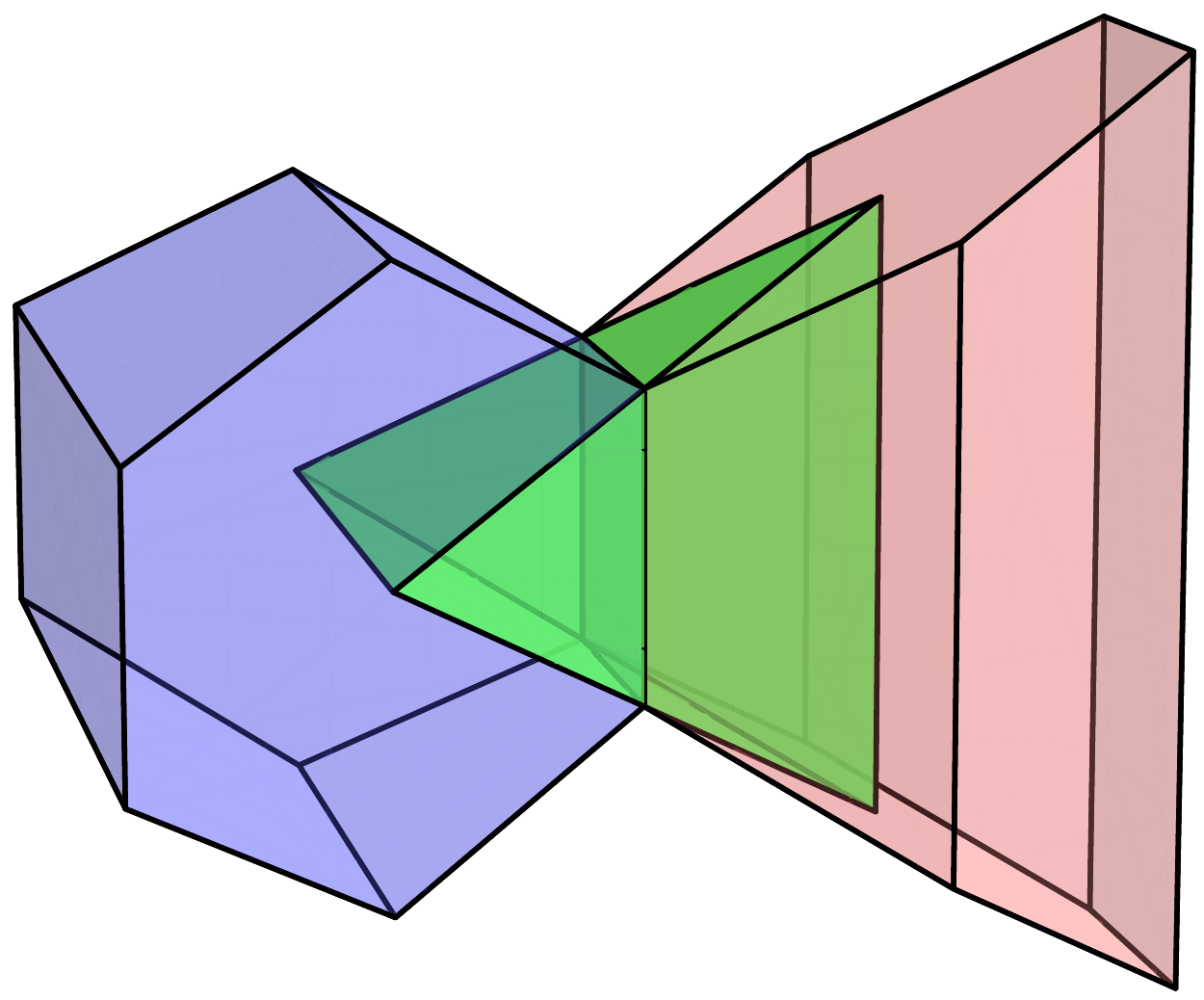}\hskip40pt
\raisebox{-15pt}{\includegraphics[width=2.4in]{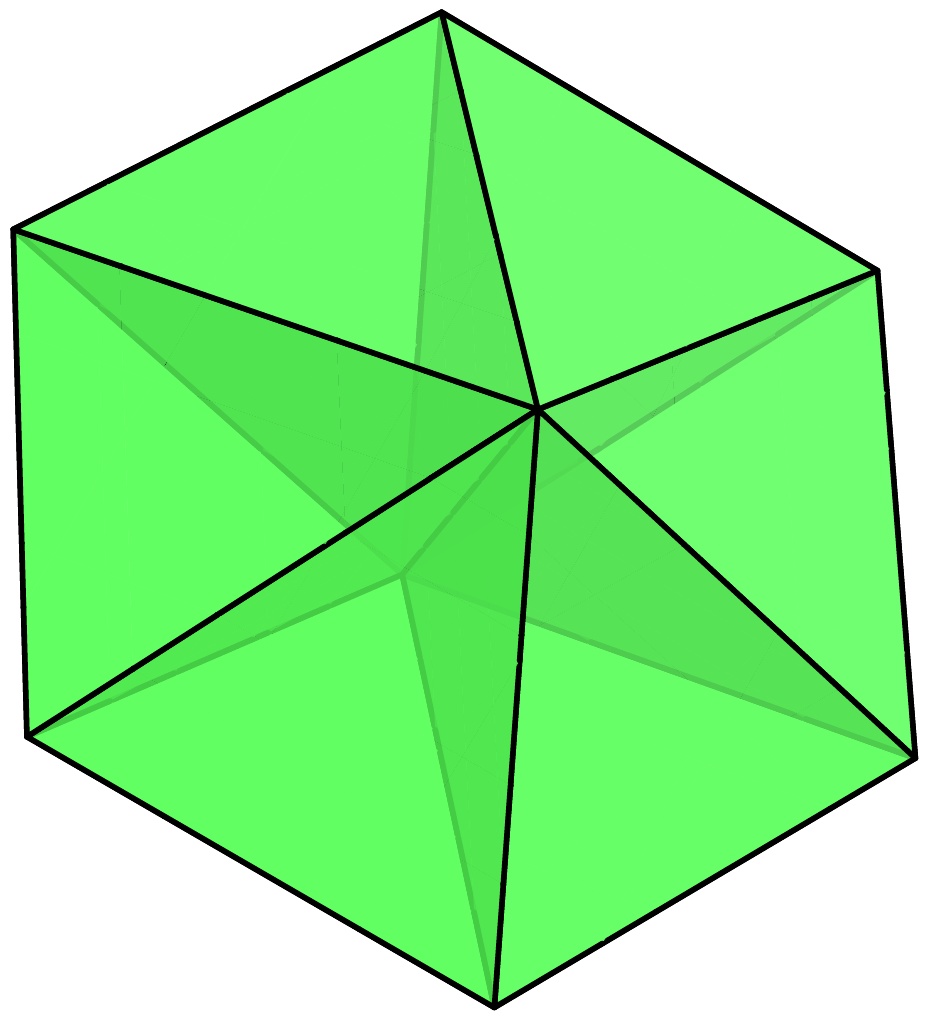}}\\
\includegraphics[width=4in]{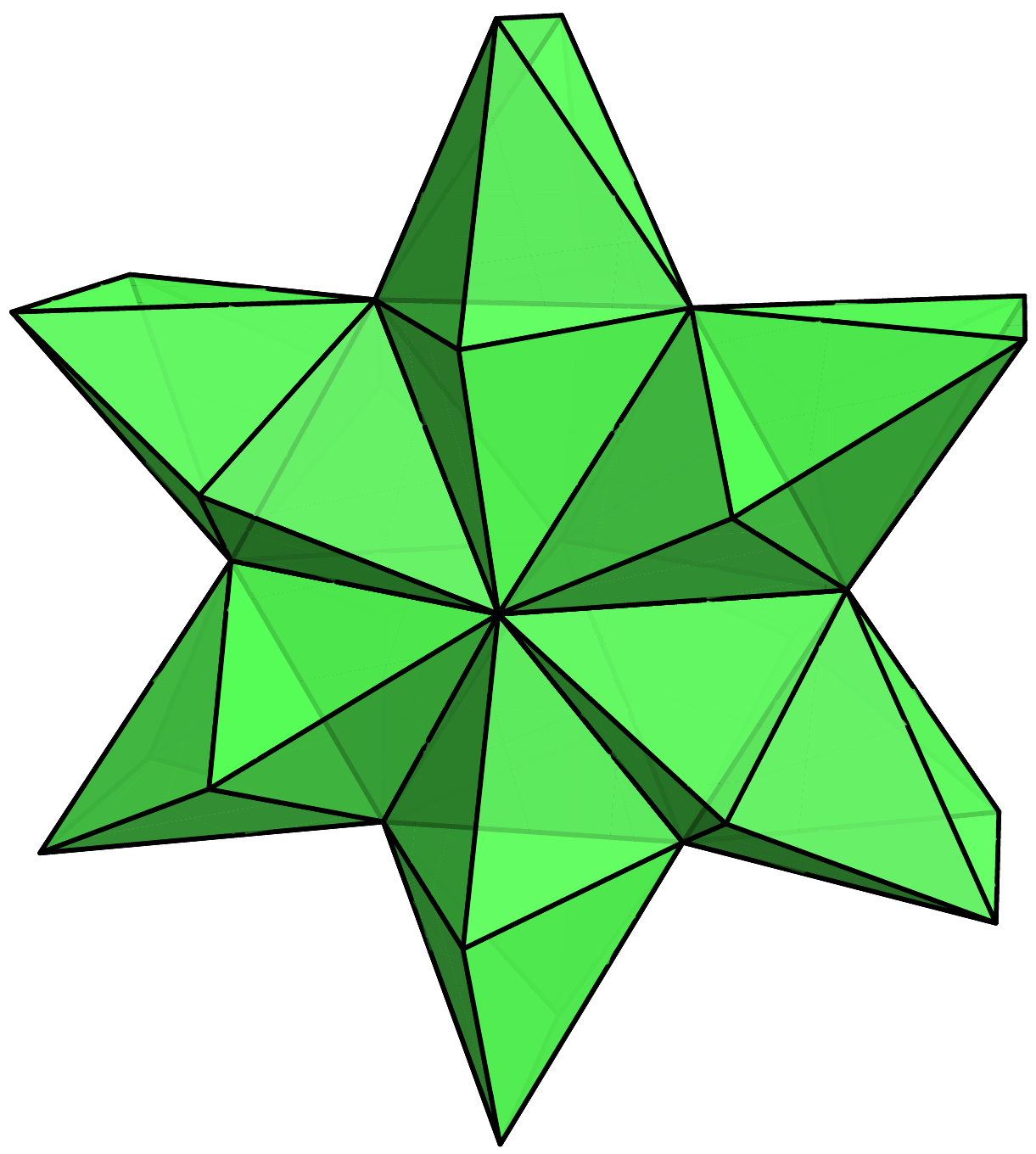}
\end{center}\vskip-8pt}}
\vskip0pt
\parbox{6.2in}{\caption{\label{fig:DualThreeFace}\small\it 
The vertices of the three-faces of the polyhedron 
$\nabla$ are the interior points to the two-faces of $\D$. The three-faces are tetrahedra, as shown in the first figure. Note however a hazard of projecting from four dimensions to three. Four of the vertices of $\D$ project onto the faces of the tetrahedron but they do not lie on the tetrahedron as they appear in the figure. Six of these tetrahedra fit together to form the polyhedron on the right. Six of these polyhedra, in turn, fit together to form the star-shaped polyhedron, with the exposed faces identified in pairs.}}
\end{center}
\end{figure}
\subsection{Triangulations}
The mirror of $Y^{8,44}$ is realised as the resolution of a hypersurface in the toric variety defined by the fan over the faces of $\D$. The toric variety defined by $\D$ is singular, and since the singularities will intersect a hypersurface, so is the hypersurface. The singularities of the hypersurface are resolved by resolving the singularities of the embedding space. This is done by subdividing the cones to refine the fan. The cones are subdivided by subdividing the three-faces of $\D$ into smaller polyhedra and the ambient variety becomes smooth if the faces of $\D$ are divided into polyhedra of minimal volume (which must then be tetrahedra of minimal volume). This process of subdividing the top-dimensional faces is known as triangulation.

We start with an $\HG$-invariant triangulation of
$\D$ by dividing the 3-faces into wedges as shown in
\fref{fig:DualThreeFace}. We may denote the blue and pink wedges that
contain the two-face $\{\n_{a,\,b},\, \n_{a+1,\, b},\,\n_{a+1,\,
  b+1},\,\n_{a,\, b+1}\}$ by $W_{a\,b}$ and $\widetilde{W}_{a\,b}$
respectively, 
\begin{equation*}
  \begin{split}
    W_{a\,b}~&=~\big\{ \n_{a,\,b},\, \n_{a+1,\, b},\,\n_{a+1,\, b+1},\,\n_{a,\, b+1},\, 
    \tilde\io_b,\, \tilde\io_{b+1}\big\} \\[1ex]
    \widetilde{W}_{a\,b}~&=~\big\{ \n_{a,\,b},\, \n_{a+1,\, b},\,\n_{a+1,\, b+1},\,\n_{a,\, b+1},\, 
    \io_a,\, \io_{a+1}\big\}~.
  \end{split}
\end{equation*}
The group $\HG$ acts on the wedges in the expected way $g_6 W_{a\,b} =
W_{a+1,\, b-1}$ and $g_4 W_{a\,b} = \widetilde{W}_{b+3,\, a}$, and so on.
The triangulation of $\D$ into the wedges yields a toric variety
with singularities along curves, and therefore a hypersurface with
point singularities. Each wedge can be cut into $3$ tetrahedra of minimal lattice volume. This further subdivision
will yield a smooth toric variety. For reasons that will become clear
shortly, we only enforce the $\goth$-symmetry at this point. Therefore, a fundamental region for the group action is a three-face of $\D$ (one-sixth of the blue or pink prism in
\fref{fig:hexagonalprisms}), consisting of $6$ wedges.
Each wedge can be triangulated in $6$ different ways. These $6$
possibilities can be distinguished by how they bisect the three
rectangular two-faces of the wedge\footnote{A wedge has three rectangular faces and each of these can be bisected in two ways so there are eight ways to bisect the faces. Two of these ways, however, do not correspond to triangulations.}, as shown in
\begin{figure}[hbt]
\centering
\framebox[6.5in][c]{
  \begin{minipage}{0.95\linewidth}
  \vskip20pt
    \begin{minipage}{0.31\linewidth}
      \centering
      \includegraphics[width=\textwidth,viewport=70 100 475 370,clip]{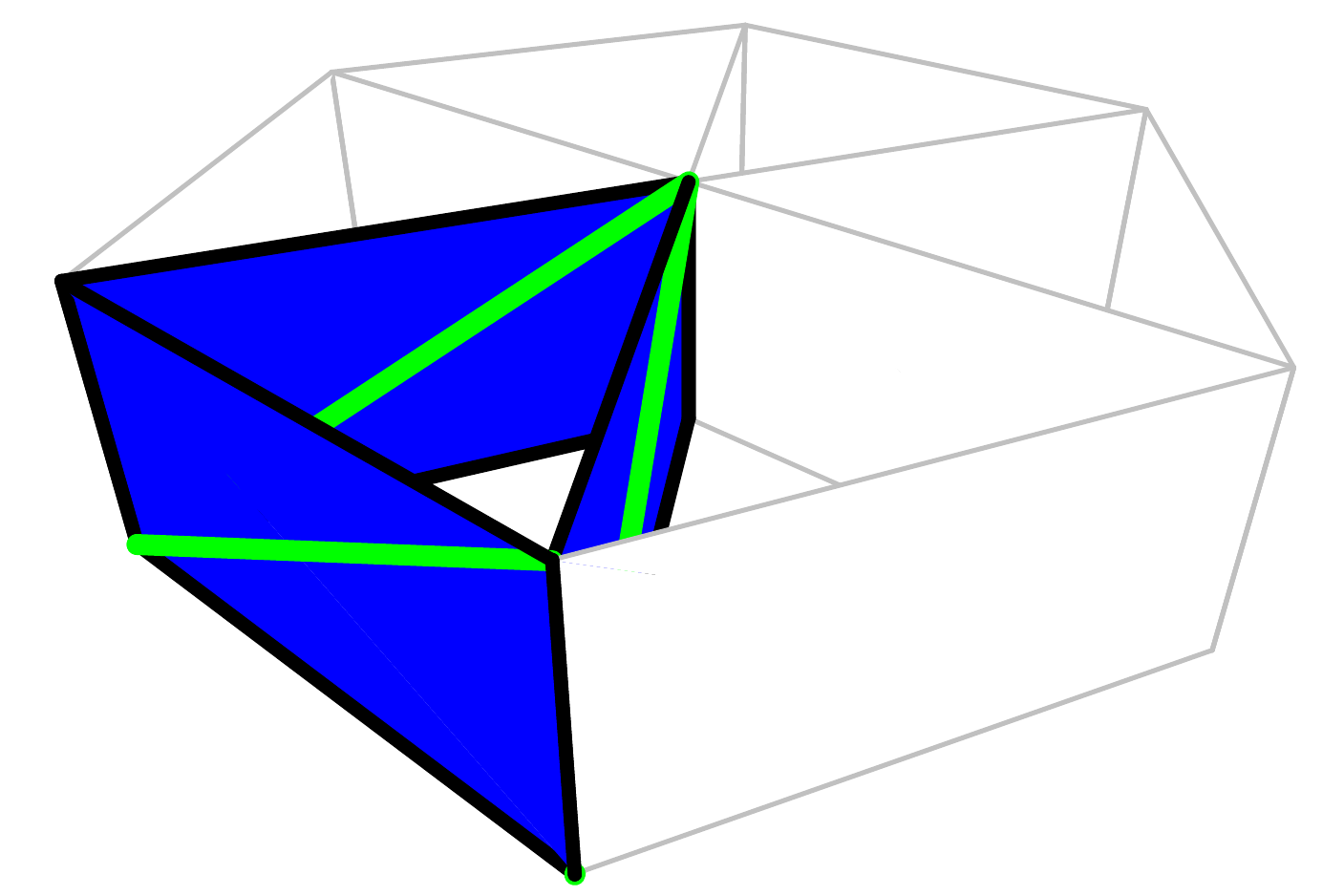}
      \\
      $(uUu)$
    \end{minipage}
    \hfill
    \begin{minipage}{0.31\linewidth}
      \centering
      \includegraphics[width=\textwidth,viewport=70 100 475 370,clip]{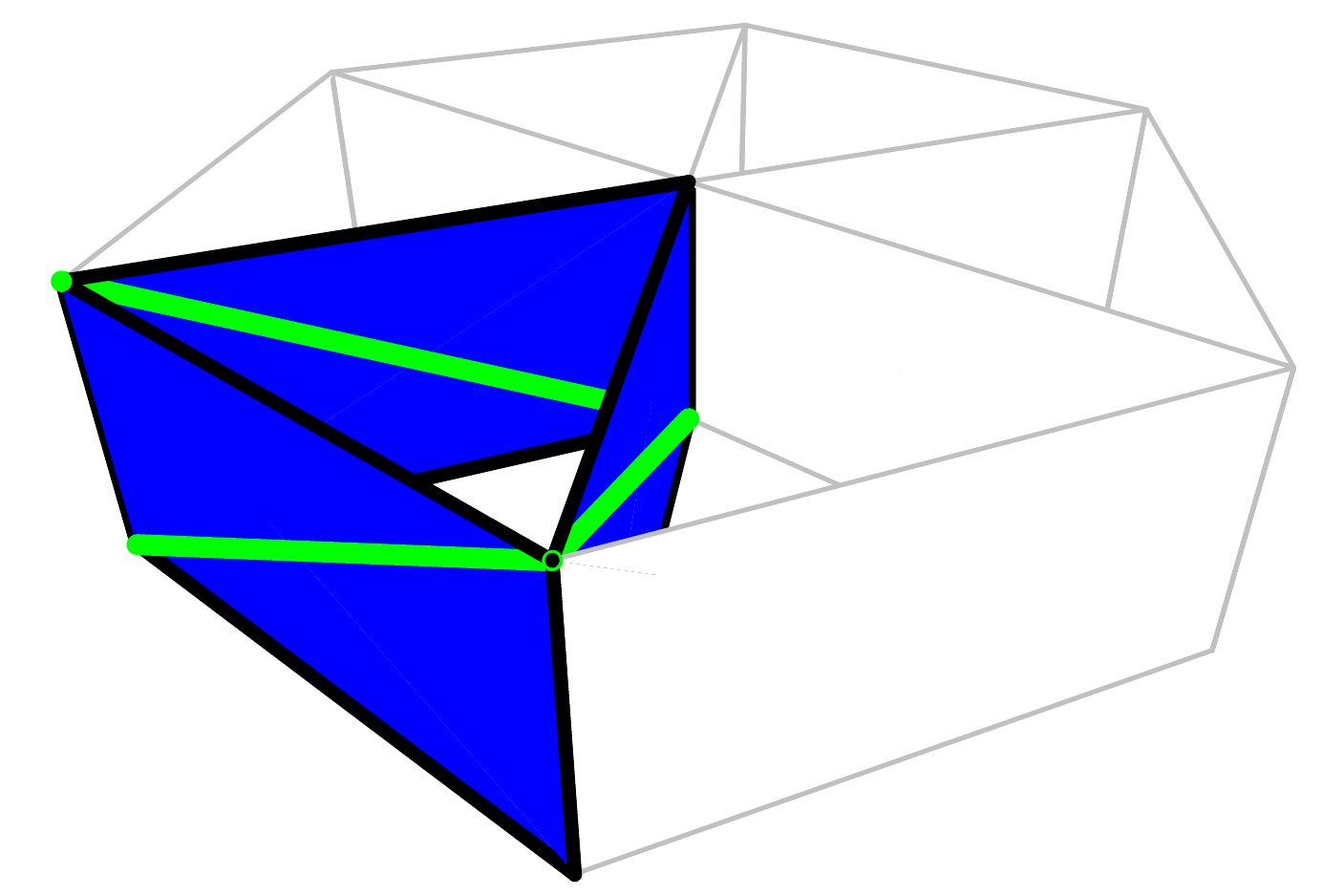}
      \\
      $(dUd)$
    \end{minipage}
    \hfill
    \begin{minipage}{0.31\linewidth}
      \centering
      \includegraphics[width=\textwidth,viewport=70 100 475 370,clip]{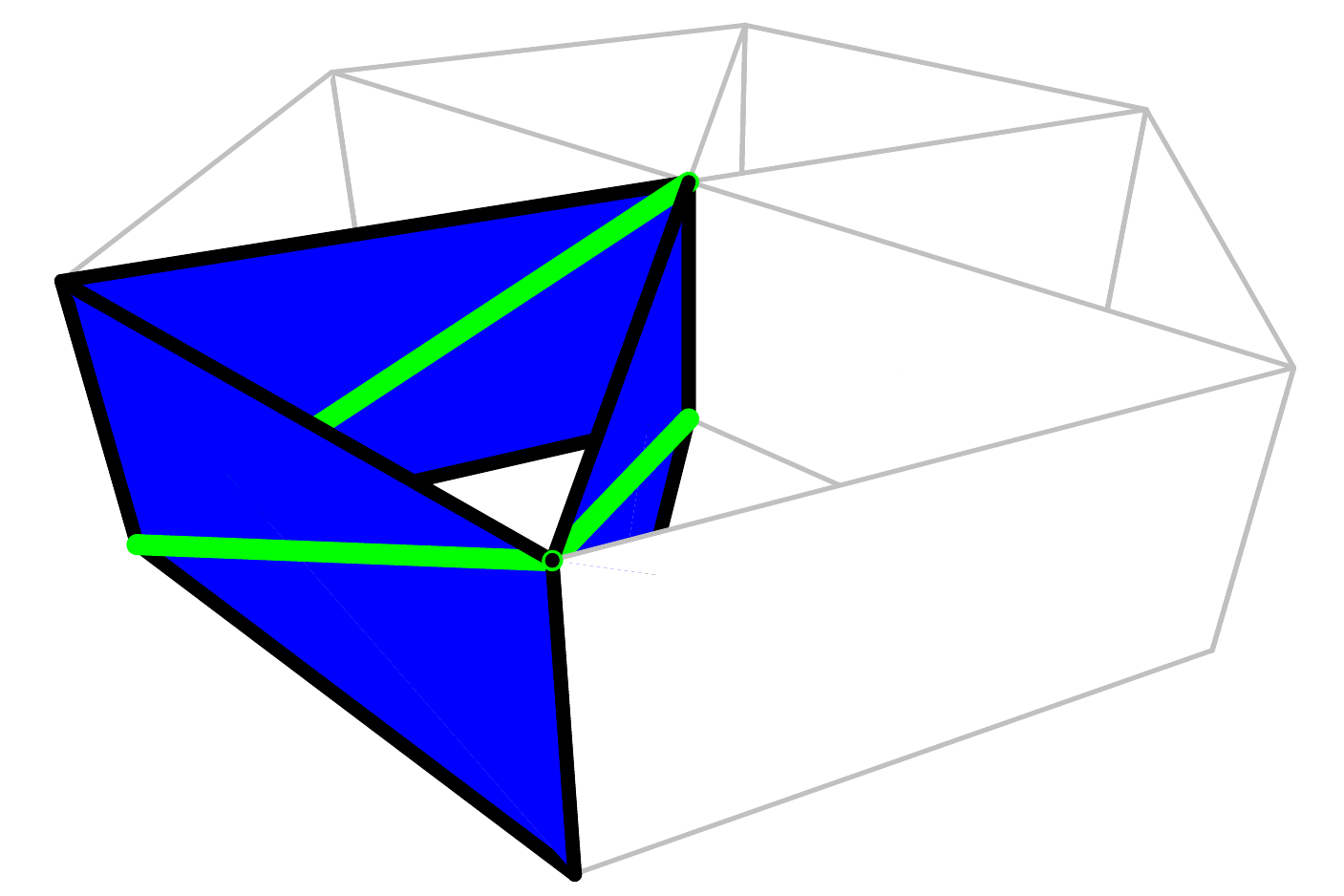}
      \\
      $(uUd)$
    \end{minipage}
    \\[3ex]
    \begin{minipage}{0.31\linewidth}
      \centering
      \includegraphics[width=\textwidth,viewport=70 100 475 370,clip]{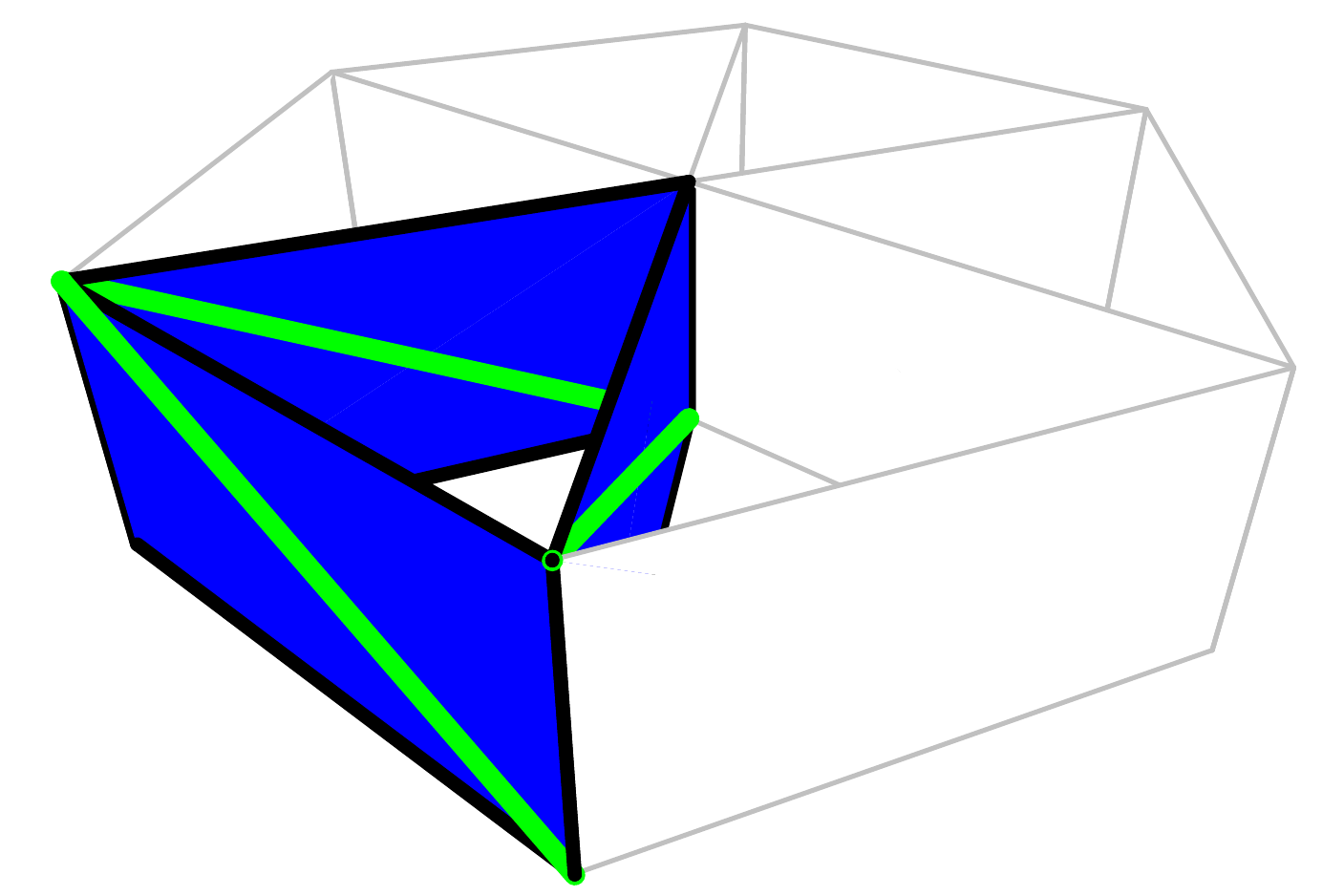}
      \\
      $(dDd)$
    \end{minipage}
    \hfill
    \begin{minipage}{0.31\linewidth}
      \centering
      \includegraphics[width=\textwidth,viewport=70 100 475 370,clip]{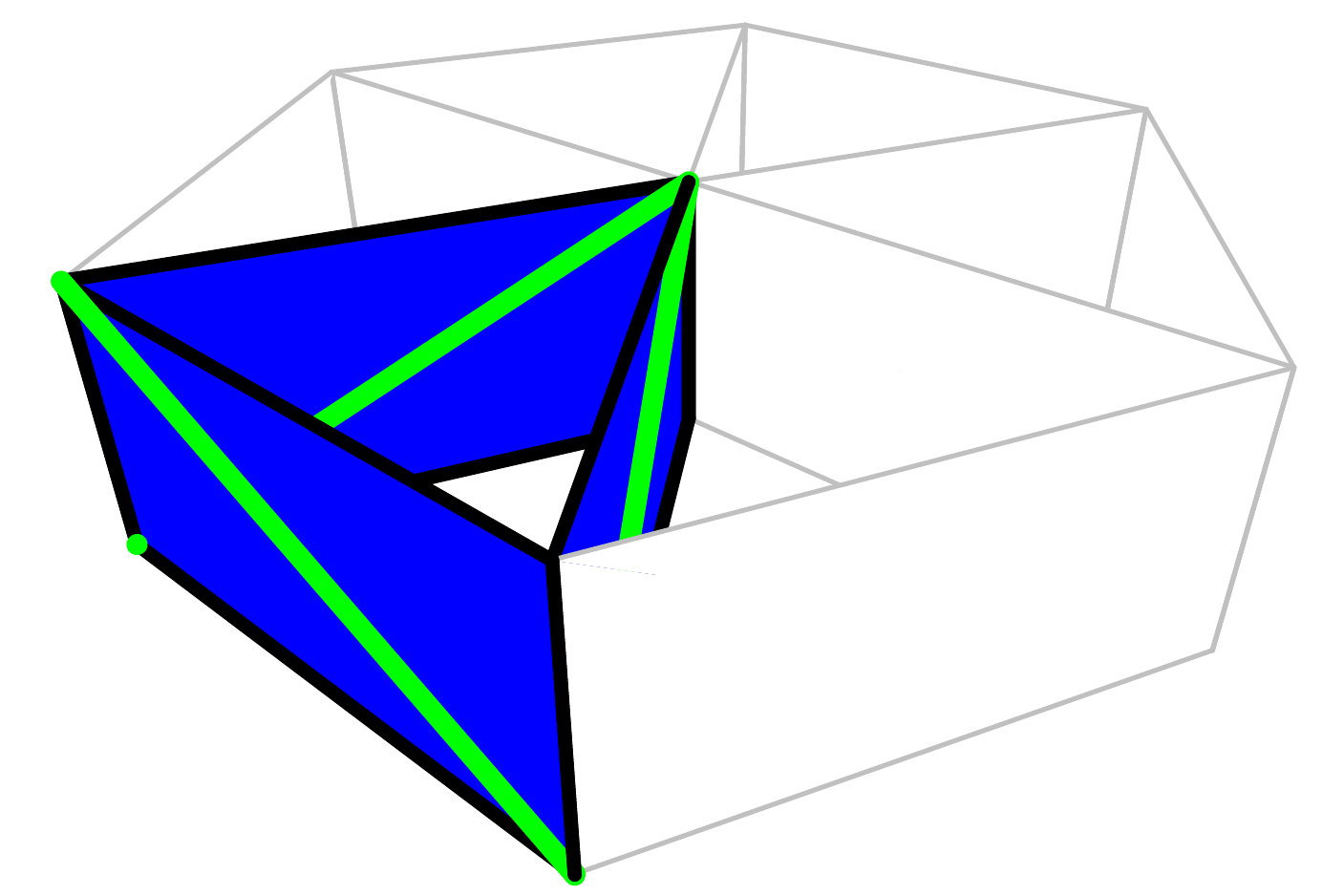}
      \\
      $(uDu)$
    \end{minipage}
    \hfill
    \begin{minipage}{0.31\linewidth}
      \centering
      \includegraphics[width=\textwidth,viewport=70 100 475 370,clip]{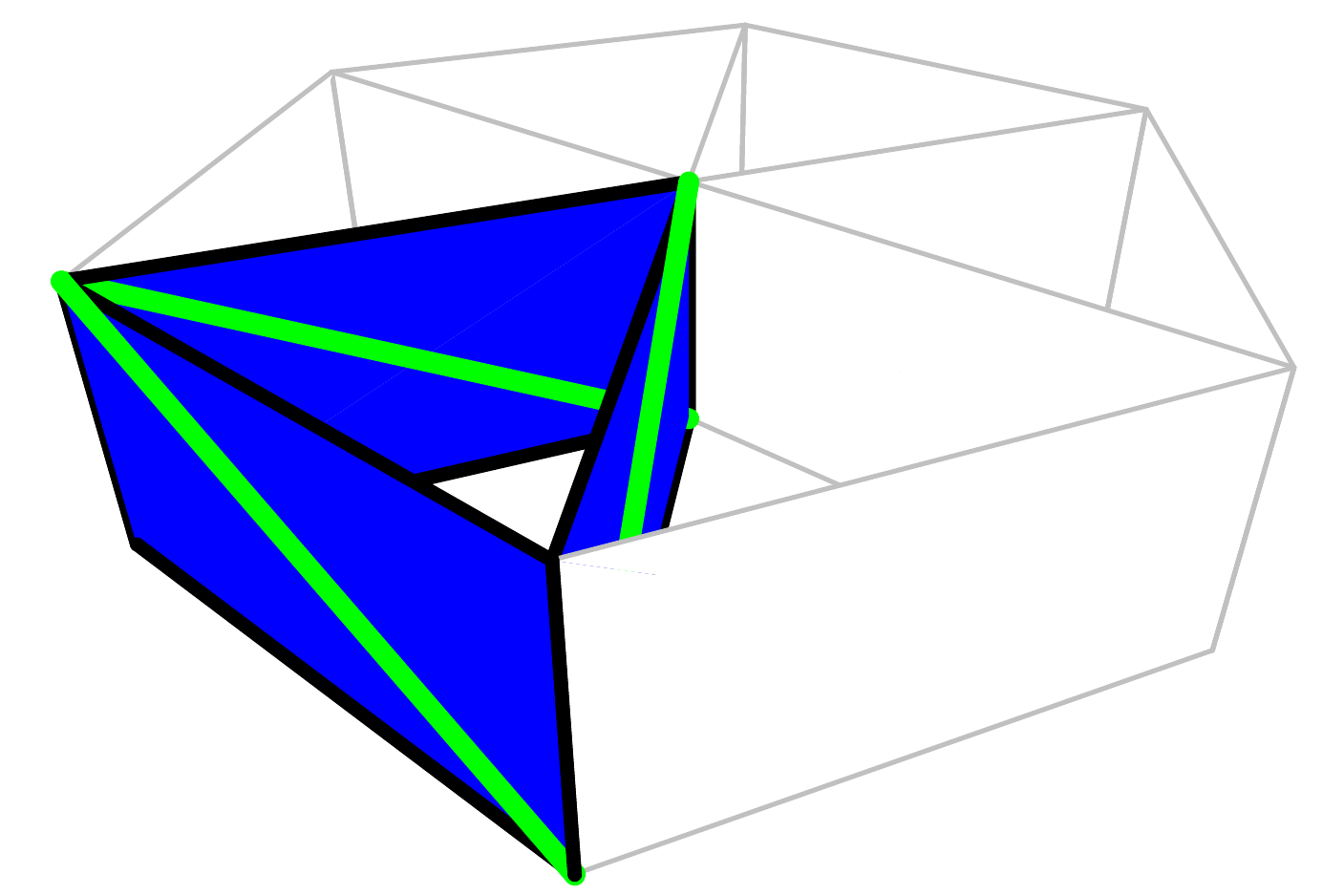}
      \\
      $(dDu)$
    \end{minipage}\\[17pt]
  \end{minipage}
}
  \parbox{4.5in}{\caption{\label{fig:triangulation}\small\it 
The $6$ different triangulations of a wedge inside a three-face of $\D$, which is a fundamental region for the triangulation.}}
\end{figure}
\fref{fig:triangulation}. We will label these choices as $(uUu)$,
$\dots$, $(dDu)$ corresponding to whether the three lines go \textbf{u}p
or \textbf{d}own.
When assembling the triangulated wedges into the hexagonal fundamental
region, we must ensure that the triangulations match along the
two-faces of the wedges. Therefore, the whole triangulation can be
written as a cyclic string of six wedge-triangulations such that only
two $u$'s or two $d$'s meet, that is, \framebox{$\cdots u)(u\cdots$} or
\framebox{$\cdots d)(d\cdots$}. This ensures a consistent
triangulation in the interior of the fundamental group. The boundary
of the fundamental region will intersect the boundaries of some of its
$\goth$-images. 
compatibility of the triangulations along the outward-facing
rectangular two-faces requires that the $i$-th and the $(i+3)$-rd
wedge have the rectangular two-face cut in the same way, both `$U$'
or both `$D$'. Up to symmetries\footnote{The symmetries of the
  fundamental region are $\text{Dih}_6$ transformations together with a reflection in a
  horizontal plane.} of the fundamental region, there are $6$ distinct triangulations of a three-face. These are shown in \tref{tab:triangulations}.
\begin{table}
\def\str{\vrule height15pt width0pt depth8pt}
\centering
  \begin{tabular}{| @{\quad\big\{}c@{, }c@{, }c@{, }c@{, }c@{, }c@{\big\}\quad}|c|}
    \hline
    \multicolumn{6}{|c|}{\str Triangulation} & Regular \\ \hline\hline
    \str $(dDu)$ & $(uUd)$ & $(dDu)$ & $(uDu)$ & $(uUd)$ & $(dDd)$ & No \\ 
    \str $(dDu)$ & $(uUd)$ & $(dDd)$ & $(dDu)$ & $(uUd)$ & $(dDd)$ & Yes  \\
    \str $(dDu)$ & $(uUd)$ & $(dDd)$ & $(dDd)$ & $(uUu)$ & $(dDd)$ & No  \\
    \str $(dDu)$ & $(uDu)$ & $(uUd)$ & $(dDd)$ & $(dDd)$ & $(uUu)$ & No  \\
    \str $(dDd)$ & $(dDd)$ & $(dDd)$ & $(dDd)$ & $(dDd)$ & $(dDd)$ & No  \\
    \str $(dDd)$ & $(dDd)$ & $(uUu)$ & $(dDd)$ & $(dDd)$ & $(uUu)$ & No  \\ \hline
  \end{tabular}
  \parbox{5in}{\caption{\label{tab:triangulations}\small\it 
      Distinct choices for the triangulation of one hexagonal
      three-face, the fundamental region for the $\goth$-action. The last column states
      whether the ensuing $\goth$-invariant triangulation of $\D$ is regular.}}
\end{table}
It is interesting that precisely one of these
yields a \emph{regular} $\goth$-invariant triangulation of
$\D$. Recall that a regular triangulation is one that is induced
by the `crease lines' of the graph of a convex support function, a
condition that is equivalent to the toric variety being K\"ahler. We
will always use the regular triangulation in the following.
It is interesting also that the regular triangulation is not the $\euf{H}$-invariant triangulation $(dDd)(dDd)(dDd)\ldots(dDd)$. Thus we learn that the mirror manifold is not $\text{Dih}_6$-invariant. Note, however, that the the regular triangulation $(dDu)(uUd)(dDd)(dDu)(uUd)(dDd)$ repeats with period 3. The upper-case characters that correspond to the triangulations of the two-faces of $\D$ are required to repeat with period three, in virtue of our observations above, but the lower-case characters are not constrained by this requirement. We learn that the mirror remains invariant under the symmetry $h_2$.

Note that $\D$ has $49$ points, none of which lie interior to a three-face. These yield $h^{11}=44$ divisor
classes after deleting the origin and modding out the $4$ linear relations between the
points. We have seen that there is a $\goth$-action on $\D$ and there is a corresponding 
$\goth$-action on the divisors. Since the group action on the toric
hypersurface is free, the coinvariant divisors form a basis for the
divisors on the quotient. In other words, one has to identify the
divisors on the covering space with their $\goth$-images. There are
$4$ orbits, and, therefore, $4\,(=h^{11}(\YG^*))$ linearly independent divisors on the
quotient. 

Consider further the $\goth$-action on the points of $\D$. These fall into five orbits which we denote by $\D_i$,
$0\leq i\leq 4$.  
The origin of $\D$, which forms an orbit of length one, we denote by $\D_0$. There remain the four orbits mentioned above. One of these, which we choose to be $\D_1$, consists of the
$12$ points of $\D$ that are internal to two-faces. These are the
points $\io_a = (v_a,\,0)$ and $\tilde{\io}_b = (0,\, v_b)$. The
remaining three orbits consist of the vertices $\n_{a b} = (v_a,\,v_b)$ 
which fall into three orbits according to the value of
$a+b$ mod 3. We take $\D_2$, $\D_3$ and $\D_4$ to consist of the
vertices $\n_{a b}$ such that $a+b$ mod $3$ takes the values $0$, $1$
and $2$, respectively. We abuse notation by identifying divisors on
the quotient with the corresponding orbits of vertices. 

As divisors we have a relation 
\begin{equation*}
\D_0~=~-\sum_{i=1}^4 \D_i~.
\end{equation*}
The calculation that finds the convex piecewise linear function that determines the regularity of the triangulation yields also the generators of the the Mori cone (the dual to the K\"ahler cone). These are given by 
\begin{equation*}
\def\str{\vrule height14pt depth6pt width0pt}
  \begin{array}{c@{\;=\big(\;}r@{,\;{}}r@{,\;{}}r@{,\;{}}r@{,\;{}}r@{\;\big)}}
    \multicolumn{1}{c}{\str} &
    \multicolumn{1}{c}{\D_0\hskip-6pt{}} &
    \multicolumn{1}{c}{\D_1\hskip-6pt{}} &
    \multicolumn{1}{c}{\D_2\hskip-6pt{}} &
    \multicolumn{1}{c}{\D_3\hskip-6pt{}} &
    \multicolumn{1}{c}{\D_4\hskip-6pt{}} 
    \\ \hline
    \str\ell_1 & 
    -1 & 0 & 0 &-1 & 2
    \\
    \str\ell_2 & 
    0 & -1 & 0 & 1 & 0
    \\
    \str\ell_3 & 
    0 & 2 & -1 & -1 & 0
    \\
    \str\ell_4 & 
    0 & -1 & 2 & 0 & -1\rlap{\hskip12pt.}
  \end{array}
\end{equation*}
\subsection{The mirror manifold}

We have previously observed that $\nabla$ is contained in $\D$.
As a result, the mirror family is defined by a polynomial which is a
specialisation of $r$, given by specialising the coefficients.  We will
denote this specialisation by $r^*$.  The equation $r^*=0$ defines
a singular variety which, generically, has 72 nodes, and the mirror $Y^*$ is
obtained as its resolution.  In \sref{sec:transgression}, in order 
to construct the $(2,2)$ manifold, we resolved the 36-nodal varieties
by finding a suitable set of divisors and blowing up along these.  In this
way we demonstrated the existence of a $\goth$-invariant, Calabi-Yau
resolution.  In the present case we do not know of suitable divisors,
however we may now avail ourselves of the techniques of toric geometry.
We have just seen that there exists a maximal triangulation of $\D$
which is regular and $\goth$-invariant, and this provides the
$\goth$-invariant, Calabi-Yau resolution of the 72-nodal varieties
corresponding~to~$r^*=0$.

Let us therefore consider the form of the polynomial $r$. The integral
points of $\D$ correspond to monomials on the embedded torus,
and $r$, restricted to the torus, is a linear combination of these.   A
four-parameter family of $\goth$-invariant Laurent polynomials is
obtained by writing
\begin{equation*}
f~=~\sum_{i=0}^4 \g_i f_i~~~\text{where}~~~f_i~=~\sum_{u\in \D_i}t^u~.
\end{equation*}
Being invariant, this family of Laurent polynomials must be equivalent to the family from \eqref{eq:finalpolys} and the $\g_i$ must be another system of coordinates on the parameter space and so expressible in terms of the $c_i$.  To determine the relations\footnote{Of course there is a scaling ambiguity in $r$, so in fact the relations between the $\g_i$ and $c_i$ are only determined up to scale}, we set $r=f$ on the embedded torus given in \eqref{eq:dP6torus}.  In this way we find the following correspondence
\begin{alignat}{3}
\g_0&= c_2 + c_3 + c_4              &c_0=\;&3\big( \z\g_2 + \z^2\g_3 + \g_4 \big)
\notag\\[5pt]
\g_1&=\frac{1}{12} \big( 4 c_2 - 2 c_3 + c_4\big) & c_1=\;&3\big( \z^2\g_2 + \z\g_3 + \g_4 \big)
\notag\\[5pt]
\g_2&=\frac{1}{36} \big(4 \z^2  c_0+4 \z c_1 + (4 c_2 + c_3 - 2 c_4)\big) \hskip25pt
                            & c_2=\;&\frac{1}{9}\big(\g_0 + 12\g_1 + 12(\g_2 + \g_3 + \g_4)\big)
\label{eq:CoeffConversion}\\[5pt]
\g_3&=\frac{1}{36} \big(4 \z c_0+4 \z^2  c_1 + (4 c_2 + c_3 - 2 c_4)\big) 
                            &c_3=\;&\frac{4}{9}\big(\g_0 - 6\g_1 + 3(\g_2 + \g_3 + \g_4)\big)
\notag\\[5pt]
\g_4&=\frac{1}{36} \big(4 c_0 + 4 c_1 + (4 c_2 + c_3 - 2 c_4)\big) 
                            &c_4=\;&\frac{4}{9}\big(\g_0 +3\g_1 - 6(\g_2 + \g_3 + \g_4)\big)~.\notag
\end{alignat}
Now $\nabla$ is obtained by deleting the vertices of  $\D$. Thus 
$\nabla=\D_0\cup\D_1$ and the polynomial $r^*$ corresponds to setting $\g_2=\g_3=\g_4=0$. In virtue of the relations above we see that this is equivalent to the conditions
\beq
c_0~=~c_1~=~0 ~~~\text{and}~~~4 c_2 + c_3 - 2 c_4~=~0~.
\label{eq:cCoeffsMirror}\eeq
We learn that the parameter space of the mirror is contained as a curve in the parameter space of $\YG$ and, moreover, that this curve lies in $\G$ and is the curve $\G^{\text{(i)}}$ of 
\tref{tab:DiscriminantComponents}. 

\begin{figure}[b]
\begin{center}
\parbox{6.1in}{
\framebox[1.9in][c]{\vrule width0pt height 1.65in depth 0.1in
\raisebox{33pt}{\includegraphics[width=1.5in]{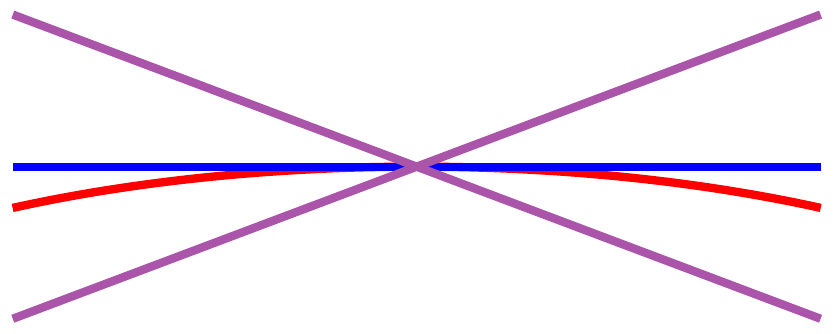}}}
\hskip10pt
\framebox[1.9in][c]{\vrule width0pt height 1.65in depth 0.1in\includegraphics[width=1.5in]{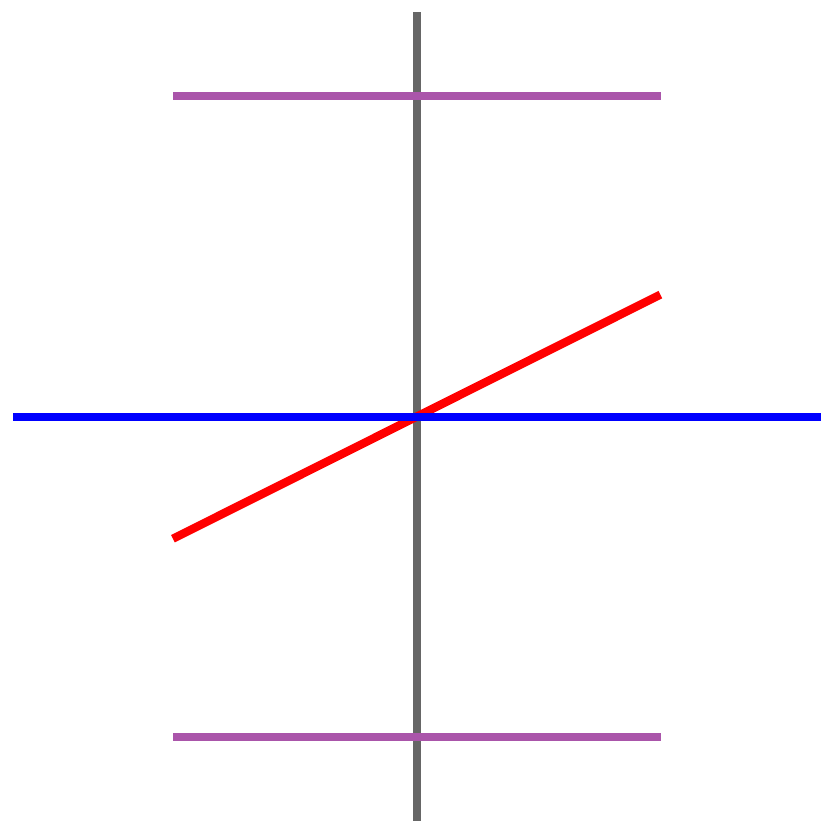}}\hskip14pt
\framebox[1.9in][c]{\vrule width0pt height 1.65in depth 0.1in\includegraphics[width=1.5in]{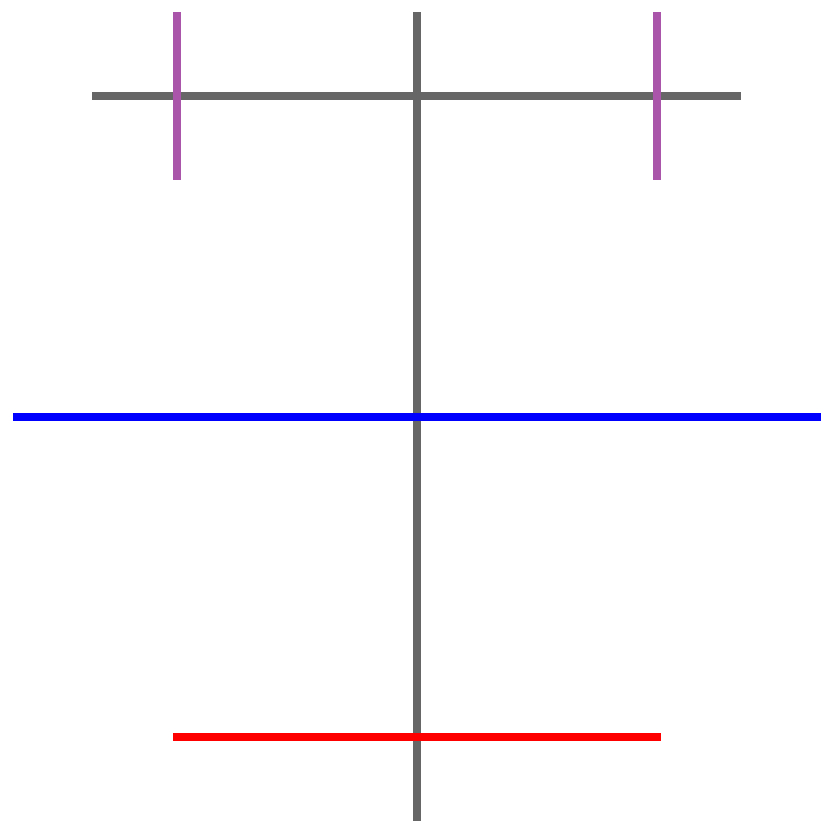}}
}
\vskip0pt
\place{3.29}{0.6}{\small $F_0$}
\place{5.39}{0.6}{\small $F_1$}
\place{5.5}{1.56}{\small $F_0$}
\parbox{5.2in}{\caption{\label{fig:LCSLresolutions}\small\it
The resolution of the point $(1,4,4)$ of \fref{fig:Dih6Parameters} requires a sequence of two blow ups which introduce the two exceptional divisors $F_0$ and~$F_1$.}}
\end{center}
\end{figure}

It is worth pursuing the forms of $f$ and $r^*$ a little further. From the relation
\begin{equation*}
f~=~\g_0 + \g_1 f_1~~\text{with}~~
f_1~=~ 
t_1+\frac{1}{t_1}+t_2+\frac{1}{t_2}+t_3+\frac{1}{t_3}+t_4+\frac{1}{t_4}
+\frac{t_1}{t_2}+\frac{t_2}{t_1}+\frac{t_3}{t_4}+\frac{t_4}{t_3}
\end{equation*}
it is compelling that the point corresponding to the large complex structure limit is the point $\g_1=0$, which in terms of the $c_j$ is $c_j = (0,0,1,4,4)$. As may be seen from \fref{fig:Dih6Parameters}, this is a point where the components of the discriminant locus have a high order contact so it may well be necessary to blow up this point, as in \fref{fig:LCSLresolutions}, in order to discuss the monodromies about the large complex structure limit adequately. In any event, we have come rather rapidly to an identification of this limit.
Returning to the polynomial $r^*$, we have, on imposing the conditions \eref{eq:cCoeffsMirror}
\begin{equation}\label{eq:f0}
r^*~=~\frac{1}{9} \g_0\, ( m_{2222} + 4m_{0011} + 4m_{0212} ) 
+ \frac{4}{3} \g_1\, ( m_{2222} - 2m_{0011} + m_{0212} )
\end{equation}
The polynomial varying with $\g_0$ factorises
\begin{equation*}
m_{2222} + 4m_{0011} + 4m_{0212}~=~s(x_1, x_3)\, s(x_2, x_4) ~~~\text{with}~~~
s(w,z)~=~w_0z_1 + w_1z_0 + w_2z_2~.
\end{equation*}
The part of $r$ that varies with $\g_1$ also simplifies
\begin{equation*}
m_{2222} - 2m_{0011} + m_{0212}~=~-\frac{1}{2}\, s(x_1, x_3)\, s(x_2, x_4) +
\frac{3}{4}\,\big( s(x_1, x_3)\, x_{22}\, x_{42} + x_{12}\, x_{32}\, s(x_2, x_4) \big)~.
\end{equation*}
The polynomial $s$ is a natural analogue of $p$ and $q$. The polynomial $p$ is a sum of monomials such that the coordinate indices sum to 0 mod 3 and the coordinate indices for the monomials of $q$ sum to 2 mod 3. For $s$ the coordinate indices sum to 1 mod 3. Consider now the locus $s=0$ in $\dP_6$ which corresponds to the locus $p=q=s=0$ in $\IP^2{\times}\IP^2$. At first sight one might be tempted to identify this as a CICY given by three bilinear equations in $\IP^2{\times}\IP^2$,
which is an elliptic curve. This however is a false conclusion owing to the fact that the intersection 
$p=q=s=0$ is not transverse. A little thought, and consultation with \tref{lines}, reveals that all six divisors 
$D_a$ lie in the locus $s=0$ and that this locus is precisely the hexagon formed by the $D_a$. We may think of this as a degenerate elliptic curve which has become a chain of six $\IP^1$'s. The hexagon less its vertices consists of the six one-dimensional orbits of the torus action and the vertices are the zero-dimensional orbits.  Therefore when $\g_1 = 0$, so that $r^* = \frac 19 s(x_1, x_3)\, s(x_2, x_4)$, we obtain a very singular variety which is invariant under the whole torus action.

The reader may be worried about an apparent contradiction between the fact that $f_0$ appears to be equal to  unity, and \eqref{eq:f0}, where $f_0 = \frac 19 s(x_1,x_3)\, s(x_2, x_4)$.  This is resolved by noticing that in writing $r = f$ only on the torus \eqref{eq:dP6torus}, it is implicit that $y_{\a 0} = 1$.  If we write $s$ in terms of the torus coordinates, we find $s(x_1,x_3) = s(x_2,x_4) =3$.  This is a result of the normalisation $y_{\a 0} = 1$.  It does not contradict the fact that $s$ vanishes on the hexagon since no point of the hexagon lies on the torus.  With $s=3$, we get $f_0 = \frac 19 s(x_1, x_3)\, s(x_2, x_4) = 1$, so there is, in fact, no contradiction.

For generic $\g_1/\g_0$, the variety described by this family of polynomials $r^*$ has 72 nodes which form six 
$\goth$-orbits. These comprise the 36 nodes that we have met previously and 36 nodes that are new and that are located at the points $\pt'_a{\times}\widetilde{\pt}'_b$, where $\pt'_a$ denotes the point on the hexagon corresponding to the intersection of the divisors $D_a$ and $D_{a+1}$, with $\widetilde{\pt}'_b$ understood analogously. The polyhedron $\D$, with its $\goth$-invariant triangulation provides a \cy resolution of these nodes. The resolution of 6 nodes on the quotient $\YG$ gives a manifold with $\chi=+6$ and Hodge numbers
$\hodgenos=(4,1)$, which we identify with the mirror, $\YG^*$,~of~$\YG$.

In \tref{tab:singmirr} we list the values of the parameter for which the variety, whose resolution is 
$\YG^*$, develops extra singularities. This occurs at the values of $\g_0/\g_1$ for which $\G^\text{(i)}$ intersects the other components of the discriminant. In each case where $\G^\text{(i)}$ intersects another component $\G'$, for which the generic singularity consists of the 3 nodes that $\G'$ has in common with 
$\G^\text{(i)}$ together with additional singularities,
the singularities associated with the intersection are the 3 nodes, together with the 3 extra nodes of 
$\G^\text{(i)}$ and the extra singularities of $\G'$.
\vskip3pt
\begin{table}[H]
\setlength{\doublerulesep}{1pt}
\def\str{\vrule height14pt depth8pt width0pt}
\begin{center}
\newcolumntype{C}{>{$\hskip1pt}c<{\hskip1pt$}}
\begin{tabular}{| C || C | C | C | C | C | C | C | C | C | C |}\hline
\str \text{Curve}& \text{(o)} &  \text{(ii)} &  \text{(iii)} &  \text{(iv)} &  \text{(v)} &  \text{(vi)} 
&  \text{(vii)} & \text{(viii)} &  \text{(ix)} &  \text{(x)} \\ 
\hline
\str \g_0/\g_1    & \hskip5pt\infty\hskip5pt{} & 4 & -12\+ & 6 & -3\+ & 5 & \infty & \infty & -4\+ & 4 \\
\hline
\str \text{Sing.} &  & 6,1\,g_4,1\,g_4^2 & 6,1\,\goth & 6,1\, g_3 & 6,1\, g_3 & 7 &   &   & 6, 1\, g_4^2 
&\hskip6pt  6 \hskip6pt{}\\
\hline
\end{tabular}
\vskip5pt
\parbox{6.2in}{\caption{\label{tab:singmirr}\small\it
The values of the parameter $\g_0/\g_1$ for which the mirror manifold is singular together with the type of singularity. The values given for $\g_0/\g_1$ are those corresponding to the intersection of\/ $\G^\text{(i)}$ with the other components of the discriminant locus listed in \tref{tab:DiscriminantComponents}. The entry
$(6,1\,g_4,1\,g_4^2)$, for example means 6 nodes, 1 $g_4$-node and 1 $g_4^2$-node. Where the intersection is at the large complex structure limit, for which $\g_0/\g_1=\infty$, the singularity is not listed.}}
\setlength{\doublerulesep}{3pt}
\end{center}
\end{table}

\newpage
\section{The Abelian Quotient} \label{sec:abelianquotient}
The manifold $Y^{8,44}$ also admits a free quotient by the Abelian group $\IZ_{12}$.  With the same conventions as used in \sref{sec:nonabelianquotient}, the action of the group generator is
\begin{equation*}
g_{12}:~x_{\a j}~\to~\z^j x_{\a+1,j}~,~~p^1\to \z^2 p^1~,~~p^2\to \z^2 p^2~,~~q^1\leftrightarrow \z q^1~,~~q^2\leftrightarrow \z q^2~,~~r~\to~r~.
\end{equation*}
The covariant polynomials are given by
\begin{equation*}
\begin{split}
p^1~&=~x_{10}\,x_{32} + x_{12}\,x_{30} + x_{11}\,x_{31}~,\qquad
q^1~=~x_{10}\,x_{31} + x_{11}\,x_{30} + x_{12}\,x_{32}~,\\[8pt]
p^2~&=~x_{20}\,x_{42} + x_{22}\,x_{40} + x_{21}\,x_{41}~,\qquad
q^2~=~x_{20}\,x_{41} + x_{21}\,x_{40} + x_{22}\,x_{42}~,\\[8pt]
&\hskip20pt r~=~ C_0\,m_{0000} + C_1\,m_{2211} + C_2\,m_{2121} + C_3\,m_{2010} + C_4\,m_{1110}~.
\end{split}\end{equation*}
It is straightforward to check that the corresponding variety is smooth, and that the induced action of $\IZ_{12}$ is fixed-point-free.  So we obtain another smooth quotient, this time with fundamental group $\IZ_{12}$.
\subsection{Group action on homology}
The representation theory of $\IZ_{12}$ is very straightforward:  there are exactly $12$ distinct one-dimensional representations, in which the generator of $\IZ_{12}$ corresponds to multiplication by one of the twelfth roots of unity.  We will denote by $R_k$ the representation in which the generator acts as multiplication by $\exp(2\p\ii k /12)$.  Then, repeating the type of argument used in \sref{sec:homology}, we find that $\IZ_{12}$ acts on $H^2(Y^{8,44})$ through the representation
\begin{equation*}
R_0 \oplus R_2 \oplus R_3 \oplus R_4 \oplus R_6 \oplus R_8 \oplus R_9 \oplus R_{10}
\end{equation*}
There is again a single invariant, corresponding to the canonical class, so the Hodge numbers of the quotient are once more $(h^{11}, h^{21}) = (1,4)$.
\subsectionhyp
[$\IZ_{12}$ flux lines]
[Z/12 flux lines]
{\boldmath $\IZ_{12}$ flux lines}
Notice that the argument of \sref{sec:gaugedeform}, that it is
impossible to deform $\cT \oplus \cO \oplus \cO$ to a non-split
$SU(5)$ bundle, depends only on the fact that $h^{11} = 1$ and,
therefore, applies to this case as well.  Therefore we must again turn
to the Hosotani mechanism.
Since $\IZ_{12}$ is Abelian, the best the Hosotani mechanism can do in this case is reduce the $E_6$ gauge symmetry to $SU(3){\otimes}SU(2){\otimes}U(1)^3$.  There are a very large number of choices which can achieve this, and these will generally give different light spectra and interactions.  While there may be promising models here, their identification would require a detailed study which is beyond the scope of this paper.
\section*{Acknowledgements}
\vskip-10pt
It is a pleasure to acknowledge instructive discussions with X.\ de la Ossa, M.\ Gross and A.\ Lukas. We are grateful also to B.\ Szendr\H{o}i for instruction in the resolution of Calabi-Yau varieties with multiple nodes. 
RD wishes to acknowledge support from the Sir Arthur Sims Travelling Scholarship Fund and the University College Old Members'-Oxford Australia Scholarship Fund and VB would like to acknowledge the hospitality of Oxford University, of the KITP, and of the Centro de Ciencias de Benasque where part of this work was
performed.
\vskip20pt
\appendix
\section{Alternative Representations}
\vskip-10pt
We have discussed the fact that the manifold $Y^{8,44}$ can be viewed as a hypersurface in 
$\cS{\times}\cS$ embedded via a section of the anticanonical bundle. The same manifold can be represented as a CICY in different ways owing to the fact that there are alternative ways of representing $\cS\cong\dP_6$. In addition to the representation \eref{Srepone} the following representation is useful
\begin{equation}
  \cS~=~\cicy{\IP^1 \\ \IP^1 \\ \IP^1}{1\\ 1\\ 1}\cicystop
  \label{Sreptwo}
\end{equation}
The identification follows from the fact that the configuration on the
right has Euler number 6 and positive canonical class. In this way we
see that the following configuration is an alternative way of
representing the manifold $Y^{8,44}$, namely
\begin{equation*}
  X^{8,44} ~=~~ 
  \cicy{\IP^1\\ \IP^1\\ \IP^1\\ \IP^1\\ \IP^1\\ \IP^1\\}
  {\one&0&\one \\
    \one&0&\one \\
    \one&0&\one \\
    0&\one&\one \\
    0&\one&\one \\
    0&\one&\one}^{8,44}
\hskip20pt
  \raise-5pt\hbox{\parbox{5cm}{
        \hbox{\hphantom{$p\;$}\includegraphics[width=2.5in]{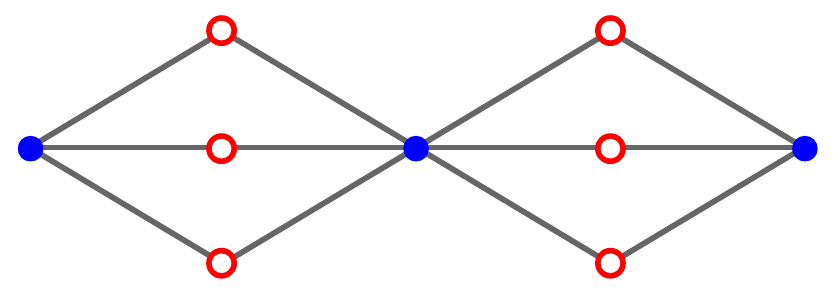}}
        \place{0}{0.6}{\small $p$}
        \place{1.33}{0.75}{\small $r$}
        \place{2.70}{0.6}{\small $q$}
        \place{0.75}{0.1}{\small $s_j$}
        \place{1.95}{0.1}{\small $t_j$}
      }}
\end{equation*}
There is also a ``hybrid'' representation, $Z^{8,44}$, that is formed by taking one copy of $\cS$ to be represented as in \eref{Srepone} and the other as in \eref{Sreptwo}. This representation is however less useful owing to the fact that the groups $\goth$ are not represented linearly on the coordinates. The three representations $X^{8,44}$, $Y^{8,44}$ and $Z^{8,44}$ are the three occurrences of the Hodge numbers
$\hodgenos=(8,44)$ in the CICY list.
 
The configuration $X^{8,44}$ is described, together with the common
$\IZ_6$ subgroup of $\text{Dic}_3$ and $\IZ_{12}$,
in~\cite[\SS{}3.5.2]{SHN}. We can therefore give a brief account here
that concentrates on the action of the enlarged groups.  We take
coordinates $s_{ja}$ on the first three $\IP^1$ and $t_{ja}$ on the
last three, where $j = 0,1,2$, $a = 0,1$, and polynomials $p$, $q$,
and $r$, as indicated by the diagram.  We can then take the action of
the two generators $g_3, g_4$ to be given by
\begin{alignat*}{5}
g_3: &~ s_{j,a} \to s_{j+1,a}~,~~~ && t_{j,a} \to t_{j+1,a}~,~~~            && p \to p~,~~~   
&& q \to q~,~~~ && r \to r~,\\[5pt]
g_4: &~ s_{j,a} \to (-1)^{a+1}\, t_{-j,a}~,~~~    && t_{j,a} \to s_{-j,a}~,~~~~  && p \to -q~,~~~  
&& q \to p~,~~~ && r \to r~.
\end{alignat*}
Note that the two generators $S$ and $U$ of~\cite[\SS{}3.5.2]{SHN}
correspond to the symmetries $g_3$ and $g_4^2$.
We construct the polynomials from the quantities
\begin{equation*}
\begin{split}
& m_{a b c} = \sum_{i=0}^2 s_{i, a}\, s_{i+1, b}\, s_{i+2, c}~, \qquad
n_{a b c} = \sum_{i=0}^2 t_{i, a}\, t_{i+1, b}\, t_{i+2, c} \\
&\hskip30pt l_{a b c d e f} = \sum_{i=0}^2 s_{i, a}\, s_{i+1, b}\, s_{i+2, c} \,
 t_{i, d}\, t_{i+1, e}\, t_{i+2, f}
\end{split}
\end{equation*}
Then, with an appropriate choice of coordinates, the most general polynomials transforming as above are
\begin{gather*}
p ~=~ \frac 13 m_{000} + m_{110}~, \qquad q ~=~ \frac 13 n_{000} + n_{110} \\[10pt]
r ~=~ \tilde c_0\,l_{111111} + \tilde c_1\,l_{001001} +  \tilde c_2\, l_{001010} + \tilde c_3\, l_{001100} + 
\tilde c_4\,(l_{111001} + l_{001111})
\end{gather*}
Note that a further invariant term $\ell_{000011} - \ell_{011000}$ is excluded since
\begin{equation*}
\ell_{000011} - \ell_{011000}~=~m_{000}\, n_{011} - m_{011}\, n_{000}~=~p\, n_{011} - m_{011}\, q
\end{equation*}
and so corresponds merely to a redefinition of $r$ by multiples of the polynomials $p$ and $q$. This being so, we see that there is a 4 parameter family of polynomials $r$.
It is straightforward to check that for generic values of the four undetermined coefficients, the resulting variety is smooth, and the group action is free.
\vskip20pt
\section{Deformation of the Tangent Bundle} \label{sec:deformations}
\vskip-10pt
We will show here that the tangent bundle of the manifold $X^{8,44}$ is rigid, that is, it has no infinitesimal deformations. This is a consequence of the fact that all the defining polynomials are multilinear in their arguments. The rigidity holds already at the level of the covering manifold. We do not have to invoke the group
$\goth$ to establish the result.
Let a CICY manifold $X$ be defined by polynomials $f^\s$, $\s=1,2,\ldots$ and let the embedding spaces have coordinates $x_{\a j}$ where $\a$ labels the projective space and $j$ runs over the coordinates within each projective space. A vector is a quantity
\begin{equation*}
V~=~\sum_{\a,\,j}V_{\a j}\,\pd{}{x_{\a j}}
\end{equation*}
and is tangent to $X$ if 
\begin{equation}
V(f^\s)~=~\sum_{\a,\,j}V_{\a j}\,\pd{f^\s}{x_{\a j}}
\label{tangvec}\end{equation}
vanishes on $X$ for each $\s$. In virtue of the Euler relation 
\begin{equation*}
\sum_j x_{\a j} \pd{f^\s}{x_{\a j}}~=~\deg_\a(\s)\, f^\s
\end{equation*}
where $\deg_\a(\s)$ denotes the homogeneity degree of $f^\s$ as a function of the $x_{\a j}$, it is consistent to identify
\begin{equation}
V_{\a j}~\simeq~V_{\a j} + \l_\a\, x_{\a j}
\label{vecident}\end{equation}
for each $\a$ and all $\l_\a$.
The condition for tangency \eref{tangvec} is deformed by imposing instead
\begin{equation}
\sum_{\a,\,j}V_{\a j}\,F^\s_{\a j}~=~0~~~\text{with}~~~
F^\s_{\a j}~=~\pd{f^\s}{x_{\a j}} + \tilde{f}^\s_{\a j}
\label{vecdefs}\end{equation}
a set of polynomials of the same multidegrees as $\pd{f^\s}{x_{\a j}}$. In order to maintain consistency with \eref{vecident} we must require that $\sum_j x_{\a j}\tilde{f}^\s_{\a j}$ vanish on $X$, for each $\a$. It is worth examining the meaning of requiring this quantity to vanish on $X$ more closely. In order for this quantity to vanish on $X$ we must have
\begin{equation*}
\sum_j x_{\a j}\,\tilde{f}^\s_{\a j}~=~\deg_\a(\s)\sum_\t m^\s{}_\t\, f^\t
\end{equation*}
for some matrix of polynomials $m^\s{}_\t$ of appropriate degrees. Now let us write
\begin{equation*}
\tilde{f}^\s_{\a j}~=~\hat{f}^\s_{\a j}+\pd{}{x_{\a j}}\sum_\t m^\s{}_\t\, f^\t
\end{equation*}
so that $\sum_j x_{\a j}\hat{f}^\s_{\a j}=0$ identically, that is these quantities vanish identically as polynomials. On substituting the decomposition above into \eref{vecdefs} we find
\begin{equation*}
F^\s_{\a j}~=~\pd{}{x_{\a j}}\left(f^\s + \sum_\t m^\s{}_\t\, f^\t \right) + \hat{f}^\s_{\a j}
\end{equation*}
and the quantity in parentheses corresponds to a redefinition of the $f^\s$. The burden of these comments is that in writing the deformation in the form \eref{vecdefs} we may, without loss of generality, demand that
$\sum_j x_{\a j}\tilde{f}^\s_{\a j}$ vanish identically. 
We now apply these considerations to the case that the polynomials $f^\s$ are all multilinear. When this is so each
deformation $\tilde{f}^\s_{\a j}$, having the same multidegrees as the derivatives 
$\partial f^\s/\partial x_{\a j}$, is independent of the coordinates $x_{\a k}$. Thus the condition 
\begin{equation*}
\sum_j x_{\a j}\tilde{f}^\s_{\a j}~=~0
\end{equation*}
forces the $\tilde{f}^\s_{\a j}$ to vanish identically.
Since this fact follows simply from the multilinearity of the configuration the same conclusion holds for the
alternative representations $X^{8,44}$ and $Z^{8,44}$ as well as for the various extended representations.
\newpage
\renewcommand{\baselinestretch}{1.05}\normalsize
\bibliographystyle{hunsrt}
\bibliography{references}
\end{document}